\documentclass[usenatbib]{mn2e}
\usepackage{epsfig,lscape}

\newcommand{\etal}{et~al.}

\newcommand{\kms}{$\mbox{km~s}^{-1}$ }
\newcommand{\kmsns}{$\mbox{km~s}^{-1}$}

\newcommand{\vsfig}[2]           
{
  \begin{center}
    \begin{minipage}[t]{0.05\textwidth}
      {\footnotesize \raisebox{40mm}{(#2)}}
    \end{minipage}
    \begin{minipage}[t]{0.42\textwidth}
      \psfig{file=./#1.ps,height=0.95\textwidth,angle=0}
    \end{minipage}
    \hfill
  \end{center}
}

\newcommand{\specdfig}[2]        
{
   \begin{center}
     \begin{minipage}[t]{0.45\textwidth}
         \psfig{file=eps/#1.eps,width=0.35\textwidth,width=0.8\textwidth,angle=270}
     \end{minipage}
     \hfill
     \begin{minipage}[t]{0.45\textwidth}
         \psfig{file=eps/#2.eps,width=0.35\textwidth,width=0.8\textwidth,angle=270}
     \end{minipage}
   \end{center}
}

\newcommand{\specsfig}[1]        
{
   \begin{center}
     \begin{minipage}[t]{0.45\textwidth}
         \psfig{file=eps/#1.eps,height=0.65\textwidth,width=1\textwidth,angle=0}
     \end{minipage}
   \end{center}
}

\newcommand{\boxfig}[1]        
{
   \begin{center}
     \begin{minipage}[t]{0.46\textwidth}
         \psfig{file=eps/#1.eps,height=0.45\textwidth,angle=0}
     \end{minipage}
   \end{center}
}

\newcommand{\twofig}[2]        
{
   \begin{center}
     \begin{minipage}[t]{0.5\textwidth}
         \psfig{file=eps/#1.eps,height=0.95\textwidth}
     \end{minipage}
     \hfill
     \begin{minipage}[t]{0.5\textwidth}
         \psfig{file=eps/#2.eps,height=0.95\textwidth}
     \end{minipage}
   \end{center}
}

\begin{document}

\title[12.2-GHz methanol maser catalogue: 330 to 10$^{\circ}$ longitude]{12.2-GHz methanol maser MMB follow-up catalogue - I. Longitude range 330$^{\circ}$ to 10$^{\circ}$}
\author[S.\ L.\ Breen \etal]{S.\ L. Breen,$^{1,2}$\thanks{Email: Shari.Breen@csiro.au} S.\ P. Ellingsen,$^2$ J.\ L. Caswell,$^1$ J.\ A.\ Green,$^1$ M.\ A.\ Voronkov,$^1$ \newauthor G.\ A.\ Fuller,$^3$  L.\ J.\ Quinn,$^3$ A.\ Avison$^3$\\
 \\
  $^1$ CSIRO Astronomy and Space Science, Australia Telescope National Facility, PO Box 76, Epping, NSW 1710, Australia;\\
  $^2$ School of Mathematics and Physics, University of Tasmania, Private Bag 37, Hobart, Tasmania 7001, Australia;\\
  $^3$ Jodrell Bank Centre for Astrophysics, Alan Turing Building, School of Physics and Astronomy, University of Manchester,\\ Manchester M13 9PL, UK}
 
 \maketitle
  
 \begin{abstract}
We present a catalogue of 12.2-GHz methanol masers detected towards 6.7-GHz methanol masers observed in the unbiased Methanol Multibeam (MMB) survey in the longitude range 330$^{\circ}$ (through 360$^{\circ}$) to 10$^{\circ}$. This is the first portion of the catalogue which, when complete, will encompass all of the MMB detections. We report the detection of 184 12.2-GHz sources towards 400 6.7-GHz methanol maser targets, equating to a detection rate of 46 per cent. Of the 184 12.2-GHz detections, 117 are reported here for the first time. We draw attention to a number of `special' sources, particularly those with emission at 12.2-GHz stronger than their 6.7-GHz counterpart and conclude that these unusual sources are not associated with a specific evolutionary stage. 


\end{abstract}


\section{Introduction}

 The 6.7- and 12.2-GHz methanol maser transitions are two of the strongest detected towards star formation regions and both are known to
trace an early evolutionary stage of high-mass star formation \citep{Ellingsen06}. Complementary observations of these two strong methanol maser
transitions are especially useful in probing the physical conditions
of the environments in which they arise \citep[see e.g.][]{Cragg01,Sutton01}, as they are 
typically found to be co-spatial to within a few milliarcseconds \citep{Mos02}. This is especially likely where the spectra at the two transitions are similar \citep{Norris93}. 

 \citet{Cragg05} showed, through maser modelling, that the physical conditions required to produce luminous class II methanol masers in the 6.7- and 12.2-GHz transitions are similar. While there is a large amount of overlap in the conditions for the two transitions, they do not cover exactly the same regions of parameter space. Given that there are a large fraction of 6.7-GHz methanol masers devoid of 12.2-GHz methanol maser counterparts \citep[e.g.][]{Caswell95b,Breen10a}, the physical conditions in these regions must be close to the point where the 12.2-GHz masers switch on or off. Due to this, important insights can be gained if observations of the two transitions are made near-simultaneously.
 
 A number of previous searches for 12.2-GHz methanol maser emission have been targeted towards sites of known 6.7-GHz methanol masers \citep[e.g.][]{Gay,Caswell95b,Blas04,Breen10a}. These searches have had varying detection rates (between 19 and 60 per cent), partially attributable to the varying sensitivities, but also due to the biases that had been introduced in the target source lists \citep{Breen12stats}.
 
 Methanol maser searches such as those by \citet{Gay,Caswell95b,Blas04,Breen10a} have shown
that the flux density of 12.2-GHz methanol maser emission is only rarely larger than the
associated 6.7-GHz methanol maser emission. This, coupled with the fact
that there have been no serendipitous detections of 12.2-GHz masers
without a 6.7-GHz counterpart, means that it is unlikely that an
unbiased Galactic survey would uncover many, {\em if any}, 12.2-GHz methanol maser sites that had no detectable 6.7 GHz methanol maser emission. Due to this, a complete sample of 12.2-GHz methanol masers can be gained by targeting a complete sample of 6.7-GHz methanol masers. Therefore, the 6.7-GHz methanol masers detected in the Methanol Multibeam (MMB) survey represent a unique opportunity to undertake a definitive search for 12.2-GHz masers within the Galactic plane.

The MMB survey aims to search the entire Galactic plane for 6.7-GHz methanol masers within latitudes of $\pm$2$^{\circ}$ \citep{Green09}. The southern hemisphere component of the survey has recently been completed and is producing survey results \citep[e.g.][]{CasMMB10,GreenMMB10,CasMMB102}. Since the 6.7-GHz  transition of methanol exclusively traces sites of high-mass star formation \citep[e.g][]{Minier03,Xu08}, the MMB survey is the most sensitive large survey yet undertaken for young high-mass stars forming within the Galactic Plane.

We have targeted all of the  southern MMB 6.7-GHz methanol masers for the presence of 12.2-GHz methanol maser counterparts. This 12.2-GHz search represents the largest, statistically robust sample of these sources; allowing us to address the role of the biases present in previous samples. Recently, \citet{Breen12stats} presented a statistical analysis of the 12.2-GHz methanol maser properties associated with MMB sources that lie south of declination --20$^{\circ}$ (Galactic longitudes $\sim$254$^{\circ}$ (through 360$^{\circ}$) to $\sim$10$^{\circ}$). Here we present the first portion of the 12.2-GHz catalogue used in the analysis of \citet{Breen12stats}, covering a longitude range of 330$^{\circ}$ (through 360$^{\circ}$) to 10$^{\circ}$. In addition to the data used in the analysis of \citet{Breen12stats}, we include a third epoch of data (2010 March). The presented longitude range is wholly included in the MMB catalogues released to date \citep{CasMMB10,GreenMMB10,CasMMB102}. The remaining 12.2-GHz sources that we have observed are to be published following the publication of the MMB targets. In addition to offering a large catalogue of 12.2-GHz methanol masers, this publication allows us to draw attention to specific sources that show intriguing properties.

\section{Observations and data reduction}

12.2-GHz methanol maser observations towards MMB sources in the longitude range 330$^{\circ}$ (through 360$^{\circ}$) to 10$^{\circ}$ degrees were carried out over three sessions; 2008 June 20-25, 2008 December 5-10 and 2010 March 19-23. An effort was made to carry out these 12.2-GHz follow-up observations as close as practicable to the final 6.7-GHz MMB methanol maser spectrum (which were mainly taken during 2008 March and August and 2009 March). This is particularly important given the intrinsically variable nature of masers and allows us to meaningfully compare the source properties.


A detailed account of the observing strategy and parameters is given in \citet{Breen12stats} and we give only a summary of the important specifications here. The 12.2-GHz observations were made with the Parkes 64-m radio telescope using the Ku-band receiver. This receiver detected two orthogonal linear polarisations and had typical system equivalent flux densities of 205 and 225~Jy for the respective polarisations throughout the observations in 2008 June, and slightly higher at 220 and 240~Jy during 2008 December and 2010 March. The Parkes multibeam correlator was configured to record 8192 channels over 16-MHz for each of the recorded linearly polarised signals. This configuration yielded a usable velocity coverage of $\sim$290 \kms and a spectral resolution of 0.08 \kmsns, after Hanning smoothing which was applied during the data processing. The Parkes radio telescope has rms pointing errors of $\sim$10 arcsec and at 12.2-GHz the telescope has a half power beam width of 1.9 arcmin. Flux density calibration is with respect to observations of PKS B1934--638, which has an assumed flux density of 1.825 Jy at 12178-MHz \citep{Sault03}, and were carried out daily. From these observations we are able to conclude that the system was extremely stable over each observing run (a period of several days) and we expect our flux density measurements to be accurate to $\sim$10 per cent.


Sources were observed at a fixed frequency of 12178~MHz which alleviated the requirement for a unique reference spectrum to be obtained for each of the sources. Instead, all spectra collected were combined to form a much more sensitive reference bandpass made up of the median value of each of the spectral channels. This strategy allows for the most efficient usage of the observing time, as well as reducing the noise contribution from the reference to the quotient spectrum. However, this method caused the introduction of baseline ripples which we have removed during processing by subtracting a running median  over 100 channels from the baseline. This method works extremely well for narrow lines such as masers, except when there is multiple strong features over a channel range comparable to the width of the running median. We have inspected each of the spectra given in Fig.~\ref{fig:12MMB} and believe that there is only one maser spectrum that has been evidently been distorted by our reduction method. This source, G\,9.621+0.196, shows strong emission over a large velocity, and has a peak flux density that is greater than the majority of sources. The distortion appears as small negative dips flanking the peak emission, appearing to mimic absorption (features that are not present in the earlier spectrum of \citet{Caswell95b}). Comparison between the spectra of other strong sources we detect with those of \citet{Caswell95b} allows us to be confident that no other spectra have been distorted in this way, and therefore do not believe that this reduction artefact is responsible for any absorption features in any further sources (i.e. all other absorption features are real). We do, however, note that subtracting a running median greatly reduces the chances of detecting absorption features since they are generally much broader and weaker, comparable to the width of the subtracted running median. Where absorption is detected, the amplitude of detected absorption features is likely to have been reduced. 

Each target was observed for at least 5 (but often 10 or more) minutes over the three epochs. The majority of sources achieving no detection in 5 minutes during the 2008 June observations were observed for a further 5 minutes in the 2008 December or 2010 March observations. The resultant 5-$\sigma$ detection limits, after the averaging of the two polarisations, chiefly lie in the range 0.55 to 0.80~Jy, comparable to the 5-$\sigma$ sensitivity of the MMB survey \citep[0.85~Jy;][]{Green09}. Data were reduced using the ATNF Spectral Analysis Package (ASAP). Alignment of velocity channels was carried out during processing, and all presented velocities are with respect to LSR. The adopted rest frequency was 12.178597 GHz \citep{Muller04}. Data from each epoch were first reduced individually and inspected for maser emission. Where multiple epochs of data were available, these epochs were subsequently averaged and inspected for maser emission, allowing us to identify additional weak sources. 

For one source, G\,353.410--0.360, no usable data was obtained during a single observation (2008 Dec) made with the Parkes radio telescope. Instead, we present data taken with the University of Tasmania Hobart 26-m radio telescope (during 2010 December), made using a cryogenically cooled receiver that detects both left and right circularly polarised signals and has a typical system equivalent flux density of $\sim$1200~Jy. The data were recorded using a 2-bit auto-correlation spectrometer configured to record 4096 channels per polarisation over a bandwidth of 8-MHz. This configuration yielded the same channel spacing and therefore spectral resolution as achieved in the Parkes observations. An integration time of 5 minutes was used and a unique reference observation was made. Flux density calibration was similarly made with respect to PKS B1934--638 and we expect that the calibration is accurate to $\sim$30 per cent.

\section{Results}\label{sect:results}

Observations conducted with the Parkes radio telescope over three epochs resulted in the detection of 184 12.2-GHz methanol masers towards 400 6.7-GHz methanol masers detected in the MMB survey in the longitude range 330$^{\circ}$ (through 360$^{\circ}$) to 10$^{\circ}$. This equates to a detection rate of 46 per cent over this longitude range. The 6.7-GHz target sources are listed in Table~\ref{tab:6MMB}. A search of the literature indicates that 117 of the 184 12.2-GHz methanol masers that we detect are new discoveries. References to the 67 sources detected previously are given in Table~\ref{tab:12MMB}.  The observations targeted a total of 401 sources within the designated longitude range but  G\,338.925+0.634 has not been included in the detection statistics because its emission features could not be separated from nearby source G\,338.926+0.634 and so we regard the two sources as one.

Table~\ref{tab:6MMB} presents the characteristics of the targeted 6.7-GHz methanol masers and the results of the 12.2-GHz observations. For some of the analysis undertaken we required the integrated flux density data for both the 6.7 and 12.2-GHz transitions.  The former was not determined in the MMB survey papers \citep{CasMMB10,GreenMMB10, CasMMB102} and we have extracted this information from the MMB `MX' spectra. `MX' refers to an observational mode used in MMB follow-up observations whereby a sources was tracked, and the pointing centre cycled through each of the receiver's seven beams. For internal consistency we also independently determined the peak flux density and velocity range of the 6.7-GHz emission at the same time as extracting the integrated flux density measurement. There are only minor discrepancies between our values and those reported in the MMB catalogues and these have most likely arisen from using data from slightly different epochs. In addition to the properties of the targets, this table also contains a summary of the 12.2-GHz methanol maser observation results, noting either `detection' or the 5-$\sigma$ detection limits for non-detections. Where 12.2-GHz observations were completed during two or more epochs, the data-averaged rms noise is usually a factor of $\sqrt{2}$ or more better than those which are listed.

Table~\ref{tab:12MMB} presents the characteristics of the 12.2-GHz sources that we detect. Given the luxury of knowing the velocity of the 6.7-GHz maser emission, we have been able to readily and reliably identify sources that are much weaker than the  5-$\sigma$ detection limits. A number of weak sources can confidently been regarded as genuine at 3- to 4-$\sigma$ levels. The details presented in Table~\ref{tab:12MMB} follow the usual practice in column 1 where the Galactic longitude and latitude is used as a source name for each source (references to previous 12.2-GHz observations are noted as superscripts after source names);  columns 2 and 3 give the source right ascension  and declination; column 4 gives the epoch of the 12.2-GHz observations that are referred to in columns 5 to 8 which give the peak flux density (Jy), velocity of the 12.2-GHz peak (\kmsns), velocity range (\kmsns) and integrated flux density (Jy \kmsns), respectively.

Spectra for each of the 12.2-GHz detections are presented in Fig.~\ref{fig:12MMB}. The epoch of the observation, either 2008 June, 2008 December or 2010 March, is annotated in the top left hand corner of each of the spectra. For some weak sources, where observations were carried out over multiple epochs, the average of the spectra is presented and these are annotated with two or three observing dates in the top left hand corner of the spectrum. For one source (G\,353.410--0.360) the presented observations were taken with the Hobart 26-m radio telescope and this spectrum is marked with `Hobart' preceding the epoch of observation. Spectra are presented in order of Galactic longitude except for where vertical alignment of spectra corresponding to nearby sources was necessary to highlight where features of a source were also detected at nearby positions. 

 The velocity range of the majority of spectra is 30 \kmsns, usually centred on the velocity of the 6.7-GHz methanol maser peak emission. This method was chosen to give an immediate feeling for the basic structure of the two methanol maser transitions, in a number of cases showing that lone features detected at 12.2-GHz are associated with the 6.7-GHz peak emission. A velocity range of 30 \kms allows the full velocity range of the 6.7-GHz methanol maser emission to be shown in the majority of cases. For some sources, the centre velocity of the spectra had to be changed from the 6.7-GHz peak velocity in order to contain the full extent of the 6.7-GHz emission. Sources that have been vertically aligned have been plotted on the same velocity axis for ease of comparison, and often resulted in the central velocity deviating from that of the 6.7-GHz maser peak. Sources not centred on the velocity of the 6.7-GHz peak are;  G\,335.060--0.427, G\,337.613--0.060, G\,340.785--0.096, G\,348.703-1.043, G\,350.340+0.141, G\,351.417+0.645, G\,351.417+0.646, G\,351.445+0.660, G\,352.083+0.167, G\,352.111+0.176, G\,0.647--0.055, G\,0.657--0.041, G\,0.667--0.034, G\,2.536+0.198, G\,8.683--0.368, G\,9.619+0.193.

Descriptions of specific sources of interest are presented in Section~\ref{sect:12GHz_ind} and these are clearly marked with an `*' in Table~\ref{tab:6MMB}.

\begin{table*}\footnotesize
 \caption{Characteristics of the 6.7-GHz methanol maser targets as well as a brief description of the 12.2-GHz results. The full complement of 12.2-GHz source properties are listed in Table~\ref{tab:12MMB}. Column 1 gives the Galactic longitude and latitude of each source and is used as an identifier (a `*' following the source name indicates those with notes in Section~\ref{sect:12GHz_ind}); column 2 gives the peak flux density (Jy) of the 6.7-GHz sources, derived from follow-up MX observations at the accurate 6.7-GHz position unless otherwise noted; columns 3 and 4 give the peak velocity and velocity ranges (\kmsns) of the 6.7-GHz emission respectively (also derived from Parkes MX observations); column 5 gives the integrated flux density of the 6.7-GHz sources (Jy \kmsns). A `--' in either of the detection limit columns (i.e. columns 6, 8 or 10) indicates that no observations were made on the given epoch. The values listed in columns 6-11 are replaced with the word `detection' where 12.2-GHz emission is observed.  A `c' in column 5 (6.7-GHz integrated flux density) indicates that the 6.7-GHz source statistics were derived from the MMB data cube, rather than follow-up MX observations which means that while the flux density and velocity ranges are a good estimate they are not as reliable as that of other sources and secondly that no integrated flux density could be computed. The presence of `conf' in the columns showing the integrated flux density indicates that a value for this characteristic could not be extracted do to confusion from nearby sources. Columns 6-11 give the 5-$\sigma$ 12.2-GHz methanol maser detection limits and observed velocity ranges for the 2008 June, 2008 December and 2010 March epochs respectively. The presence of a `t' in the place of any value indicates that an explanation is located in Section~\ref{sect:12GHz_ind}.} 
  \begin{tabular}{lllllclllclllcll} \hline
 \multicolumn{1}{c}{\bf Methanol maser} &\multicolumn{4}{c}{\bf 6.7-GHz properties} & \multicolumn{6}{c}{\bf 12.2-GHz observations}\\
    \multicolumn{1}{c}{\bf ($l,b$)}& {\bf S$_{6.7}$} & {\bf Vp$_{6.7}$} & {\bf Vr$_{6.7}$ } & {\bf  I$_{6.7}$} & \multicolumn{2}{c}{\bf 2008 June} & \multicolumn{2}{c}{\bf 2008 December} & \multicolumn{2}{c}{\bf 2010 March}\\	
      \multicolumn{1}{c}{\bf (degrees)}  &{\bf (Jy)} &&& & \multicolumn{1}{c}{\bf 5-$\sigma$} & \multicolumn{1}{c}{\bf Vr} & \multicolumn{1}{c}{\bf 5-$\sigma$} & \multicolumn{1}{c}{\bf Vr} & \multicolumn{1}{c}{\bf 5-$\sigma$} & \multicolumn{1}{c}{\bf Vr}\\  \hline \hline
G\,330.070+1.064	&	8.7	&	--41.8	&	--55.5,--35.3	&	8.8 &\multicolumn{6}{c}{detection}	\\
G\,330.226+0.290*	&	1.1	&	--90.5	&	--92.2,--74.5	&	1.5	&\multicolumn{6}{c}{detection}\\
G\,330.283+0.493	&	4.4	&	--88.8	&	--90.0,--87.5	&	5.2	&\multicolumn{6}{c}{detection}\\
G\,330.875--0.383	&	0.3	&	--69.7	&	--69.7,--69.5	&	0.1	&	$<$0.75	&	--165,175	&			$<$0.80			&	--200,140	& --\\
G\,330.878--0.367	&	0.8	&	--59.2	&	--59.4,--57.8	&	0.4	&	$<$0.75			&	--165,175	&			$<$0.80			&	--200,140	& --\\
G\,330.953--0.182	&	6.8   	&	--87.5	&	--90.1,--87.1	&	4.4 	&	$<$0.50			&	--185,155	&			--	&					& --\\
G\,330.998+0.093	&	0.7	&	--28.6	&	--29.4,--28.4	&	0.4	&	$<$0.75			&	--165,175	&			$<$0.80			&	--200,140	& --\\
G\,331.059+0.375	&	2.9	&	--82.5	&	--82.9,--69.7	&	4.1	&	$<$0.70			&	--165,175	&			$<$0.80			&	--200,140	&	--	\\
G\,331.120--0.118	&	2.5	&	--93.1	&	--94.5,--91.2	&	2.2&\multicolumn{6}{c}{detection}	\\
G\,331.132--0.244*	&	20	&	--84.3	&	--92.1,--81.0	&	29	&\multicolumn{6}{c}{detection}\\
G\,331.134+0.156	&	1.7	&	--72.1	&	--83.5,--72.0	&	1.1	&	$<$0.75			&	--165,175	&			$<$0.80			&	--200,140 & --\\
G\,331.278--0.188*	&	89	&	--78.1	&	--87.7,--76.9	&	221	&\multicolumn{6}{c}{detection}\\
G\,331.342--0.346*	&	86	&	--65.1	&	--76.4,--60.7	&	127	&\multicolumn{6}{c}{detection}\\
G\,331.425+0.264	&	16	&	--88.8	&	--91.5,--77.3	&	16	&\multicolumn{6}{c}{detection}\\
G\,331.437--0.304	&	5.9	&	--89.8	&	--90.7,--85.3	&	5.7	&\multicolumn{6}{c}{detection}\\
G\,331.442--0.187	&	61	&	--87.8	&	--93.3,--83.9	&	135	&\multicolumn{6}{c}{detection}\\
G\,331.542--0.066	&	6.9	&	--86	&	--90.2,--85.2	&	3.5	&$<$0.75			&	--160,180	&			$<$0.60			&	--200,140	&	$<$0.83	& --160,180\\
G\,331.543--0.066	&	15	&	--84	&	--85.0,--79.9	&	14	&\multicolumn{6}{c}{detection}\\
G\,331.556--0.121	&	52	&	--97.1	&	--105.1,--93.8	&	64	&\multicolumn{6}{c}{detection}\\
G\,331.710+0.603	&	18	&	--73.3	&	--75.6,--69.7	&	17&	$<$0.75			&	--160,180	&			$<$0.80			&	--200,140	&	--\\
G\,331.900--1.186*	&	2.4	&	--46.3	&	--59.2,--45.7	&	2.9&\multicolumn{6}{c}{detection}	\\
G\,332.094--0.421	&	12	&	--58.5	&	--62.0,--58.0	&	14	&\multicolumn{6}{c}{detection}\\
G\,332.295--0.094	&	8.0	&	--42.6	&	--59.2,--42	&	13	&\multicolumn{6}{c}{detection}\\
G\,332.296--0.094	&	2.2	&	--46.6	&	--46,--47		&	1.7	&	$<$0.70	&--160,180 & -- & & --		\\ 
G\,332.295+2.280	&	152	&	--24.0	&	--27.0,--20.6	&	117	&\multicolumn{6}{c}{detection}\\
G\,332.351--0.436	&	8.5	&	--52.8	&	--54.2,--43.2	&	6.7	&	$<$0.70			&	--160,180	&			$<$0.80			&	--200,140	&--\\
G\,332.352--0.117	&	6.3	&	--44.6	&	--56.0,--41.4	&	6.3	&	$<$0.75			&	--160,180	&			$<$0.60			&	--200,140	&--\\
G\,332.364+0.607	&	0.4	&	--48.5	&	--48.6,--48.2	&	0.2	&	$<$0.70			&	--160,180	&			$<$0.80			&	--200,140	&--\\
G\,332.560--0.148	&	1.5	&	--54.8	&	--56.3,--54.4	&	1.5	&	$<$0.70			&	--160,180	&			$<$0.80			&	--200,140	&--\\
G\,332.583+0.147	&	4.7	&	--39.9	&	--40.7,--38.8	&	3.8	&	$<$0.70			&	--160,180	&			$<$0.80			&	--200,140	&--\\
G\,332.604--0.168	&	0.4	&	--49.8	&	--56.3,--41.9	&	0.6	&	$<$0.70			&	--160,180	&			$<$0.55			&	--200,140	&--\\
G\,332.653--0.621	&	5.7	&	--50.6	&	--51.1,--44.5	&	5.0	&	$<$0.50			&	--160,180	&			$<$0.80			&	--200,140	&--\\
G\,332.701--0.588	&	0.4	&	--62.8	&	--62.8,--62.7	&	0.1	&	$<$0.70			&	--160,180	&			--	&				&     $<$0.80 & --150,190 \\
G\,332.726--0.621	&	5.1	&	--47.4	&	--56.7,--43.7	&	7.3	&\multicolumn{6}{c}{detection}\\
G\,332.813--0.701	&	7.0	&	--53.2	&	--58.9,--50.4	&	7.3	&\multicolumn{6}{c}{detection}\\
G\,332.826--0.549	&	2.0	&	--61.8	&	--62.1,--54.1	&	3.3	&	$<$0.75			&	--160,180	&			--	&				&$<$0.83 & --150,190\\
G\,332.854+0.817	&	1.1	&	--49.0	&	--49.2,--45.0	&	0.6	&	$<$0.70			&	--160,180	&			--	&				&          $<$0.83 & --150,190\\
G\,332.942--0.686	&	7.9	&	--52.0	&	--55.7,--51.1	&	9.1	&	$<$0.70			&	--160,180	&			--	&				& $<$0.80 & --150,190\\
G\,332.960+0.135	&	1.9	&	--54.2	&	--54.7,--54.0	&	1.4	&	$<$0.70			&	--160,180	&			$<$0.55			&	--200,140	&\\
G\,332.963--0.679	&	26	&	--45.9	&	--49.8,--38.0	&	28	&	$<$0.70			&	--160,180	&			--	&				& $<$0.79 & --150,190\\
G\,332.975+0.773	&	1.4	&	--50.9	&	--55.4,--50.7	&	0.6	&	$<$0.70			&	--160,180	&			--	&				& $<$0.82 & --150,190\\
G\,332.987--0.487	&	8.3	&	--55.7	&	--67.5,--51.2	&	8.6	&	$<$0.75			&	--160,180	&			--	&				&$<$0.81 & --150,190\\
G\,333.029--0.015	&	5.4	&	--54.1	&	--62.4,--52.6	&	7.5	&\multicolumn{6}{c}{detection}\\
G\,333.029--0.063	&	1.2	&	--40.1	&	--40.6,--39.5	&	0.8	&	$<$0.70			&	--160,180	&			--	&				& $<$0.80 & --150,190	\\
G\,333.068--0.447*	&	18	&	--54.4	&	--56.4,--53.5	&	9.3	&\multicolumn{6}{c}{detection}\\
G\,333.109--0.500	&	1.5	&	--60.8	&	--61.1,--52.5	&	0.6	&	$<$0.70			&	--160,180	&			--	&				& $<$0.81 & --150,190\\
G\,333.121--0.434	&	17	&	--48.5	&	--56.4,--47.8	&	19	&	$<$0.80			&	--160,180	&			--	&				& $<$0.82 & --150,190\\
\end{tabular}\label{tab:6MMB}
\end{table*}

\begin{table*}\addtocounter{table}{-1}
  \caption{-- {\emph {continued}}}
  \begin{tabular}{lllllclllclllcll} \hline
 \multicolumn{1}{c}{\bf Methanol maser} &\multicolumn{4}{c}{\bf 6.7-GHz properties} & \multicolumn{6}{c}{\bf 12.2-GHz observations}\\
    \multicolumn{1}{c}{\bf ($l,b$)}& {\bf S$_{6.7}$} & {\bf Vp$_{6.7}$} & {\bf Vr$_{6.7}$ } & {\bf  I$_{6.7}$} & \multicolumn{2}{c}{\bf 2008 June} & \multicolumn{2}{c}{\bf 2008 December} & \multicolumn{2}{c}{\bf 2010 March}\\	
      \multicolumn{1}{c}{\bf (degrees)}  &{\bf (Jy)} &&& & \multicolumn{1}{c}{\bf 5-$\sigma$} & \multicolumn{1}{c}{\bf Vr} & \multicolumn{1}{c}{\bf 5-$\sigma$} & \multicolumn{1}{c}{\bf Vr} & \multicolumn{1}{c}{\bf 5-$\sigma$} & \multicolumn{1}{c}{\bf Vr}\\  \hline \hline
G\,333.126--0.440	&	8.2	&	--43.9	&	--44.5,--42.4	&	4.0	&	$<$0.80			&	--160,180	&			--	&				& $<$0.83 & --150,190\\
G\,333.128--0.440	&	1.4	&	--44.0	&	--47.8,--45.4	&	1.3	&	$<$0.80			&	--160,180	&			$<$0.65			&	--200,140	& $<$0.82 & --150,190\\
G\,333.135--0.431*	&	1.0	& 	--52.5	& 	--53.0,--52.0	& 0.4		&	$<$1.0			& 	--160,180	&	$<$0.81	& --200,140	& $<$1.0	& --150,190\\
G\,333.128--0.560	&	8.7	&	--52.7	&	--60.9,--52.0	&	14&	$<$0.70			&	--160,180	&			--	&				&	$<$0.81 & --150,190\\
G\,333.130--0.560	&	24	&	--56.7	&	--64.4,--56.3	&	15	&	$<$0.70			&	--160,180	&			--	&				& $<$0.82 & --150,190\\
G\,333.163--0.101	&	9.8	&	--95.1	&	--95.8,--90.0	&	5.4	&\multicolumn{6}{c}{detection}\\
G\,333.184--0.091	&	7.4	&	--81.9	&	--91.4,--81.0	&	7.0	&\multicolumn{6}{c}{detection}\\
G\,333.234--0.060	&	0.7	&	--85.3	&	--87.0,-85.2	&	0.3	& $<$0.75			&	--160,180	&			--	&				& $<$0.80 & --150,190\\
G\,333.234--0.062	&	1.3	&	--91.8	&	--92.5,--79.6	&	2.2	&	$<$0.75			&	--160,180	&			--	&				& $<$0.80 & --150,190\\
G\,333.315+0.105	&	178	&	--41.3	&	--51.4,--36.9	&	41&	$<$0.70			&	--160,180	&			--	&				&	$<$0.81 & --150,190\\
G\,333.387+0.032	&	3.5	&	--73.8	&	--74.2,--60.2	&	3.9	&$<$0.70			&	--160,180	&			--	&				&	$<$0.78	& --150,190\\
G\,333.466--0.164	&	30	&	--42.2	&	--50.9,--37.1	&	35	&	$<$0.75			&	--160,180	&			--	&				& $<$0.79 & --150,190\\
G\,333.562--0.025	&	33	&	--35.3	&	--43.9,--33.1	&	56&\multicolumn{6}{c}{detection}	\\
G\,333.646+0.058*	&	8.4	&	--87.3	&	--89.0,--80.8	&	10&\multicolumn{6}{c}{detection}	\\
G\,333.683--0.437	&	25	&	--5.6	&	--7.7,--0.4	&	20	&\multicolumn{6}{c}{detection}\\
G\,333.761--0.226	&	3.4	&	--55.2	&	--56.5,--51.3	&	2.4	&	$<$0.70			&	--160,180	&			--	&				&$<$0.79 & --150,190\\
G\,333.851+0.527	&	0.8	&	--40.3	&	--49.7,--40.1	&	0.4	&	$<$0.70			&	--160,180	&			--	&				& $<$0.80 & --150,190\\
G\,333.900--0.099	&	0.7	&	--59.4	&	--59.6,--56.6	&	0.7&	$<$0.75			&	--160,180	&			--	&				&	$<$0.79 & --150,190\\
G\,333.931--0.135	&	15	&	--36.9	&	--37.3,--36.3	&	6.1	&	$<$0.75			&	--160,180	&			--	&			&	$<$0.81 & --150,190\\
G\,334.138--0.023	&	2.9	&	--31.1	&	--34.3,--30.5	&	3.0	&\multicolumn{6}{c}{detection}\\
G\,334.307--0.079	&	2.0	&	--36.7	&	--37.8,--33.0	&	1.8	&\multicolumn{6}{c}{detection}\\
G\,334.635--0.015	&	28	&	--30.1	&	--32.0,--27.0	&	28&\multicolumn{6}{c}{detection}	\\
G\,334.933--0.307	&	3.3	&	--102.8	&	--111.2,--100.8	&	3.5	&\multicolumn{6}{c}{detection}\\
G\,334.935--0.098	&	4.6	&	--21.0	&	--22.4,--16.8	&	4.9	&\multicolumn{6}{c}{detection}\\
G\,335.060--0.427	&	42	&	--46.9	&	--47.7,--25.8	&	48	&\multicolumn{6}{c}{detection}\\
G\,335.426--0.240*	&	66	&	--50.5	&	--53.5,--39.7	&	22	&\multicolumn{6}{c}{detection}\\
G\,335.556--0.307*	&	22	&	--116.3	&	--119.7,--110.6	&	35	&\multicolumn{6}{c}{detection}\\
G\,335.585--0.285	&	35	&	--48.6	&	--50.5,--43.5	&	47	&	$<$0.80			&	--160,180	&			$<$0.65			&	--200,140	& --\\
G\,335.585--0.289	&	51	&	--51.3	&	--56.5,--49.8	&	39	&\multicolumn{6}{c}{detection}\\
G\,335.585--0.290	&	29	&	--47.5	&--48,--45			&  c	&	$<$0.85			&	--160,180	&			$<$0.65			&	--200,140	& --			\\
G\,335.726+0.191	&	49	&	--44.4	&	--54.9,--43.3	&	39&\multicolumn{6}{c}{detection}	\\
G\,335.789+0.174*	&	166	&	--47.5	&	--56.5,--44.2	&	378	&\multicolumn{6}{c}{detection}\\
G\,335.824--0.177	&	0.9	&	--26.1	&	--26.2,--26.0	&	0.2		&	--	&				&			$<$0.55	&			--200,140	&  $<$0.82 & --150,190\\
G\,336.018--0.827	&	87	&	--53.2	&	--54.6,--39.1	&	181	&\multicolumn{6}{c}{detection}\\
G\,336.358--0.137	&	13	&	--73.5	&	--81.2,--70.9	&	21&\multicolumn{6}{c}{detection}	\\
G\,336.409--0.257	&	6.1	&	--85.9	&	--86.1,--83.9	&	5.3		&	--	&				&			$<$0.60	&			--200,140	& $<$0.81 & --150,190\\
G\,336.433--0.262	&	24	&	--93.0	&	--95.8,--86.1	&	42&\multicolumn{6}{c}{detection}	\\
G\,336.464--0.157	&	0.7	&	--85.8	&	--86.3,--77.9	&	0.6		&	--	&				&			$<$0.55	&			--200,140	& $<$0.80 & --150,190\\
G\,336.496--0.271	&	17	&	--24.1	&	--27.1,--19.1	&	20&\multicolumn{6}{c}{detection}	\\
G\,336.526--0.156	&	0.7	&	--94.9	&	--95.1,--88.2	&	0.3	&	--	&				&			$<$0.55	&			--200,140	& $<$0.78 & --150,190\\
G\,336.703--0.099	&	3.2	&	--35.6	&	--37.3,--35.3	&	0.8	&	$<$0.75			&	--160,180			&	--			&		& $<$0.80 & --150,190\\
G\,336.809+0.119	&	5.5	&	--84.4	&	--85.9,--78.4	&	11	&	$<$0.75			&	--160,180			&	--			&		& $<$0.79 & --150,190	\\
G\,336.822+0.028	&	28	&	--76.8	&	--79.9,--74.2	&	21	&\multicolumn{6}{c}{detection}\\
G\,336.825+0.139	&	3.4	&	--88.4	&	--89.5,--86.8	&	4.5	&\multicolumn{6}{c}{detection}\\
G\,336.830--0.375	&	20	&	--22.8	&	--27.3,--17.4	&	38	&	$<$0.75			&	--160,180			&	$<$0.80			&	--200,140	& --\\
G\,336.864+0.005*	&	59	&	--76.1	&	--81.7,--72.9	&	44.1&\multicolumn{6}{c}{detection}	\\
G\,336.881+0.008*	&	2.1	&	--68.1	&	--70.3,--67.6	& 2.1		&$<$1.5	&		--160,180	& $<$1.5	& --200,140 & --\\ 
G\,336.916--0.024	&	4.1	&	--127.4	&	--127.9,--114.2	&	2.4	&\multicolumn{6}{c}{detection}\\
G\,336.941--0.156	&	23	&	--67.2	&	--79.4,--62.3	&	48&\multicolumn{6}{c}{detection}	\\
G\,336.957--0.225*	&	1.6	&	--68.0	&	--85.4,--63.9	&	2.9	&\multicolumn{6}{c}{detection}\\
G\,336.958--0.977	&	0.8	&	--49.3	&	--50.5,--47.8	&	0.5		&	$<$0.75			&	--160,180			&	--			&		&--\\
G\,336.983--0.183	&	12	&	--80.7	&	--89.4,--78.5	&	16&\multicolumn{6}{c}{detection}	\\
G\,336.994--0.027	&	26	&	--125.8	&	--126.8,--115.3	&	20	&	$<$0.70			&	--160,180			&	--			&		&--\\
G\,337.052--0.226	&	14	&	--77.6	&	--78.4,--74.9	&	15	&\multicolumn{6}{c}{detection}\\
G\,337.097--0.929	&	6.8	&	--40.4	&	--42.5,--37.4	&	5.0	&\multicolumn{6}{c}{detection}\\
G\,337.132--0.068*	&	2	&	--60.6	&	--62.3,--59.5	&	1.8	&\multicolumn{6}{c}{detection}\\
G\,337.153--0.395	&	17	&	--49.4	&	--50.9,--47.8	&	15	&\multicolumn{6}{c}{detection}\\
G\,337.176--0.032	&	4.5	&	--64.8	&	--74.8,--62.9	&	7.0	&\multicolumn{6}{c}{detection}\\
G\,337.201+0.114	&	7.2	&	--57.5	&	--64.5,--51.9	&	23	&\multicolumn{6}{c}{detection}\\
G\,337.202--0.094*	&	1.4	&	--67.6	&	--83.2,--65.9	&	3.6	&	\multicolumn{6}{c}{detection}\\
G\,337.258--0.101	&	7.7	&	--69.3	&	--73.7,--63.6	&	17	&	$<$0.70			&	--160,180			&	--			&		&--\\
G\,337.263--0.070	&	1.1	&	--39.4	&	--41.3,--39.1	&	0.9	&	--	&				&			$<$0.60	&			--200,140	&--\\
G\,337.300--0.874	&	2.2	&	--87.4	&	--97.5,--87.4	&	4.3	&	$<$0.70			&	--160,180			&	--			&		&--	\\
\end{tabular}
\end{table*}

\begin{table*}\addtocounter{table}{-1}
  \caption{-- {\emph {continued}}}
  \begin{tabular}{lllllclllclllcll} \hline
 \multicolumn{1}{c}{\bf Methanol maser} &\multicolumn{4}{c}{\bf 6.7-GHz properties} & \multicolumn{6}{c}{\bf 12.2-GHz observations}\\
    \multicolumn{1}{c}{\bf ($l,b$)}& {\bf S$_{6.7}$} & {\bf Vp$_{6.7}$} & {\bf Vr$_{6.7}$ } & {\bf  I$_{6.7}$} & \multicolumn{2}{c}{\bf 2008 June} & \multicolumn{2}{c}{\bf 2008 December} & \multicolumn{2}{c}{\bf 2010 March}\\	
      \multicolumn{1}{c}{\bf (degrees)}  &{\bf (Jy)} &&& & \multicolumn{1}{c}{\bf 5-$\sigma$} & \multicolumn{1}{c}{\bf Vr} & \multicolumn{1}{c}{\bf 5-$\sigma$} & \multicolumn{1}{c}{\bf Vr} & \multicolumn{1}{c}{\bf 5-$\sigma$} & \multicolumn{1}{c}{\bf Vr}\\  \hline \hline
G\,337.388--0.210	&	23	&	--56.2	&	--67.0,--52.0	&	43&\multicolumn{6}{c}{detection}	\\
G\,337.404--0.402	&	77	&	--39.6	&	--43.3,--36.5	&	76&\multicolumn{6}{c}{detection}	\\
G\,337.517--0.348	&	0.7	&	9.4	&	9.2,9.8	&	0.3		&	--	&				&			$<$0.55	&			--200,140	&--\\
G\,337.613--0.060	&	21	&	--41.6	&	--53.6,--38.0	&	47&\multicolumn{6}{c}{detection}	\\
G\,337.632--0.079*	&	14	&	--56.9	&	--63.2,--54.0	&	19	&\multicolumn{6}{c}{detection}\\
G\,337.686+0.137	&	0.7	&	--75.1	&	--75.4,--73.8	&	0.5	&	$<$0.75			&	--160,180			&	$<$0.80			&	--200,140	& --\\
G\,337.703--0.053*	&	13	&	--44.1	& 	--52,--43.2	&10	&	--	&	&	$<$0.80	&	--200,140 & --\\ 
G\,337.705--0.053*	&	170	&	--54.6	&	--58.1,--49	&	236	&\multicolumn{6}{c}{detection}\\
G\,337.710+0.089	&	3.3	&	--72.6	&	--78.7,--72.4	&	1.5	&	$<$0.75			&	--160,180			&	$<$0.75			&	--200,140	&--\\
G\,337.720+0.065	&	0.7	&	--63.9	&	--69.9,--63.8	&	0.4	&	$<$0.75			&	--160,180			&	$<$0.75			&	--200,140	&--	\\
G\,337.844--0.375	&	5.8	&	--38.6	&	--42.2,--37.3	&	3.4	&\multicolumn{6}{c}{detection}\\
G\,337.920--0.456	&	32	&	--37.8	&	--40.7,--35.8	&	27&\multicolumn{6}{c}{detection}	\\
G\,337.966--0.169	&	9.3	&	--59.4	&	--67.0,--54.2	&	16	&\multicolumn{6}{c}{detection}\\
G\,337.997+0.136	&	6.0	&	--32.3	&	--35.0,--31.2	&	2.6	&	$<$0.75			&	--160,180			&	$<$0.80			&	--200,140	&--\\
G\,338.069+0.011	&	2.6	&	--39.9	&	--40.3,--39.7	&	1.3	&	$<$0.80			&	--160,180			&	$<$0.80			&	--200,140	&--\\
G\,338.075+0.012*	&	14	&	--44	&	--53.9,--42.8	&	19	&\multicolumn{6}{c}{detection}\\
G\,338.075+0.009	&	4.4	&	--39.1	&	--41.9,--35.6	&	10	&	--			&		&			$<$0.80	&			--200,140	&--\\
G\,338.140+0.178	&	7.2	&	--34.5	&	--42.6,--33.7	&	4.7	&	$<$0.70			&	--160,180			&	$<$0.80			&	--200,140	&--\\
G\,338.160--0.064	&	2.5	&	--66.2	&	--68.4,--62.2	&	4.4	&\multicolumn{6}{c}{detection}\\
G\,338.280+0.542	&	5.4	&	--56.7	&	--62.2,--56.5	&	3.0		&	$<$0.75			&	--160,180			&	$<$0.55			&	--200,140	&--	\\
G\,338.287+0.120	&	26	&	--39.9	&	--43.8,--35.7	&	14&	$<$0.75			&	--160,180			&	$<$0.55			&	--200,140	&--	\\
G\,338.325--0.409	&	2.8	&	--25.7	&	--34.2,--25.0	&	1.9	&	$<$0.75			&	--160,180			&	$<$0.80			&	--200,140	&--\\
G\,338.388+0.162	&	3.6	&	--29.8	&	--37.1,--26.2	&	8.9	&\multicolumn{6}{c}{detection}\\
G\,338.392--0.403	&	5.3	&	--33.5	&	--34.0,--32.6	&	2.2	&	$<$0.70			&	--160,180	&			$<$0.80			&	--200,140	&--\\
G\,338.396--0.007	&	2.7	&	--48.9	&	--54.2,--44.5	&	2.5	&	$<$0.75			&	--160,180	&			$<$0.75			&	--200,140	&--\\
G\,338.432+0.058	&	30	&	--30.2	&	--33.5,--22.5	&	20	&	$<$0.75			&	--160,180	&			$<$0.80			&	--200,140	&--\\
G\,338.461--0.245	&	69	&	--51.7	&	--63.3,--48.7	&	145	&\multicolumn{6}{c}{detection}\\
G\,338.472+0.289	&	0.9	&	--29.4	&	--34.8,--29.2	&	0.7	&	--	&				&			$<$0.55	&			--200,140	&--\\
G\,338.497+0.207	&	3.3	&	--28.1	&	--30.7,--27.9	&	2.2	&	--	&				&			$<$0.55	&			--200,140	&--\\
G\,338.561+0.218*	&	38	&	--39.1	&	--43.8,--29.6	&	87	&\multicolumn{6}{c}{detection}\\
G\,338.566+0.110	&	2.7	&	--74.7	&	--78.5,--74.2	&	2.0	&	--	&				&			$<$0.55	&			--200,140	&--\\
G\,338.850+0.409	&	1.6	&	--55.7	&	--59.6,--54.4	&	1.4		&	--	&				&			$<$0.55	&			--200,140	&--\\
G\,338.875--0.084	&	19	&	--41.3	&	--42.2,--35.1	&	13	&\multicolumn{6}{c}{detection}\\
G\,338.902+0.394	&	1.7	&	--26.0	&	--35.4,--19.1	&	3.4	&\multicolumn{6}{c}{detection}\\
G\,338.920+0.550	&	76	&	--61.3	&	--67.8,--58.3	&	68&\multicolumn{6}{c}{detection}	\\
G\,338.925+0.557	&	5	&	--62.4	&	--66,--59	&	conf	 &	--	&				&			$<$0.75	&			--200,140	&--\\
G\,338.925+0.634*	&	65	&	--60.8	&	--68.5,--52.1	&t	&\multicolumn{6}{c}{detection}\\
G\,338.926+0.634*	&	3.8	&	--64.5	&	--65.5,--59.5	&	102	&\multicolumn{6}{c}{detection}\\
G\,338.935--0.062	&	29	&	--41.9	&	--43.1,--41.2	&	14	&\multicolumn{6}{c}{detection}\\
G\,339.053--0.315	&	129	&	--111.6	&	--124.3,--110.1	&	117	&\multicolumn{6}{c}{detection}\\
G\,339.064+0.152	&	3.6	&	--87.5	&	--90.5,--82.3	&	8.2	&\multicolumn{6}{c}{detection}\\
G\,339.204--0.018	&	2.0	&	--14.3	&	--16.7,--11.6	&	4.2	&	--	&				&			$<$0.55	&	--200,140	&--\\
G\,339.282+0.136	&	5.7	&	--70.0	&	--71.7,--68.4	&	7.1	&\multicolumn{6}{c}{detection}\\
G\,339.294+0.139	&	7.0	&	--74.6	&	--75.9,--66.2	&	7.6	&	--	&				&			$<$0.60	&			--200,140	&  $<$0.81 	& --150,190 \\
G\,339.476+0.185*	&	7.2	&	--87.6	&	--98.1,--86.5	&	5.7	&\multicolumn{6}{c}{detection}\\
G\,339.477+0.043	&	2.7	&	--9.5	&	--20.5,--5.8	&	2.5	&	--	&				&			$<$0.55	&			--200,140	& $<$0.81 & --150,190\\
G\,339.582--0.127	&	13	&	--30.4	&	--31.6,--29.6	&	11	&\multicolumn{6}{c}{detection}\\
G\,339.622--0.121*	&	86	&	--32.9	&	--39.3,--31.6	&	141	&\multicolumn{6}{c}{detection}\\
G\,339.681--1.208	&	70	&	--21.4	&	--41.9,--20.7	&	69	&	--	&				&			$<$0.55	&			--200,140	&--\\
G\,339.682--1.207	&	23	& 	--34.4	&	--35.0,--33.0	&	c & 	--	&				&			$<$0.55	&			--200,140	&--					\\ 
G\,339.762+0.054	&	8.0	&	--51.0	&	--54.0,--49.5	&	9.1	&\multicolumn{6}{c}{detection}\\
G\,339.884--1.259	&	1510	&	--38.7	&	--45.0,--27.5	&	2568	&\multicolumn{6}{c}{detection}\\
G\,339.909+0.240	&	0.5	&	--12.1	&	--12.5,--11.9	&	0.3	&	--	&				&			$<$0.55	&			--200,140	&--\\
G\,339.949--0.539	&	71	&	--97.8	&	--107.9,--89.5	&	195	&\multicolumn{6}{c}{detection}\\
G\,339.980--0.538	&	7.0	&	--89.2	&	--89.4,--89.0	&	2.1	&	--	&				&			$<$0.55	&			--200,140	&--\\
G\,339.986--0.425	&	88	&	--89.4	&	--91.8,--86.2	&	109	&\multicolumn{6}{c}{detection}\\
G\,340.034--1.110	&	6.8	&	--27.5	&	--28.8,--24.3	&	3.4		&	--	&			&			$<$0.55	&			--200,140	&--\\
G\,340.054--0.244	&	48	&	--59.4	&	--62.6,--58.5	&	39&\multicolumn{6}{c}{detection}	\\
G\,340.118--0.021	&	1.2	&	--123.3	&	--123.8,--118.1	&	0.8&\multicolumn{6}{c}{detection}	\\
G\,340.182--0.047	&	3.1	&	--131.1	&	--132.1,--117.6	&	3		&	--	&				&			$<$0.55	&			--200,140	&--\\
G\,340.249--0.046	&	12	&	--126.3	&	--135.3,--119.9	&	19	&	--	&				&			$<$0.60	&			--200,140	&--\\
G\,340.249--0.372	&	0.7	&	--51.2	&	--51.5,--51.2	&	0.2	&	--	&				&			$<$0.60	&			--200,140	&--\\
G\,340.518--0.152	&	7.3	&	--48.2	&	--50.2,--43.6	&	6.4	&\multicolumn{6}{c}{detection}\\
\end{tabular}
\end{table*}

\begin{table*}\addtocounter{table}{-1}
  \caption{-- {\emph {continued}}}
  \begin{tabular}{lllllclllclllcll} \hline
 \multicolumn{1}{c}{\bf Methanol maser} &\multicolumn{4}{c}{\bf 6.7-GHz properties} & \multicolumn{6}{c}{\bf 12.2-GHz observations}\\
    \multicolumn{1}{c}{\bf ($l,b$)}& {\bf S$_{6.7}$} & {\bf Vp$_{6.7}$} & {\bf Vr$_{6.7}$ } & {\bf  I$_{6.7}$} & \multicolumn{2}{c}{\bf 2008 June} & \multicolumn{2}{c}{\bf 2008 December} & \multicolumn{2}{c}{\bf 2010 March}\\	
      \multicolumn{1}{c}{\bf (degrees)}  &{\bf (Jy)} &&& & \multicolumn{1}{c}{\bf 5-$\sigma$} & \multicolumn{1}{c}{\bf Vr} & \multicolumn{1}{c}{\bf 5-$\sigma$} & \multicolumn{1}{c}{\bf Vr} & \multicolumn{1}{c}{\bf 5-$\sigma$} & \multicolumn{1}{c}{\bf Vr}\\  \hline \hline
G\,340.543--0.162	&	1.0	&	--51.6	&	--52.1,--51.4	&	0.5&	--	&				&			$<$0.55	&			--200,140	&	$<$0.78	& --150,190\\
G\,340.655--0.235	&	1.2	&	--21.6	&	--21.7,--21.4	&	0.3	&	$<$0.75			&	--160,180			&	$<$0.80			&	--200,140	&--\\
G\,340.785--0.096	&	158	&	--108.1	&	--111.7,--84.7	&	268	&\multicolumn{6}{c}{detection}\\
G\,340.970--1.022	&	10	&	--31.4	&	--33.9,--19.7	&	9.1	&	$<$0.70			&	--155,185			&	$<$0.80			&	--200,140	&--\\
G\,341.124--0.361	&	1.6	&	--37.2	&	--43.7,--36.1	&	1.3	&	$<$0.70			&	--155,185			&	$<$0.75			&	--200,140	&--\\
G\,341.218--0.212	&	192	&	--37.9	&	--50.4,--35.1	&	336	&\multicolumn{6}{c}{detection}\\
G\,341.238--0.270	&	4.3	&	--51.3	&	--52.3,--49.3	&	7.2	&\multicolumn{6}{c}{detection}\\
G\,341.276+0.062	&	7.7	&	--70.6	&	--77.3,--70.0	&	9.3	&\multicolumn{6}{c}{detection}\\
G\,341.367+0.336	&	1.0	&	--80.5	&	--81.0,--79.8	&	0.4	&	$<$0.75			&	--155,185			&	$<$0.75			&	--200,140	&--\\
G\,341.973+0.233	&	1.4	&	--11.9	&	--12.4,--11.2	&	1.0	&	$<$0.70			&	--155,185			&	$<$0.55			&	--200,140	&--\\
G\,341.990--0.103	&	1.3	&	--37.0	&	--41.8,--36.8	&	0.5	&	$<$0.70			&	--155,185			&	$<$0.80			&	--200,140	&--\\
G\,342.251+0.308	&	1.9	&	--122.7	&	--123.8,--120.4	&	2.3	&	$<$0.70			&	--155,185			&	$<$0.70			&	--200,140	&--\\
G\,342.338+0.305*	&	0.9	&	--112.1	&	--112.5,--108.8	&	1.4	&\multicolumn{6}{c}{detection}\\
G\,342.368+0.140	&	3.5	&	--7.2	&	--15.4,1.0	&	9.1	&	$<$0.70			&	--155,185			&	--	&				&\\
G\,342.446--0.072*	&	2.1	&	--20.2	&	--32.2,--14.3	&	3.0	&\multicolumn{6}{c}{detection}\\
G\,342.484+0.183	&	81	&	--42.2	&	--44.3,--38.4	&	83	&\multicolumn{6}{c}{detection}\\
G\,342.954--0.019	&	2.0	&	--3.1	&	--13.2,--2.1	&	4.5	&	$<$0.75			&	--155,185			&	--	&				& $<$0.80 & --150,190\\
G\,343.354--0.067*	&	18	&	--117.7	&	--133.1,--116.6	&	13	&\multicolumn{6}{c}{detection}\\
G\,343.502--0.472	&	13	&	--39.0	&	--4.03,--31.8	&	9.3&\multicolumn{6}{c}{detection}	\\
G\,343.756--0.163	&	5.1	&	--30.8	&	--32.4,--23.0	&	4.2		&	$<$0.50			&	--155,185			&	--	&				&--\\
G\,343.929+0.125	&	9.0	&	14.5	&	7.7,18.8	&	16	&\multicolumn{6}{c}{detection}\\
G\,344.227--0.569	&	90	&	--19.7	&	--32.1,--10.6	&	128	&\multicolumn{6}{c}{detection}\\
G\,344.419+0.044	&	2.3	&	--64.9	&	--65.0,--63.0	&	1.5	&	$<$0.75			&	--155,185			&	$<$1.45	&			--200,140	&--\\
G\,344.421+0.045	&	17	&	--71.4	&	--72.3,--70.6	&	11&\multicolumn{6}{c}{detection}	\\
G\,344.581--0.024	&	3.5	&	1.5	&	--4.5,5.8	&	2.8	&	$<$0.70			&	--155,185			&	--	&				&--\\
G\,345.003--0.223	&	236	&	--23.1	&	--24.9,--20.2	&	227	&\multicolumn{6}{c}{detection}\\
G\,345.003--0.224	&	84	&	--26.1	&	--32.7,--25	&	120	&\multicolumn{6}{c}{detection}\\
G\,345.010+1.792*	&	268	&	--21	&	--23.5,--16	&	513	&\multicolumn{6}{c}{detection}\\
G\,345.012+1.797	&	32	&	--12	&	--16,--10.1	&	53&\multicolumn{6}{c}{detection}	\\
G\,345.131--0.174	&	3.1	&	--28.9	&	--34.0,--28.6	&	2.1	&	$<$0.70			&	--155,185			&	--	&				&--\\
G\,345.198--0.030	&	2.5	&	--0.5	&	--3.8,0.8	&	2.8	&\multicolumn{6}{c}{detection}\\
G\,345.205+0.317	&	0.8	&	--63.5	&	--64.0,--60.4	&	0.6	&	$<$0.75		&	--155,185	&			--	&		&	--\\
G\,345.407--0.952	&	2.0	&	--14.2	&	--15.0,--14.0	&	1.5	&	$<$0.80			&	--155,185			&	--	&				&--\\
G\,345.424--0.951	&	1.8	&	--13.1	&	--21.1,--6.9	&	2.5	&	$<$0.75			&	--155,185			&	--	&				&--\\
G\,345.441+0.205	&	2.3	&	0.9	&	--12.3,1.3	&	4.4	&	$<$0.75			&	--155,185			&	--	&				&--\\
G\,345.487+0.314	&	2.8	&	--22.0	&	--23.0,--21.5	&	2.6	&	$<$0.75	`	&	--155,185	&			--	&		&	--\\
G\,345.505+0.348	&	300	&	--17.8	&	--24.2,--10.0	&	733	&\multicolumn{6}{c}{detection}\\
G\,345.498+1.467	&	1.2	&	--14.2	&	--14.4,--13.5	&	0.7	&	$<$0.75			&	--155,185			&	--	&				& $<$0.80 	& --140,200\\
G\,345.576--0.225	&	0.6	&	--126.9	&	--126.9,--126.6	&	0.2	&	$<$0.80			&	--155,185			&	--	&				&--\\
G\,345.807--0.044	&	1.0	&	--1.6	&	--2.6,--1.2	&	0.9&\multicolumn{6}{c}{detection}	\\
G\,345.824+0.044	&	3.2	&	--10.3	&	--11.8,--8.9	&	1.7	&	$<$0.70			&	--155,185			&	--	&				& $<$0.80 	& --140,200\\
G\,345.949--0.268	&	1.5	&	--22.0	&	--22.3,--21.7	&	0.6	&	$<$0.75			&	--155,185			&	$<$0.80	&			--200,140	&--\\
G\,345.985--0.020	&	5.7	&	--83.2	&	--84.3,--81.8	&	4.4	&	$<$0.70			&	--155,185			&	$<$0.80	&			--200,140	&--\\
G\,346.036+0.048	&	9.0	&	--6.4	&	--14.5,--4.3	&	8.5	&	$<$0.75			&	--155,185			&	$<$0.55	&			--200,140	&--\\
G\,346.231+0.119	&	1.5	&	--95.0	&	--96.6,--92.5	&	1.6	&	--			&		&			$<$0.55	&			--200,140	&--	\\
G\,346.480+0.221	&	30	&	--18.9	&	--20.5,--13.7	&	22	&\multicolumn{6}{c}{detection}\\
G\,346.481+0.132	&	2.1	&	--5.5	&	--11.6,--4.3	&	2.7	&	$<$0.75			&	--155,185			&	$<$0.75			&	--200,140	&--\\
G\,346.517+0.117	&	0.8	&	--11.0	&	--11.2,--1.7	&	0.9	&	$<$0.70			&	--155,185			&	$<$0.75			&	--200,140	&--\\
G\,346.522+0.085	&	1.9	&	5.7	&	5.6,6.1	&	0.6	&	$<$0.70			&	--155,185			&	$<$0.55			&	--200,140	&--\\
G\,347.230+0.016	&	0.9	&	--68.9	&	--69.1,--68.7	&	0.3		&	$<$0.70			&	--155,185			&	$<$0.80			&	--200,140	&--\\
G\,347.583+0.213	&	3.2	&	--102.3	&	--103.5,--94.4	&	6.3	&\multicolumn{6}{c}{detection}\\
G\,347.628+0.149	&	19	&	--96.5	&	--97.8,--91.7	&	6.9	&	$<$0.75			&	--155,185			&	$<$0.80	&			--200,140	&--\\
G\,347.631+0.211	&	5.8	&	--91.9	&	--93.4,--89.3	&	5.4	&	$<$0.70			&	--155,185			&	$<$0.75	&			--200,140	&--\\
G\,347.817+0.018	&	2.5	&	--24.1	&	--26.4,--23.0	&	4.3	&	$<$0.75			&	--155,185			&	$<$0.75	&			--200,140	&--\\
G\,347.863+0.019*	&	6.4	&	--34.7	&	--37.6,--28.1	&	8.7	&\multicolumn{6}{c}{detection}\\
G\,347.902+0.052	&	5.4	&	--27.4	&	--30.8,--26.6	&	6.4	&\multicolumn{6}{c}{detection}\\
G\,348.027+0.106	&	3.1	&	--121.2	&	--122.4,--114.6	&	4.5&	$<$0.70			&	--155,185			&	$<$0.80	&			--200,140	&--	\\
G\,348.195+0.768	&	4.5	&	--0.8	&	--2.1,--0.5	&	1.9	&\multicolumn{6}{c}{detection}\\
G\,348.550--0.979*	&	36	&	--10	&	--19,--7	&	81	&\multicolumn{6}{c}{detection}\\
G\,348.550--0.979n*	&	21	&	--20	&	--23,--14	&	49&\multicolumn{6}{c}{detection}	\\
G\,348.579--0.920	&	0.7	&	--14.1	&	--15.0,--13.4	&	0.3	&	$<$0.75			&		&			$<$0.80	&			--200,140	&--\\
G\,348.617--1.162	&	48	&	--11.4	&	--21.4,--7.4	&	77	&\multicolumn{6}{c}{detection}\\
G\,348.654+0.244	&	0.8	&	16.9	&	16.5,17.1	&	0.3&	--			&		&			$<$0.55	&			--200,140	&	$<$0.81 	& --140,200\\
\end{tabular}
\end{table*}

\begin{table*}\addtocounter{table}{-1}
  \caption{-- {\emph {continued}}}
  \begin{tabular}{lllllclllclllcll} \hline
 \multicolumn{1}{c}{\bf Methanol maser} &\multicolumn{4}{c}{\bf 6.7-GHz properties} & \multicolumn{6}{c}{\bf 12.2-GHz observations}\\
    \multicolumn{1}{c}{\bf ($l,b$)}& {\bf S$_{6.7}$} & {\bf Vp$_{6.7}$} & {\bf Vr$_{6.7}$ } & {\bf  I$_{6.7}$} & \multicolumn{2}{c}{\bf 2008 June} & \multicolumn{2}{c}{\bf 2008 December} & \multicolumn{2}{c}{\bf 2010 March}\\	
      \multicolumn{1}{c}{\bf (degrees)}  &{\bf (Jy)} &&& & \multicolumn{1}{c}{\bf 5-$\sigma$} & \multicolumn{1}{c}{\bf Vr} & \multicolumn{1}{c}{\bf 5-$\sigma$} & \multicolumn{1}{c}{\bf Vr} & \multicolumn{1}{c}{\bf 5-$\sigma$} & \multicolumn{1}{c}{\bf Vr}\\  \hline \hline
G\,348.723--0.078	&	2.6	&	11.5	&	8.9,12.0	&	3.5	&\multicolumn{6}{c}{detection}\\
G\,348.703--1.043	&	65	&	--3.5	&	--17.5,--2.6	&	213	&\multicolumn{6}{c}{detection}\\
G\,348.727--1.037	&	83	&	--7.4	&	--12.1,--6.0	&	149	&\multicolumn{6}{c}{detection}\\
G\,348.884+0.096	&	12	&	--74.5	&	--77.1,--73.7	&	17	&\multicolumn{6}{c}{detection}\\
G\,348.892--0.180	&	2.7	&	1.5	&	1.3,1.7	&	0.8	&	--			&		&			$<$0.55	&			--200,140	&--\\
G\,349.067--0.017	&	2.3	&	11.6	&	4.6,15.8	&	5.5	&	--			&		&			$<$0.55	&			--200,140	&--\\
G\,349.092+0.105	&	33	&	--76.5	&	--77.9,--73.4	&	22	&	--			&		&			$<$0.60	&			--200,140	&--\\
G\,349.092+0.106	&	11	&	--81.4	&	--82.2,--78.1	&	12&\multicolumn{6}{c}{detection}	\\
G\,349.151+0.021	&	3.4	&	14.5	&	11.5,24.8	&	1.6	&	--			&		&			$<$0.60	&			--200,140	&--\\
G\,349.579--0.679	&	1.9	&	--24.9	&	--25.7,--24.0	&	1.8	& 	--			&		&			$<$0.55	&			--200,140	&--\\
G\,349.799+0.108*	&	3.0	&	--64.7	&	--65.2,--57.3	&	7.8	&\multicolumn{6}{c}{detection}\\
G\,349.884+0.231	&	7.0	&	16.2	&	13.5,17.3	&	7.6	& 	--			&		&			$<$0.55	&			--200,140	&--\\
G\,350.011--1.342	&	2.4	&	--25.8	&	--26.1,--25.3	&	0.9	& 	--			&		&			$<$0.60	&			--200,140	&--\\
G\,350.015+0.433	&	7.2	&	--30.3	&	--36.3,--28.7	&	7.8		&	--			&		&			$<$0.55	&			--200,140	&--\\
G\,350.104+0.084	&	12	&	--68	&	--68.9,--67.6	&	6.0	&	--			&		&			$<$0.75	&			--200,140	&  $<$0.83	& --140,200  \\
G\,350.105+0.083*	&	14	&	--74	&	--79.3,--60.3	&	58&\multicolumn{6}{c}{detection}	\\
G\,350.116+0.084	&	13	&	--68	&	--68.9,--67.0	&	7.8		&	--			&		&			$<$0.60	&			--200,140	& $<$0.83 & --140,200\\
G\,350.116+0.220	&	2.8	&	4.2	&	3.5,7.4	&	1.2	&	\multicolumn{6}{c}{detection}	\\
G\,350.189+0.003	&	1.1	&	--62.5	&	--64.4,--62.2	&	0.4	&	--			&		&			$<$0.60	&			--200,140	&	--\\
G\,350.299+0.122	&	31	&	--62.1	&	--69.1,--60.2	&	39	&\multicolumn{6}{c}{detection}\\
G\,350.340+0.141	&	2.3	&	--58.3	&	--59.4,--57.1	&	1.2	&\multicolumn{6}{c}{detection}\\
G\,350.344+0.116	&	21	&	--65.4	&	--66.1,--55.5	&	21	&\multicolumn{6}{c}{detection}\\
G\,350.356--0.068	&	1.4	&	--67.6	&	--69.3,--66.2	&	1.2		&	--			&		&			$<$0.60	&		--200,140	&--	\\
G\,350.470+0.029	&	1.4	&	--6.3	&	--11.8,--5.9	&	1.1	&\multicolumn{6}{c}{detection}\\
G\,350.520--0.350	&	1.7	&	--24.6	&	--24.8,--13.8	&	0.9	&	--			&		&			$<$0.55	&			--200,140	&--\\
G\,350.686--0.491	&	18	&	--13.7	&	--14.9,--13.6	&	11	&\multicolumn{6}{c}{detection}\\
G\,350.776+0.138	&	0.6	&	38.7	&	33.5,42.5	&	0.5&	--			&		&			$<$0.60	&			--200,140	&	--\\
G\,351.161+0.697	&	17	&	--5.2	&	--12.2,--2.2	&	12	&	--			&		&			$<$0.75	&			--200,140	&	$<$0.80 & --140,200\\
G\,351.242+0.670	&	0.8	&	2.4	&	2.3,2.7	&	0.2	&	--			&		&			$<$0.80	&			--200,140	&--	\\
G\,351.251+0.652	&	1.0	&	--7.1	&	--10.6,--7.0	&	0.4	&	--			&		&		$<$0.65	&			--200,140	&--\\
G\,351.382--0.181	&	20	&	--59.7	&	--68.4,--58.7	&	10&\multicolumn{6}{c}{detection}	\\
G\,351.417+0.645*	&	3423	&	--10.4	&	--12.1,--6.1	&	2932	&\multicolumn{6}{c}{detection}\\
G\,351.417+0.646*	&	t	&	--11.2	&	--12,--7	&		&\multicolumn{6}{c}{detection}\\
G\,351.445+0.660*	&	t	&	--9.2	&	--14,1	&		&\multicolumn{6}{c}{detection}\\
G\,351.581--0.353	&	47	&	--94.2	&	--100.4,--87.0	&	50&	$<$0.75	&			--150,190			&	$<$0.80			&	--200,140	&--	\\
G\,351.611+0.172*	&	4	&	--43.6	&	--46.1,--31.8	&	12	&\multicolumn{6}{c}{detection}\\
G\,351.688+0.171	&	41	&	--36.1	&	--47.6,--35.3	&	56	&\multicolumn{6}{c}{detection}\\
G\,351.775--0.536	&	231	&	--1.3	&	--3.6,2.6	&	206	&\multicolumn{6}{c}{detection}\\
G\,352.083+0.167	&	6.8	&	--66.1	&	--67.0,--63.8	&	4.1&\multicolumn{6}{c}{detection}	\\
G\,352.111+0.176	&	7.5	&	--54.8	&	--60.9,--49.4	&	11&\multicolumn{6}{c}{detection}	\\
G\,352.133--0.944	&	16	&	--7.7	&	--18.5,--5.3	&	30	&	$<$0.75			&	--150,190	&			$<$0.80			&	--200,140	& $<$0.80 & --140,200\\
G\,352.517--0.155	&	9.7	&	--51.2	&	--52.3,--48.4	&	11&	$<$0.70			&	--150,190	&			$<$0.80			&	--200,140	&--	\\
G\,352.525--0.158	&	0.6	&	--53.0	&	--55.4,--52.5	&	0.4		&	$<$0.75			&	--150,190	&			$<$0.80			&	--200,140	&--\\
G\,352.584--0.185	&	6.4	&	--85.7	&	--94.7,--78.4	&	5.3	&	$<$0.75			&	--150,190	&		$<$0.80			&	--200,140	&--\\
G\,352.604--0.225	&	3.3	&	--81.7	&	--84.9,--79.4	&	2.9	&	$<$0.75			&	--150,190	&			$<$0.80			&	--200,140	&--\\
G\,352.624--1.077	&	18	&	--5.9	&	--1.9,--6.9	&	15	&	$<$0.75			&	--150,190	&			$<$0.80			&	--200,140	&	--\\
G\,352.630--1.067	&	184	&	--2.9	&	--7.5,--2.1	&	101	&	$<$0.75			&	--150,190	&			$<$0.80			&	--200,140	&	--\\
G\,352.855--0.201	&	1.3	&	--51.3	&	--53.5,--50.0	&	1.4	&	$<$0.75			&	--150,190	&			$<$0.80			&	--200,140	&--\\
G\,353.216--0.249	&	0.6	&	--10.7	&	--23.1,--10.4	&	0.5&	$<$0.75			&	--150,190	&			$<$0.80			&	--200,140	&--	\\
G\,353.273+0.641	&	8.3	&	--4.4	&	--6.4,--2.8	&	7.8	&	$<$0.75			&	--150,190	&			$<$0.80			&	--200,140	&--\\
G\,353.363--0.166	&	2.8	&	--79.0	&	--80.7,--77.2	&	0.9	&	$<$0.70		&	--150,190	&			$<$0.80	&			--200,140	&--	\\
G\,353.370--0.091	&	1.4	&	--44.7	&	--61.9,--41.3	&	2.4	&\multicolumn{6}{c}{detection}\\
G\,353.378+0.438	&	1.0	&	--15.7	&	--16.2,--14.4	&	1.2	&	$<$0.75			&	--150,190	&			$<$0.55			&	--200,140	& $<$0.80 & --140,200\\
G\,353.410--0.360*	&	115	&	--20.3	&	--22.5,--18.6	&	106	&	\multicolumn{6}{c}{detection}\\
G\,353.429--0.090	&	13	&	--61.8	&	--65.6,--45.5	&	16	&	$<$0.75			&	--150,190	&			$<$0.80			&	--200,140	&--\\
G\,353.464+0.562	&	12	&	--50.3	&	--52.3,--49.6	&	9.6&	$<$0.70			&	--150,190	&			$<$0.75			&	--200,140	&--	\\
G\,353.537--0.091	&	2.5	&	--56.6	&	--58.3,--54.0	&	2.7	&\multicolumn{6}{c}{detection}\\
G\,354.206--0.038	&	1.1	&	--37.1	&	--37.2,--35.5	&	0.4	&	$<$0.75			&	--150,190	&			$<$0.80			&	--200,140	&--\\
G\,354.308--0.110	&	3.4	&	18.8	&	11.2,19.1	&	5.3	&	$<$0.75			&	--150,190	&			$<$0.55			&	--200,140	&--\\
G\,354.496+0.083	&	8.4	&	27.0	&	17.6,27.3	&	6.3	&\multicolumn{6}{c}{detection}\\
G\,354.615+0.472*	&	167	&	--24.4	&	--27.0,--13.1	&	277	&\multicolumn{6}{c}{detection}\\
G\,354.701+0.299	&	1.3	&	102.8	&	98.1,102.9	&	1		&	$<$0.75			&	--150,190	&			$<$0.80			&	--200,140	&--\\
G\,354.724+0.300	&	13	&	93.9	&	91.3,94.4	&	13	&\multicolumn{6}{c}{detection}\\

\end{tabular}
\end{table*}

\begin{table*}\addtocounter{table}{-1}
  \caption{-- {\emph {continued}}}
  \begin{tabular}{lllllclllclllcll} \hline
 \multicolumn{1}{c}{\bf Methanol maser} &\multicolumn{4}{c}{\bf 6.7-GHz properties} & \multicolumn{6}{c}{\bf 12.2-GHz observations}\\
    \multicolumn{1}{c}{\bf ($l,b$)}& {\bf S$_{6.7}$} & {\bf Vp$_{6.7}$} & {\bf Vr$_{6.7}$ } & {\bf  I$_{6.7}$} & \multicolumn{2}{c}{\bf 2008 June} & \multicolumn{2}{c}{\bf 2008 December} & \multicolumn{2}{c}{\bf 2010 March}\\	
      \multicolumn{1}{c}{\bf (degrees)}  &{\bf (Jy)} &&& & \multicolumn{1}{c}{\bf 5-$\sigma$} & \multicolumn{1}{c}{\bf Vr} & \multicolumn{1}{c}{\bf 5-$\sigma$} & \multicolumn{1}{c}{\bf Vr} & \multicolumn{1}{c}{\bf 5-$\sigma$} & \multicolumn{1}{c}{\bf Vr}\\  \hline \hline
G\,355.184--0.419	&	1.3	&	--1.4	&	--1.8,0.3	&	c	&	--			&		&			$<$0.70	&			--200,140	&--\\
G\,355.343+0.148*	&	1.4	&	6.0	&	5.0,6.5	&	0.9	&	\multicolumn{6}{c}{detection}\\
G\,355.344+0.147	&	10	&	20.0	&	18.7,20.5	&	6.1&	--			&		&			$<$0.60	&			--200,140	&	$<$0.80 & --140,200	\\
G\,355.346+0.149	&	7.4	&	10.5	&	8.2,12.2	&	5.6	&	--			&		&			$<$0.60	&			--200,140	& $<$0.75 & --140,200\\
G\,355.538--0.105	&	1.3	&	3.8	&	--3.0,4.7	&	2.4	&	--			&		&			$<$0.60	&			--200,140	&--\\
G\,355.545--0.103	&	1.2	&	--28.2	&	--30,--27.9	&	1.0		&	--			&		&			$<$0.80	&			--200,140	&--\\
G\,355.642+0.398	&	1.4	&	--7.9	&	--8.3,--7.6	&	0.8&	--			&		&			$<$0.55	&			--200,140	&--	\\
G\,355.666+0.374	&	2.5	&	--3.3	&	--4.2,0.6	&	2.1	&	$<$0.70			&	--150,190			&	$<$0.80			&	--200,140	&--\\
G\,356.054--0.095	&	0.5	&	16.7	&	16.5,17.0	&	0.2	&	$<$0.70			&	--150,190			&	$<$0.80			&	--200,140	&--\\
      G\,356.662--0.263	&	8.4	&	--53.8	&	--56.4,--43.8	&	11	&	$<$0.75			&	--150,190			&	$<$0.55			&	--200,140	&--\\
G\,357.558--0.321	&	2.7	&	--3.9	&	--5.0,--0.5	&	1.5	&	$<$0.75			&	--150,190			&	$<$0.80			&	--200,140	&	--\\
G\,357.559--0.321	&	2	&	16.2	&	15.6,17.8	&	1.9	&\multicolumn{6}{c}{detection}\\
G\,357.922--0.337	&	1	&	--4.6	&	--6.0,--4.4	&	0.8	&	$<$0.75			&	--150,190			&	--	&				&	$<$0.58 & --140,200\\
G\,357.924--0.337	&	2.3	&	--2.1	&	--4.3,2.9	&	1.7	&\multicolumn{6}{c}{detection}\\
G\,357.965--0.164*	&	2.8	&	--8.6	&	--8.9,4.6	&	3.7	&	$<$0.75			&	--150,190			&	$<$0.80			&	--200,140	&	--\\
G\,357.967--0.163*	&	47&	--3.1	&	--5.9,0.4	&	104	&\multicolumn{6}{c}{detection}\\
G\,358.263--2.061	&	19	&	5.0	&	1.0,6.2	&	17&\multicolumn{6}{c}{detection}	\\
G\,358.371--0.468*	&	44	&	1.3	&	--0.9,13.2	&	31&\multicolumn{6}{c}{detection}	\\
G\,358.386--0.483	&	7.0	&	--6.0	&	--6.2,--5.8	&	1.9	&	$<$0.85			&	--150,190			&	$<$0.90			&	--200,140	&--\\
G\,358.460--0.391	&	0.9	&	3.3	&	--0.5,3.5	&	0.5	&	$<$0.70			&	--150,190			&	$<$0.80			&	--200,140	&   $<$0.56 & --140,200 \\
G\,358.460--0.393	&	11	&	--7.3	&	--8.4,5.6	&	9.7	&	$<$0.70			&	--150,190			&	$<$0.80 	&		--200,140		& $<$0.56 & --140,200	\\
G\,358.721--0.126	&	3.0	&	10.6	&	9.3,14.9	&	4.2	&\multicolumn{6}{c}{detection}\\
G\,358.809--0.085*	&	6.9	&	--56.2	&	--59.1,--51.2	&	6.6	&\multicolumn{6}{c}{detection}\\
G\,358.841--0.737	&	11	&	--20.6	&	--29.3,--17.6	&	6.7	&\multicolumn{6}{c}{detection}\\
G\,358.906+0.106	&	1.9	&	--17.9	&	--20.3,--17.2	&	1.7	&\multicolumn{6}{c}{detection}\\
G\,358.931--0.030	&	5.9	&	--15.9	&	--22.2,--14.7	&	9.0	&	$<$0.75			&	--150,190			&	$<$0.80			&	--200,140	&--\\
G\,358.980+0.084	&	1.1	&	6.2	&	5.5,7.0	&	c	&	$<$0.75			&	--150,190			&	$<$0.75			&	--200,140	&--\\
G\,359.138+0.031	&	16	&	--3.9	&	--5.0,0.5	&	14&	$<$0.75			&	--150,190			&	$<$0.80			&	--200,140	&--	\\
G\,359.436--0.104	&	14	&	--52.3	&	--57.5,--50.5	&	18	&\multicolumn{6}{c}{detection}\\
G\,359.436--0.102	&	73	&	--46.8	&	--53.0,--45.5	&	50		&	$<$0.75			&	--150,190			&	$<$0.80			&		&--\\
G\,359.615--0.243*	&	39	&	19.3	&	18.1,25.1	&	56	&\multicolumn{6}{c}{detection}\\
G\,359.938+0.170	&	2.3	&	--0.5	&	--0.8,--0.3	&	0.7	&	$<$0.70			&	--150,190			&	$<$0.80		&	--200,140	&--\\
G\,359.970--0.457	&	2.4	&	23.1	&	22.8,23.2	&	0.7	&	$<$0.75			&	--150,190			&	$<$0.80			&	--200,140	&--\\
G\,0.092--0.663	&	19	&	23.8	&	9.8,25.4	&	34	&\multicolumn{6}{c}{detection}\\
G\,0.167--0.446	&	1.3	&	13.8	&	11.9,16.5	&	0.6	&	$<$0.70			&	--150,190			&	$<$1.25			&	--200,140	&--\\
G\,0.212--0.001	&	3.3	&	49.5	&	42.1,50.3	&	4.8	&\multicolumn{6}{c}{detection}\\
G\,0.315--0.201	&	68	&	19.4	&	14.6,28.0	&	71&\multicolumn{6}{c}{detection}	\\
G\,0.316--0.201	&	0.6	&	21.0	&	20.8,22.7	&	0.4	&	$<$0.55			&	--150,190			&	--			&		&--\\
G\,0.376+0.040		&	0.6	&	37.1	&	37.0,37.3	&	0.2	&	$<$0.75			&	--150,190	&			$<$0.80			&	--200,140	&--\\
G\,0.409--0.504	&	2.6	&	25.4	&	22.5,26.6	&	1.9	&	$<$0.75			&	--150,190			&	$<$0.80			&	--200,140	&--\\
G\,0.475--0.010	&	3.1	&	28.8	&	25.8,33.2	&	4.8 &	$<$0.75			&	--150,190			&	$<$0.80			&	--200,140	&	--\\
G\,0.496+0.188		&	24	&	0.9	&	--12.4,2.1& 51	&\multicolumn{6}{c}{detection}\\
G\,0.546--0.852*	&	62	&	11.8	&	8.0,20.3	&	201	&\multicolumn{6}{c}{detection}\\
G\,0.645--0.042	&	54	&	49.5	&	46,53	&		&$<$0.60		&	--150,190	&			--	&				&--\\
G\,0.647--0.055	&	2.0	&		&	49,52	&		&$<$0.60		&	--150,190	&			--	&				&--\\
G\,0.651--0.049*	&	21	&	48.3	&	46,49	&		&\multicolumn{4}{c}{detection}\\
G\,0.657--0.041	&	1.8	&		&	48,56	&		&$<$0.60		&	--150,190	&			--	&				&--\\
G\,0.665--0.036	&	2.6	&	60.4	&	58,62	&		&$<$0.60		&	--150,190	&			--	&				&--\\
G\,0.666--0.029*	&	34	&	70.5	&	68,73	&		&\multicolumn{4}{c}{detection}\\
G\,0.667--0.034*	&	0.4	&		&	49,56	&		&\multicolumn{4}{c}{detection}\\
G\,0.672--0.031	&	7.3	&	58.2	&	55,59	&		&$<$0.60		&	--150,190	&			--	&				&--\\
G\,0.673--0.029	&	0.4	&		&	65.5,66.5	&		&$<$0.60		&	--150,190	&			--	&				&--\\
G\,0.677--0.025	&	4.9	&	73.3	&	70,77	&		&$<$0.60		&	--150,190	&			--	&				&--\\
G\,0.695--0.038	&	32	&	68.6	&	64,75	&		&	$<$0.60		&	--150,190	&			--	&				&--\\
G\,0.836+0.184*		&	6.6	&	3.5	&	--2.2,4.1	&	3.7	&\multicolumn{6}{c}{detection}\\
G\,1.008--0.237	&	14	&	1.6	&	1.0,7.4	&	10	&	$<$0.75			&	--150,190	&			$<$0.80			&	--200,140	&--\\
G\,1.147--0.124*	&	3.0	&	--15.3	&	--20.7,--14.7	&	4	&	$<$0.50			&	--150,190	&		&	$<$0.80	&	--200,140\\
G\,1.329+0.150		&	2.1	&	--12.2	&	--12.3,--11.9	&	0.6	&	$<$0.70			&	--150,190	&			$<$0.80			&	--200,140	&--\\
G\,1.719--0.088	&	7.8	&	--8.1	&	--8.4,--2.3	&	6.7	&\multicolumn{6}{c}{detection}\\
G\,2.143+0.009		46.5	&	7.1	&	62.6	&	53.4,64.7	&	18&	$<$0.70			&	--150,190	&			$<$0.80			&	--200,140	&--	\\
G\,2.521--0.220	&	&	1.0	&	--6.1	&	--7.2,4.4	&	0.8	&	$<$0.70			&	--150,190	&			$<$0.75			&	--200,140	&--\\
G\,2.536+0.198		&	29	&	3.1	&	2.2,20.3	&53	&\multicolumn{6}{c}{detection}\\
G\,2.591--0.029	&	1.8	&	--8.3	&	--9.0,--4.0	&	2.1	&	$<$0.70			&	--150,190	&			$<$0.80			&	--200,140	&--\\
\end{tabular}
\end{table*}

\begin{table*}\addtocounter{table}{-1}
  \caption{-- {\emph {continued}}}
  \begin{tabular}{lllllclllclllcll} \hline
 \multicolumn{1}{c}{\bf Methanol maser} &\multicolumn{4}{c}{\bf 6.7-GHz properties} & \multicolumn{6}{c}{\bf 12.2-GHz observations}\\
    \multicolumn{1}{c}{\bf ($l,b$)}& {\bf S$_{6.7}$} & {\bf Vp$_{6.7}$} & {\bf Vr$_{6.7}$ } & {\bf  I$_{6.7}$} & \multicolumn{2}{c}{\bf 2008 June} & \multicolumn{2}{c}{\bf 2008 December} & \multicolumn{2}{c}{\bf 2010 March}\\	
      \multicolumn{1}{c}{\bf (degrees)}  &{\bf (Jy)} &&& & \multicolumn{1}{c}{\bf 5-$\sigma$} & \multicolumn{1}{c}{\bf Vr} & \multicolumn{1}{c}{\bf 5-$\sigma$} & \multicolumn{1}{c}{\bf Vr} & \multicolumn{1}{c}{\bf 5-$\sigma$} & \multicolumn{1}{c}{\bf Vr}\\  \hline \hline
G\,2.615+0.134		&	1.2	&	94.1	&	93.7,103.7	&	1.9	&\multicolumn{6}{c}{detection}\\
G\,2.703+0.040		&	9.0	&	93.5	&	92.2,97.8	&	16	&\multicolumn{6}{c}{detection}\\
G\,3.253+0.018		&	3.5	&	2.2	&	--1,2.9.0	&	5.3	&\multicolumn{6}{c}{detection}\\
 G\,3.312--0.399	&	1.2	&	0.4	&	0.2,9.0	&	0.8	&	$<$0.70			&	--150,190	&			$<$0.75			&	--200,140	&--\\
G\,3.442--0.348	&	1.1	&	--35.2	&	--35.3,--35.0	&	0.3	&	$<$0.70			&	--150,190	&			$<$0.75			&	--200,140	&--\\
G\,3.502--0.200*	&	1.6	&	43.9	&	43.6,45.3	&	0.9	&\multicolumn{6}{c}{detection}\\
G\,3.910+0.001		&	5.0	&	17.8	&	16.9,24.0	&	4.1	&	$<$0.70			&	--145,195	&			$<$0.75			&	--200,140	&--\\
G\,4.393+0.079		&	6.7	&	1.9	&	0.2,9.2	&	11	&\multicolumn{6}{c}{detection}\\
     G\,4.434+0.129		&	3.3	&	--1.0	&	--1.3,13.8	&	2.2	&	$<$0.70			&	--145,195	&			$<$0.75			&	--200,140	&--\\
G\,4.569--0.079	&	0.4	&	9.5	&	9.2,9.6	&	0.1	&	$<$0.70			&	--145,195	&			$<$0.75			&	--200,140	&--\\
G\,4.586+0.028		&	1.2	&	26.1	&	14.2,26.9	&	1.4	&	$<$0.75			&	--145,195	&			$<$0.75	&			--200,140	&--\\
G\,4.676+0.276		&	2.1	&	4.5	&	--5.1,5.5	&	3.4	&	$<$0.75			&	--145,195	&			$<$0.80	&			--200,140	&--\\
G\,4.866--0.171	&	0.4	&	5.3	&	5.2,5.5	&	0.1	&	--			&		&			$<$0.55	&			--200,140	&--\\
G\,5.618--0.082	&	3.4	&	--27.1	&	--27.5,--18.3	&	4.7	&	\multicolumn{6}{c}{detection}\\
G\,5.630--0.294	&	1.3	&	10.5	&	9.4,21.6	&	3	&\multicolumn{6}{c}{detection}\\
G\,5.657+0.416		&	1.7	&	20.0	&	13.0,25.1	&	2.8	&	$<$0.70	&			--145,195	&			$<$0.75	&			--200,140	&--\\
G\,5.677--0.027	&	0.8	&	--11.7	&	--12.5,--11.4	&	0.4	&	$<$0.70			&	--145,195	&			$<$0.75			&	--200,140	&--\\
G\,5.885--0.393	&	0.5	&	6.0	&	6.7,7.5	&	1	&	$<$0.75			&	--145,195	&			$<$0.80			&	--200,140	&--\\
G\,5.900--0.430	&	6.2	&	10.0	&	0.7,10.6	&	6.1	&	$<$0.75			&	--145,195	&			$<$0.55			&	--200,140	& $<$0.69 & --140,200\\
G\,6.189--0.358	&	229	&	--30.1	&	--37.7,--27.8	&	193	&\multicolumn{6}{c}{detection}\\
G\,6.368--0.052	&	1.5	&	144.1	&	135.8,147.7	&	2.7	&	$<$0.70			&	--145,195	&			$<$0.75			&	--200,140	&--\\
G\,6.539--0.108	&	0.6	&	13.1	&	5.9,14.9	&	0.7	&	$<$0.70			&	--145,195	&			$<$0.75			&	--200,140	&--\\
G\,6.588--0.192	&	8.0	&	5.1	&	3.1,7.2	&	9.6	&\multicolumn{6}{c}{detection}\\
G\,6.610--0.082	&	23	&	0.8	&	0.4,10.8	&	15	&\multicolumn{6}{c}{detection}\\
G\,6.795--0.257	&	91	&	16.3	&	12.2,31.4	&	268	&\multicolumn{6}{c}{detection}\\
G\,6.881+0.093		&	3.1	&	--2.2	&	--3.6,--2.0	&	1.6	&	$<$0.70	&			--145,195	&			$<$0.75	&			--200,140	& $<$0.58 & --140,200\\
G\,7.166+0.131		&	2.6	&	85.7	&	74.4,90.2	&	4.1&		\multicolumn{6}{c}{detection}\\
G\,7.601--0.139	&	8.7	&	154.6	&	151.3,156.3	&	8.4	&	$<$0.75			&	--145,195	&			$<$1.25			&	--200,140	&--\\
G\,7.632--0.109	&	6.5	&	157.0	&	146.7,158.5	&	8.7	&\multicolumn{6}{c}{detection}\\
G\,8.139+0.226		&	11	&	19.9	&	19.0,21.0	&	7.6&	$<$0.55	&			--145,195	&			$<$0.80	&			--200,140	&	--\\
G\,8.317--0.096	&	4.5	&	47.0	&	44.8,50.5	&	5.6	&\multicolumn{6}{c}{detection}\\
G\,8.669--0.356*	&	10	&	39.2	&	36.0,39.9	&	6.5	&\multicolumn{6}{c}{detection}\\
G\,8.683--0.368*	&	109	&	43.1	&	40.2,45.6	&	101	&\multicolumn{6}{c}{detection}\\
G\,8.832--0.028	&	159	&	--3.8	&	--6.2,5.6	&	358	&\multicolumn{6}{c}{detection}\\
G\,8.872--0.493	&	34	&	23.3	&	22.8,26.8	&	30&	$<$0.70			&	--140,200	&			$<$0.75			&	--200,140	&	--\\
G\,9.215--0.202	&	12	&	45.5	&	36.7,49.8	&	14	&\multicolumn{6}{c}{detection}\\
G\,9.621+0.196	*	&	5240	&	1.3	&	--3.2,8.9	&	2211	&\multicolumn{6}{c}{detection}\\
G\,9.619+0.193	*	&	76	&	5.3	&	5.1,6.7	&	36	&\multicolumn{6}{c}{detection}\\
G\,9.986--0.028	&	68	&	42.2	&	40.6,51.4	&	85	&\multicolumn{6}{c}{detection}\\ \hline
\end{tabular}

\end{table*}
\clearpage

\begin{table*}
\caption{Characteristics of the 12.2-GHz methanol maser emission that we detect. Column 1 gives the Galactic longitude and latitude of each source and is used as an identifier; and columns 2 and 3 give the equatorial coordinates for each of the sources and have been derived from interferometric observations of the 6.7-GHz methanol masers detected in the MMB survey \citep{CasMMB10,GreenMMB10,CasMMB102}. References for previously detected 12.2-GHz sources follow the source name and are as follows; 1: \citet{Breen10a}; 2: \citet{Caswell95b}; 3: ~\citet{Gay}; 4: \citet{Koo88}; 5: \citet{Catarzi93}; 6: \citet{Kemball88}, 7: \citet{Batrla87}; 8: \citet{Cas93}; 9: \citet{Norris1987}; 10: \citet{MacLeod93}; and are presented as superscripts after the source name (and are followed by a `*' if we fail to detect any emission). Column 4 gives the epoch of the observed data (sometimes indicating that the presented data is the average of multiple epochs) presented in columns 5 - 8 which give the peak flux density (Jy) or 5-$\sigma$ detection limit, velocity of the 12.2-GHz peak (\kmsns), velocity range (\kmsns) and integrated flux density (Jy \kmsns), respectively. The presence of a `--' in any of the columns showing source flux density indicates that no observation was made at the given epoch. The presence of a `t' in any of the columns indicates that the characteristics associated with a source are discussed in Section~\ref{sect:12GHz_ind}.  The presence of `conf' in the columns showing the integrated flux density indicates that a value for this characteristic could not be extracted do to confusion from nearby sources. Detection limits that are followed by `$^{*}$' indicate that while no detection was made at the given epoch, emission is detected in the average spectrum. Flux densities derived from averaged spectrums are followed by a `$^{\alpha}$'. For one source, observations from the Hobart radio telescope are presented and this is indicated by the presence of a `$^{\beta}$' following the flux density measurement.} 
  \begin{tabular}{lllllccclclllclllcll} \hline
\multicolumn{1}{c}{\bf Methanol maser} &\multicolumn{1}{c}{\bf RA} & \multicolumn{1}{c}{\bf Dec}  &{\bf Epoch}&  {\bf S$_{12.2}$} & {\bf Vp$_{12.2}$} & {\bf Vr$_{12.2}$} & \multicolumn{1}{c}{\bf I$_{12.2}$} \\
    \multicolumn{1}{c}{\bf ($l,b$)}&  \multicolumn{1}{c}{\bf (J2000)} & \multicolumn{1}{c}{\bf (J2000)}   & & {\bf (Jy)} & {\bf (\kmsns)} & {\bf  (\kmsns)} & \multicolumn{1}{c}{\bf (Jy \kmsns)}\\
    \multicolumn{1}{c}{\bf (degrees)}  & \multicolumn{1}{c}{\bf (h m s)}&\multicolumn{1}{c}{\bf ($^{o}$ $'$ $``$)}\\  \hline \hline
G\,330.070+1.064			&	16 00 15.43	&	--51 34 25.6	&	2008 Dec	&	3.5	&	--42.1	&	--42.5,--41.8	&	1.5	\\
G\,330.226+0.290			&	16 04 18.94	&	--52 03 12.8	&	2008 Jun	&	1.4	&	--92.2	&	--92.4,--90.3	&	0.3	\\
						&				&				&	2008 Dec	&	$<$0.75	&		&	--200,140	& \\	
G\,330.283+0.493			&	16 03 43.06	&	--51 51 48.3	&	2008 Jun	&	1.6	&	--88.1	&	--89.1,--87.9	&	1.1	\\	
G\,331.120--0.118			&	16 10 23.04	&	--51 45 20.6	&	2008 Jun	&	0.5	&	--93.5	&	--93.6,--92.9	&	0.2	\\
						&				&				&	2008 Dec	&	0.7	&	--93.6	&	--93.7,--92.7	&	0.5	\\
						&				&				&	2010 Mar	& 	0.7 & --93.2 & --93.6,--92.7 & 0.4\\
G\,331.132--0.244			&	16 10 59.77	&	--51 50 22.4	&	2008 Jun	&	$<$0.75$^{*}$	&		&	--165,175	&		\\
						&				&				&	2008 Dec	&	$<$0.60$^{*}$	&		&	--200,140	&	\\
						&				&				&	Avg 2008 Jun,Dec	&	0.3$^{\alpha}$	&	--84.6	&	--84.9,--84.6	&	0.1\\
						&				&				&	2010 Mar	&	0.9 & --84.2 & --91.5, --82.4 & 0.3 \\	
G\,331.278--0.188$^{1,2,8,9}$	&	16 11 26.59	&	--51 41 56.7	&	2008 Jun	&	76	&	--78.8	&	--83.2,--77.8	&	56	\\	
G\,331.342--0.346			&	16 12 26.45	&	--51 46 16.4	&	2008 Jun	&$<$0.75$^{*}$	&		&	--160,180	&		\\
						&				&				&	2008 Dec	&$<$0.60$^{*}$	&		&	--200,140	&	\\
						&				&				&	Avg 2008 Jun,Dec	&	0.4$^{\alpha}$	&	--71.0	&	--71.8,--71.0	&	0.2	\\
						&				&				&	2010 Mar	& 0.5 &--71.2 & --72.2,--71.1 & 0.1 \\	
G\,331.425+0.264			&	16 10 09.36	&	--51 16 04.5	&	2008 Dec	&	6.7	&	--88.6	&	--90.5,--78.5	&	4.8 	\\
G\,331.437--0.304			&	16 12 42.00	&	--51 40 30.9	&	2008 Jun	&	$<$0.75	&		&	--160,180	&		\\
						&				&				&	2008 Dec	&	0.5	&	--89.9	&	--90,--89.5	&	0.1 	\\
G\,331.442--0.187$^{3}$		&	16 12 12.49	&	--51 35 10.1	&	2008 Dec	&	6.4	&	--87.7	&	--92.9,--84.5	&	4.7\\
G\,331.543--0.066			&	16 12 09.14	&	--51 25 45.3	& 	2008 Jun	&	$<$0.75		&	&	--160,180	&		\\
						&				&				&	2008 Dec	&	$<$0.60	&		&	--200,140	& \\	
						&				&				&	2010 Mar	&	 0.7 & --84.5 & --84.6,--84.3 &0.2\\
G\,331.556--0.121$^{2}$		&	16 12 27.21	&	--51 27 38.2	&	2008 Dec	&	8.9	&	--104.4	&	--105,--96.7	&	8.5 	\\
G\,331.900--1.186			&	16 18 49.13	&	--51 59 28.2	&	2008 Jun	&	$<$0.75$^{*}$	&		&	--160,180	&		\\
						&				&				&	2008 Dec	&	$<$0.55$^{*}$	&		&	--200,140	& \\
						&				&				&	2010 Mar	&	 $<$0.57 & & --150,190	\\	
						&				&				&	Avg all	&	0.4$^{\alpha}$	&	--46.8	&	--59.1,--46.8	&	0.1	\\
G\,332.094--0.421			&	16 16 16.45	&	--51 18 25.7	&	2008 Jun	&	1.2	&	--58.5	&	--60.4,--58.5	&	0.3	\\	
G\,332.295--0.094			&	16 15 45.40	&	--50 55 53.7	&	2008 Jun	&	3.1	&	--43.4	&	--45.1,--42.2	&	1.7\\	
G\,332.295+2.280$^{3}$		&	16 05 41.72	&	--49 11 30.3	&	2008 Jun	&	64	&	--24	&	--24.7,--22.1	&	31	\\	
G\,332.726--0.621$^{1}$		&	16 20 02.95	&	--51 00 32.0	&	2008 Jun	&	0.5	&	--47.4	&	--47.5,--47.3	&	0.1	\\
						&				&				&	2010 Mar	&  $<$0.58 & & --150,190	\\	
G\,332.813--0.701			&	16 20 48.12	&	--51 00 15.6	&	2008 Jun	&	0.4	&	--58.2	&	--58.7,--58.0	&	0.2	\\
						&				&				&	2008 Dec	&	0.4	&	--58.6	&	--58.9,--58.4	&	0.1	\\
						&				&				&	2010 Mar	& 0.7 & --58.2 & --58.3,--58.1 &0.1\\
G\,333.029--0.015			&	16 18 44.17	&	--50 21 50.6	&	2008 Jun	&0.9	&	--53.8	&	--61,--53.7	&	0.6	\\	
G\,333.068--0.447$^{2}$		&	16 20 48.97	&	--50 38 40.4	&	2008 Jun	&$<$0.75	&		&	--160,180	&		\\
						&				&				&	2008 Dec	&	0.4	&	--54.4	&	--54.6,--54.4	&	0.1	\\
						&				&				&	2010 Mar	&	0.7 & --54.5 & --54.6,--54.4 & 0.1 \\
G\,333.163--0.101$^{2}$		&	16 19 42.67	&	--50 19 53.2	&	2008 Dec	&	5.5	&	--95.2	&	--95.5,--94.8	&	2.1\\
G\,333.184--0.091			&	16 19 45.62	&	--50 18 35.0	&	2008 Jun	&	0.5	&	--81.7	&	--81.9,--81.4	&	0.2	\\
						&				&				&	2008 Dec	&	0.7	&	--81.7	&	--82,--81.4	&	0.2	\\
						&				&				&	2010 Mar	& 0.6 & --81.7 & --81.8,--81.6 & 0.1\\
G\,333.562--0.025			&	16 21 08.80	&	--49 59 48.0	&	2008 Dec	&	15	&	--34.9	&	--44.1,--34.1	&	14 	\\
G\,333.646+0.058			&	16 21 09.14	&	--49 52 45.9	&	2008 Jun	&	0.6	&	--87.5	&	--87.7,--87.5	&	0.1	\\
						&				&				&	2008 Dec	&	$<$0.75	&		&	--200,140	&	\\
						&				&				&	2010 Mar	&	 0.5 & --87.6 & --87.7,--87.3 & 0.1\\	
\end{tabular}\label{tab:12MMB}

\end{table*}
\clearpage

\begin{table*}\addtocounter{table}{-1}
  \caption{-- {\emph {continued}}}
  \begin{tabular}{lllllccclclllclllcll} \hline
\multicolumn{1}{c}{\bf Methanol maser} &\multicolumn{1}{c}{\bf RA} & \multicolumn{1}{c}{\bf Dec}  &{\bf Epoch}&  {\bf S$_{12.2}$} & {\bf Vp$_{12.2}$} & {\bf Vr$_{12.2}$} & \multicolumn{1}{c}{\bf I$_{12.2}$} \\
    \multicolumn{1}{c}{\bf ($l,b$)}&  \multicolumn{1}{c}{\bf (J2000)} & \multicolumn{1}{c}{\bf (J2000)}   & & {\bf (Jy)} & {\bf (\kmsns)} & {\bf  (\kmsns)} & \multicolumn{1}{c}{\bf (Jy \kmsns)}\\
    \multicolumn{1}{c}{\bf (degrees)}  & \multicolumn{1}{c}{\bf (h m s)}&\multicolumn{1}{c}{\bf ($^{o}$ $'$ $``$)}\\  \hline \hline
						
G\,333.683--0.437			&	16 23 29.78	&	--50 12 08.6	&	2008 Dec	&	8.4	&	--5.5	&	--7.6,--4.6	&	4.8\\
G\,334.138--0.023			&	16 23 39.56	&	--49 35 15.2	&	2008 Jun	&	1.0	&	--31.2	&	--31.3,--31.1	&	0.2	\\	
G\,334.307--0.079			&	16 24 38.12	&	--49 30 24.5	&	2008 Jun	&	1.2	&	--36.5	&	--36.9,--36.2	&	0.5	\\	
G\,334.635--0.015			&	16 25 45.73	&	--49 13 37.4	&	2008 Jun	&	1.7	&	--30.2	&	--30.6,--29.8	&	0.8	\\	
G\,334.933--0.307			&	16 28 18.81	&	--49 12 57.6	&	2008 Jun	&	2.8	&	--102.8	&	--103.7,--101.8	&	2.1	\\	
G\,334.935--0.098			&	16 27 24.25	&	--49 04 11.3	&	2008 Jun	&	0.7	&	--17.9	&	--18.9,--17.2	&	0.5	\\
						&				&				&	2010 Mar	& 0.6 & --17.7 & --18.1,--17.7 & 0.2\\	
G\,335.060--0.427			&	16 29 23.13	&	--49 12 27.1	&	2008 Dec	&	18	&	--47	&	--47.3,--46.4	&	5.7 	\\
G\,335.426--0.240			&	16 30 05.58	&	--48 48 44.8	&	2008 Dec	&	18	&	--50.7	&	--53.3,--50.3	&	8.0	\\
G\,335.556--0.307$^{2}$		&	16 30 55.96	&	--48 45 50.0	&	2008 Dec	&	5.3	&	--114.7	&	--116.7,--114.4	&	1.8 \\
G\,335.585--0.290			& 	16 30 58.67	&	--48 43 50.7	&	2008 Jun	&	1.1	&	--53.5	&	--54.3,--52.8	&	0.6	\\
						&				&				&	2008 Dec	&	1.1	&	--53.3	&	--54.4,--53.2	&	0.5	\\
						&				&				&	2010 Mar	& 0.9 & --53.7 & --53.8,--53.4 & 0.3\\
G\,335.726+0.191$^{2}$		&	16 29 27.37	&	--48 17 53.2	&	2008 Dec	&	8.0	&	--44.5	&	--44.8,--43.5	&	5.8 	\\
G\,335.789+0.174$^{2,8,9}$	&	16 29 47.33	&	--48 15 51.7	&	2008 Dec	& 	162	&	--46.2	&	--51,--45.4	&	146	\\
G\,336.018--0.827$^{3}$		&	16 35 09.30	&	--48 46 46.8	&	2008 Jun	&	7.7	&	--40.2	&	--53.4,--39.5	&	6.5	\\
G\,336.358--0.137$^{2}$		&	16 33 29.17	&	--48 03 43.9	&	2008 Dec	&	2.6	&	--74.7	&	--79.7,--73.5	&	1.5	\\
G\,336.433--0.262$^{2,8}$	&	16 34 20.22	&	--48 05 32.2	&	2008 Dec	&	15	&	--93.1	&	--93.6,--87.1	&	9.6	\\
G\,336.496--0.271			&	16 34 38.02	&	--48 03 03.9	&	2008 Jun	&	7.6	&	--24	&	--24.9,--21.3	&	5.7	\\	
G\,336.822+0.028$^{2}$		&	16 34 38.28	&	--47 36 32.2	&	2008 Dec	&	4.8	&	--76.7	&	--77.3,--76.5	&	1.3	 \\
G\,336.825+0.139			&	16 34 09.82	&	--47 31 52.9	&	2008 Jun	&	1.4	&	--88.4	&	--89.1,--87.3	&	0.7		\\	
G\,336.864+0.005$^{2}$		&	16 34 54.44	&	--47 35 37.3	&	2008 Jun	&	$<$0.75	&		&	--160,180	&		\\
						&				&				&	2008 Dec	&	0.5	&	--75.3	&	--75.4,--74.8	&	0.1 	\\
G\,336.916--0.024			&	16 35 14.27	&	--47 34 30.1	&	2008 Jun	&	2.6	&	--127.2	&	--127.5,--115.0	&	1.2	\\	
G\,336.941--0.156			&	16 35 55.19	&	--47 38 45.4	&	2008 Dec	&	11	&	--66.6	&	--76.7,--64.5	&	12	 \\
G\,336.957--0.225			&	16 36 16.99	&	--47 40 48.6	&	2008 Jun	&	1.0	&	--68	&	--68.2,--67.4	&	0.5	\\	
G\,336.983--0.183			&	16 36 12.41	&	--47 37 58.2	&	2008 Dec	&	2.5	&	--79.8	&	--85.7,--75.2	&	2.7	\\
G\,337.052--0.226			&	16 36 40.12	&	--47 36 38.2	&	2008 Dec	&	0.4	&	--76.7	&	--77,--76.4	&	0.1	 \\
G\,337.097--0.929			&	16 39 57.77	&	--48 02 49.1	&	2008 Jun	&	1.4	&	--40.6	&	--40.9,--40.2	&	0.6	\\	
G\,337.132--0.068			&	16 36 17.67	&	--47 26 41.0	&	2008 Jun	&	$<$0.70	&		&	--160,180	&		\\
						&				&				&	2008 Dec	&	0.7	&	--59.7	&	--60.2,--59.6	&	0.2	\\
G\,337.153--0.395			&	16 37 48.86	&	--47 38 56.5	&	2008 Dec	&	1.2	&	--48.3	&	--50,--48.2	&	0.4	 \\
G\,337.176--0.032			&	16 36 18.84	&	--47 23 19.8	&	2008 Jun	&	0.5	&	--64.8	&	--65.1,--64.7	&	0.2	\\
						&				&				&	2008 Dec	&	0.5	&	--64.7	&	--65.3,--64.7	&	0.2	\\	
						&				&				&	2010 Mar	&	0.8 & --64.8 & --72.3,--64.6 & 0.5\\
G\,337.201+0.114			&	16 35 46.56	&	--47 16 16.7	&	2008 Dec	&	4.4	&	--54.6	&	--57.7,--53.3	&	4.0	\\
G\,337.202--0.094			&	16 36 41.22	&	--47 24 40.2	&	2008 Jun	&	$<$0.80$^{*}$		&	&	--160,180		&	\\
						&				&				&	2010 Mar	&	$<$0.80$^{*}$ & & --150,190\\
						&				&				&	Avg all	& 0.4$^{\alpha}$	& --65.9 & --68.2,--65.9 & 0.2\\
G\,337.388--0.210			&	16 37 56.01	&	--47 21 01.2	&	2008 Dec &	2.9	&	--55.5	&	--66,--54.9	&	4.8 \\	
G\,337.404--0.402$^{2}$		&	16 38 50.51	&	--47 28 00.3	&	2008 Dec	&	4.2	&	--39.4	&	--42.5,--38.5	&	2.1 	\\
G\,337.613--0.060$^{2,8,9}$	&	16 38 09.54	&	--47 04 59.9	&	2008 Dec	&	5.7	&	--42.2	&	--53.6,--40.5	&	7.9 	\\
G\,337.632--0.079$^{2}$		&	16 38 19.13	&	--47 04 53.5	&	2008 Jun	&	$<$0.75	&		&	--160,180	&		\\
						&				&				&	2008 Dec	&	0.5	&	--55.5	&	--56.8,--54.6	&	0.2	\\
						&				&				&	2010 Mar	&	 0.6 & --55.6 & --56.8,--53.1 & 0.2\\
G\,337.705--0.053$^{2,8,9}$	&	16 38 29.63	&	--47 00 35.5	&	2008 Dec	&	94	&	--54.6	&	--55.6,--50.8	&	37 	\\
G\,337.844--0.375			&	16 40 26.67	&	--47 07 13.4	&	2008 Jun	&	0.9	&	--38.6	&	--39.1,--38.1	&	0.3	\\
G\,337.920--0.456$^{2}$		&	16 41 06.05	&	--47 07 02.1	&	2008 Dec	&	0.6	&	--37.8	&	--37.9,--37.7	&	0.2\\	
						&				&				&	2010 Mar	&	 0.7 & --37.8 & --38.9,--37.6 & 0.2 \\
G\,337.966--0.169			&	16 40 01.09	&	--46 53 34.5	&	2008 Jun	&	$<$0.75	&		&	--160,180	&		\\
						&				&				&	2008 Dec	&	0.6	&	--59	&	--59.2,--58.9	&	0.2 \\
G\,338.075+0.012$^{2,8}$	&	16 39 39.04	&	--46 41 28.0	&	2008 Jun	&	5.9	&	--53.1	&	--53.6,--52.6	&	3.0\\	
						&				&				&	2008 Dec	&	7.5	&	--53.1	&	--54,--52.5	&	4.0 \\
G\,338.160--0.064			&	16 40 18.70	&	--46 40 37.8	&	2008 Jun	&	0.8	&	--66.1	&	--66.5,--65.9	&	0.2	\\
						&				&				&	2008 Dec	&	0.6	&	--66.2	&	--66.3,--65.9	&	0.2	 \\
G\,338.388+0.162			&	16 40 12.37	&	--46 21 25.6	&	2008 Jun	&	1.8	&	--30.1	&	--30.2,--29.3	&	1.1	\\	
G\,338.461--0.245$^{2,8}$	&	16 42 15.50	&	--46 34 18.4	&	2008 Dec	&	9.4	&	--52.2	&	--52.6,--51.2	&	5.9	 \\
G\,338.561+0.218			&	16 40 37.96	&	--46 11 25.8	&	2008 Dec	&	26	&	--40.2	&	--41.4,--37.7	&	34 	\\
G\,338.875--0.084$^{2}$		&	16 43 08.25	&	--46 09 12.8	&	2008 Dec	&	1.0	&	--41.4	&	--41.5,--40.4	&	0.5	 \\
G\,338.902+0.394			&	16 41 10.06	&	--45 49 05.4	&	2008 Dec	&	1.0	&	--26	&	--26.5,--25.7	&	0.6 	\\
G\,338.920+0.550			&	16 40 34.01	&	--45 42 07.1	&	2008 Dec	&	0.6	&	--66.8	&	--66.9,--66.7	&	0.1\\
						&				&				&	2010 Mar	&	 $<$0.77 &	&	--150,190	\\
G\,338.925+0.634			&	16 40 13.56	&	--45 38 33.2	&	2008 Dec	&	33	&	--60.6	&	--61.9,--52.8	&	22 \\
\end{tabular}

\end{table*}
\clearpage

\begin{table*}\addtocounter{table}{-1}
  \caption{-- {\emph {continued}}}
  \begin{tabular}{lllllccclclllclllcll} \hline
\multicolumn{1}{c}{\bf Methanol maser} &\multicolumn{1}{c}{\bf RA} & \multicolumn{1}{c}{\bf Dec}  &{\bf Epoch}&  {\bf S$_{12.2}$} & {\bf Vp$_{12.2}$} & {\bf Vr$_{12.2}$} & \multicolumn{1}{c}{\bf I$_{12.2}$} \\
    \multicolumn{1}{c}{\bf ($l,b$)}&  \multicolumn{1}{c}{\bf (J2000)} & \multicolumn{1}{c}{\bf (J2000)}   & & {\bf (Jy)} & {\bf (\kmsns)} & {\bf  (\kmsns)} & \multicolumn{1}{c}{\bf (Jy \kmsns)}\\
    \multicolumn{1}{c}{\bf (degrees)}  & \multicolumn{1}{c}{\bf (h m s)}&\multicolumn{1}{c}{\bf ($^{o}$ $'$ $``$)}\\  \hline \hline
						
G\,338.926+0.634			&	16 40 13.95	&	--45 38 29.7	&	2008 Dec	&	t	&		&		&	 \\	
G\,338.935--0.062$^{2}$		&	16 43 16.01	&	--46 05 40.2	&	2008 Dec	&	10	&	--42.2	&	--43.1,--41.4	&	3.1	\\
G\,339.053--0.315			&	16 44 48.99	&	--46 10 13.0	&	2008 Dec	&	47	&	--111.8	&	--124,--110.2	&	40 	\\
G\,339.064+0.152			&	16 42 49.56	&	--45 51 23.8	&	2008 Dec	&	0.5	&	--85.8	&	--87.6,--85.7	&	0.5	 \\
G\,339.282+0.136			&	16 43 43.11	&	--45 42 08.0	&	2008 Dec	&	1.3	&	--70.9	&	--71.1,--69.9	&	0.7	\\
G\,339.476+0.185			&	16 44 13.79	&	--45 31 24.8	&	2008 Dec	&	0.5	&	--98.1	&	--98.1,--87.4	&	0.1	\\
						&				&				&	2010 Mar	& $<$0.80		&  & --150,190\\
G\,339.582--0.127			&	16 45 58.82	&	--45 38 47.2	&	2008 Dec	&	1.6	&	--30.5	&	--30.7,--30.2	&	0.4 \\
G\,339.622--0.121$^{2,8}$	&	16 46 05.99	&	--45 36 43.3	&	2008 Dec	&	3.4	&	--32.8	&	--39.1,--31.7	&	2.3	 \\
G\,339.762+0.054			&	16 45 51.56	&	--45 23 32.6	&	2008 Dec	&	2.6	&	--50.4	&	--51.7,--49.5	&	1.7	 \\
G\,339.884--1.259$^{2,8,9}$	&	16 52 04.67	&	--46 08 34.1	&	2008 Dec	&	846	&	--38.7	&	--39.4,--32.4	&	750	\\
G\,339.949--0.539			&	16 49 07.98	&	--45 37 58.3	&	2008 Dec	&	12	&	--97.6	&	--100.4,--94	&	11 	\\
G\,339.986--0.425			&	16 48 46.31	&	--45 31 51.3	&	2008 Dec	&	64	&	--89.4	&	--89.9,--87.1	&	54	 \\
G\,340.054--0.244$^{2}$		&	16 48 13.89	&	--45 21 43.3	&	2008 Dec	&	8.3	&	--59.8	&	--60.1,--59.3	&	2.8	\\
G\,340.118--0.021			&	16 47 30.00	&	--45 10 12.1	&	2008 Dec	&	0.5	&	--123.8	&	--124,--123.5	&	0.2	\\
G\,340.518--0.152			&	16 49 31.36	&	--44 56 54.6	&	2008 Dec	&	2	&	--48.3	&	--50,--48.1	&	1.1	\\
G\,340.785--0.096$^{2,8,9}$	&	16 50 14.84	&	--44 42 26.3	&	2008 Dec	&	42	&	--105.3	&	--109,--89	&	41	 \\
G\,341.218--0.212$^{2,8,9}$	&	16 52 17.84	&	--44 26 52.1	&	2008 Dec	&	1.2	&	--38.7	&	--48,--37.1	&	1.2	 \\
G\,341.238--0.270$^{2}$		&	16 52 37.41	&	--44 28 07.7	&	2008 Jun	&	1.1	&	--51.2	&	--52,--49.5	&	0.8\\	
G\,341.276+0.062			&	16 51 19.41	&	--44 13 44.5	&	2008 Jun	&	0.8	&	--73.5	&	--73.8,--70.4	&	0.6	\\	
G\,342.338+0.305			&	16 54 00.71	&	--43 15 12.3	&	2008 Jun	&	$<$0.70	&		&	--155,185	&	\\
						&				&				&	2008 Dec	&	0.8	&	--112.2	&	--112.2,--112.1	&	0.1	\\
						&				&				&	2010 Mar	&	 0.5 & --112.2	& --112.2,--112.0	&0.1\\
G\,342.446--0.072			&	16 55 59.94	&	--43 24 22.5	&	2008 Jun	&	0.5	&	--20.1	&	--20.2,--19.4	&	0.1	\\	
G\,342.484+0.183			&	16 55 02.30	&	--43 12 59.8	&	2008 Dec	&	8.1	&	--42.2	&	--43,--41	&	5.7	 \\
G\,343.354--0.067			&	16 59 04.24	&	--42 41 34.6	&	2008 Jun	&	0.6	&	--127.7	&	--127.8,--127.6	&	0.1	\\
						&				&				&	2008 Dec	&	0.5	&	--116.8	&	--116.9,116.8	&	0.1	 \\
G\,343.502--0.472			&	17 01 18.40	&	--42 49 36.8	&	2008 Jun	&	0.4	&	--42.5	&	--42.6,--41.9	&	0.2	\\	
G\,343.929+0.125$^{2}$		&	17 00 10.91	&	--42 07 19.3	&	2008 Jun	&	2.4	&	13.7	&	13.4,14.5	&	1.2	\\	
G\,344.227--0.569$^{2,10}$	&	17 04 07.78	&	--42 18 39.5	&	2008 Jun	&	2.4	&	--19.7	&	--30.8,--18.7	&	3.1		\\	
G\,344.421+0.045$^{2}$		&	17 02 08.77	&	--41 46 58.5	&	2008 Jun	&0.8	&	--71.5	&	--71.8,--71.2	&	0.3	\\	
G\,345.003--0.223$^{8,9}$	&	17 05 10.89	&	--41 29 06.2	&	2008 Jun	&1.0	&	--22.2	&	--23.1,--21.7	&	0.8		\\
G\,345.003--0.224$^{8,9}$	&	17 05 11.23	&	--41 29 06.9	&	2008 Jun	&0.9	&	--28.5	&	--30.3,--28.3	&	0.4	\\
G\,345.010+1.792$^{2,8,9}$	&	16 56 47.58	&	--40 14 25.8	&	2008 Jun	&296	&	--21.8	&	--23.5,--16.8	&	305	\\
G\,345.012+1.797$^{2,8,9}$	&	16 56 46.82	&	--40 14 08.9	&	2008 Jun	&12	&	--10.8	&	--13.1,--10.4	&	12	\\
G\,345.198--0.030			&	17 04 59.49	&	--41 12 45.7	&	2008 Jun	&1.2	&	--0.5	&	--1.5,--0.2	&	0.7		\\
G\,345.505+0.348$^{2,10}$	&	17 04 22.91	&	--40 44 21.7	&	2008 Jun	&8.4	&	--17.7	&	--18.5,--15.5	&	5.1	\\
G\,345.807--0.044			&	17 06 59.85	&	--40 44 08.2	&	2008 Jun	&$<$0.75		&	&	--155,185	&		\\
						&				&				&	2008 Dec	&	0.3	&	--1.5	&	--2.1,--1.3	&	0.1	\\	
						&				&				&	2010 Mar	&       $<$0.82	& & --140,200    \\
G\,346.480+0.221			&	17 08 00.11	&	--40 02 15.9	&	2008 Jun	&8.7	&	--19	&	--19.7,--13.7	&	6.6	\\	
G\,347.583+0.213$^{2}$		&	17 11 26.72	&	--39 09 22.5	&	2008 Jun	&0.8	&	--102.7	&	--102.8,--101.7	&	0.5	\\	
G\,347.863+0.019$^{2}$		&	17 13 06.23	&	--39 02 40.0	&	2008 Jun	&1.6	&	--35.3	&	--36.5,--23.6	&	1.3	\\	
G\,347.902+0.052$^{2}$		&	17 13 05.11	&	--38 59 35.5	&	2008 Jun	&1.3	&	--27.3	&	--27.8,--27	&	0.7	\\	
G\,348.195+0.768			&	17 11 00.20	&	--38 20 05.5	&	2008 Jun	&0.5	&	--0.8	&	--0.9,--0.8	&	0.1	\\
						&				&				&	2008 Dec	&	0.4	&	--0.9	&	--1,--0.8	&	0.1	\\
						&				&				&	2010 Mar	& $<$0.79	& & --140,200	\\
G\,348.550--0.979$^{2,8}$	&	17 19 20.41	&	--39 03 51.6	&	2008 Jun	&	8.9	&	--10.1	&	--10.9,--9.4	&	3.7	\\	
G\,348.550--0.979n$^{2,8}$	&	17 19 20.45	&	--39 03 49.4	&	2008 Jun	&	1.0	&	--22.1	&	--22.2,--21.8	&	0.3	\\	
G\,348.617--1.162			&	17 20 18.65	&	--39 06 50.8	&	2008 Dec	&	28	&	--11.5	&	--21.3,--9.9	&	39 \\
G\,348.723--0.078			&	17 16 04.77	&	--38 24 08.8	&	2008 Dec	&	0.9	&	11.6	&	9.7,11.7	&	0.9 	\\
G\,348.703--1.043$^{2,8}$	&	17 20 04.06	&	--38 58 30.9	&	2008 Dec	&	34	&	--3.5	&	--16.3,--2.9	&	26 	\\
G\,348.727--1.037$^{2,8}$	&	17 20 06.54	&	--38 57 09.1	&	2008 Dec	&	13	&	7.4	&	--9.1,--6.6	&	8.9	 \\
G\,348.884+0.096			&	17 15 50.13	&	--38 10 12.4	&	2008 Dec	&	3.3	&	--74.4	&	--76.6,--74.1	&	2.7	 \\
G\,349.092+0.106			&	17 16 24.59	&	--37 59 45.8	&	2008 Dec	&	0.5	&	--80.4	&	--81.6,--80.2	&	0.2	\\
G\,349.799+0.108			&	17 18 27.74	&	--37 25 03.5	&	2008 Dec	&	1.3	&	--59.5	&	--59.7,--59.1	&	0.6	\\
G\,350.105+0.083$^{2}$		&	17 19 27.01	&	--37 10 53.3	&	2008 Dec	&	1.4	&	--74.1	&	--74.8,--66.9	&	1.1	\\
						&				&				&	2010 Mar	&	 0.7 & --74.1 & --74.1, --66.8 & 0.2\\
G\,350.116+0.220			&	17 18 55.11	&	--37 05 38.1	&	2008 Dec	&	$<$0.60$^{*}$	&	&		--200,140	& \\
						&				&				&	2010 Mar	&	 $<$0.82$^{*}$ & &--140,200\\
						&				&				&	Avg all	&	0.4$^{\alpha}$	& 4.4		& 4.3,4.5 &0.1\\
G\,350.299+0.122			&	17 19 50.87	&	--36 59 59.9	&	2008 Dec	&	0.6	&	--63.6	&	--63.7,--63.5	&	0.1 	\\
G\,350.340+0.141			&	17 19 53.43	&	--36 57 18.8	&	2008 Dec	&	0.6	&	--58.5	&	--58.5,--58.4	&	0.1 	\\
\end{tabular}

\end{table*}
\clearpage

\begin{table*}\addtocounter{table}{-1}
  \caption{-- {\emph {continued}}}
  \begin{tabular}{lllllccclclllclllcll} \hline
\multicolumn{1}{c}{\bf Methanol maser} &\multicolumn{1}{c}{\bf RA} & \multicolumn{1}{c}{\bf Dec}  &{\bf Epoch}&  {\bf S$_{12.2}$} & {\bf Vp$_{12.2}$} & {\bf Vr$_{12.2}$} & \multicolumn{1}{c}{\bf I$_{12.2}$} \\
    \multicolumn{1}{c}{\bf ($l,b$)}&  \multicolumn{1}{c}{\bf (J2000)} & \multicolumn{1}{c}{\bf (J2000)}   & & {\bf (Jy)} & {\bf (\kmsns)} & {\bf  (\kmsns)} & \multicolumn{1}{c}{\bf (Jy \kmsns)}\\
    \multicolumn{1}{c}{\bf (degrees)}  & \multicolumn{1}{c}{\bf (h m s)}&\multicolumn{1}{c}{\bf ($^{o}$ $'$ $``$)}\\  \hline \hline
						
G\,350.344+0.116			&	17 20 00.03	&	--36 58 00.1	&	2008 Dec	&	7.2	&	--65.4	&	--65.7,--64.9	&	3.0 	\\
G\,350.470+0.029			&	17 20 43.24	&	--36 54 46.6	&	2008 Dec	&	0.7	&	--6.2	&	--6.4,--6.1	&	0.1 	\\
G\,350.686--0.491$^{3}$		&	17 23 28.63	&	--37 01 48.8	&	2008 Dec	&	13	&	--13.8	&	--15.1,--13.7	&	4.0 	\\
G\,351.382--0.181			&	17 24 09.58	&	--36 16 49.3	&	2008 Dec	&	2	&	--59.7	&	--60.5,--59.4	&	0.7 	\\
G\,351.417+0.645$^{2,7,8}$	&	17 20 53.37	&	--35 47 01.2	&	2008 Dec	&	976	&	--10.4	&	--12.0,0.4	&	866 \\
						&				&				&	2010 Mar	& 742		& --10.4	& --12.0,--6.5 & 656	\\
G\,351.417+0.646$^{2,7,8}$	&	17 20 53.18	&	--35 46 59.3	&	2008 Dec	&	692	&	--11.2	&	t	&	t 	&\\
						&				&				&	2010 Mar	& 515	& --11.2	& t&t\\
G\,351.445+0.660$^{2,7,8}$	&	17 20 54.61	&	--35 45 08.6	&	2008 Dec	&	37	&	--9.3	&	--9.6,--0.2	&	19 \\
G\,351.611+0.172			&	17 23 21.25	&	--35 53 32.4	&	2010 Mar	&	2	&	--37.3	&	--43.7,--35.7	&	2		\\	
G\,351.688+0.171			&	17 23 34.52	&	--35 49 46.3	&	2008 Dec	&	21	&	--36.2	&	--44.5,--35.3	&	14.3	 \\
G\,351.775--0.536$^{2,10}$	&	17 26 42.57	&	--36 09 17.6	&	2008 Dec	&	12	&	1.8	&	0.8,2.3	&	4.7	 \\
G\,352.083+0.167			&	17 24 41.22	&	--35 30 18.6	&	2008 Jun	&	2.8	&	--66.2	&	--66.5,--65.7	&	1.1	\\	
G\,352.111+0.176			&	17 24 43.56	&	--35 28 38.4	&	2008 Jun	&	3.1	&	--54.1	&	--55,--53	&	2.8	\\	
G\,353.370--0.091			&	17 29 14.27	&	--34 34 50.2	&	2008 Jun	&	$<$0.75	&		&	--150,190	&		\\
						&				&				&	2008 Dec	&	0.6	&	--44.6	&	--44.7,--44.4	&	0.2	 \\
G\,353.410--0.360$^{2,8,9}$	&	17 30 26.18	&	--34 41 45.6	& 	2010 Dec$^{\beta}$	&	21	&	--20.5	&	--22.5,--19.5	&	21	\\
G\,353.537--0.091			& 	17 29 41.25	&	--34 26 28.4	&	2008 Jun	&	0.6	&	--56.6	&	--56.9,--55.1	&	0.2	\\
						&				&				&	2008 Dec	&	0.8	&	--56.7	&	--56.9,--56.5	&	0.2	\\
G\,354.496+0.083	      		&	17 31 31.77	&	--33 32 44.0	&	2008 Jun	&	2.1	&	27	&	25.6,27.3	&	1.1	\\	
G\,354.615+0.472$^{2,8,9}$	&	17 30 17.13	&	--33 13 55.1	&	2008 Dec	&	33	&	--16.5	&	--24.8,--15.5	&	16 \\
G\,354.724+0.300			&	17 31 15.55	&	--33 14 05.7	&	2008 Jun	&	3.3	&	93.9	&	92.3,94.1	&	1.7	\\	
G\,355.343+0.148			&	17 33 28.84	&	--32 48 00.2	&	2008 Dec	&	$<$0.55$^{*}$	&	&		--200,140	&	\\
						&				&				&	2010 Mar	&$<$0.82$^{*}$ && --140,200\\
						&				&				&	Avg all	& 0.3$^{\alpha}$& 5.3 & 5.3,5.8	& 0.1\\
G\,357.559--0.321			&	17 40 57.33	&	--31 10 56.9	&	2008 Jun	&0.9	&	16.1	&	15.9,16.2	&	0.2	\\
						&				&				&	2008 Dec	&	0.7	&	16	&	15.9,16.2	&	0.2 \\
G\,357.924--0.337			&	17 41 55.17	&	--30 52 50.2	&	2008 Jun	&0.7	&	--2.4	&	--2.6,--1.7	&	0.3	\\	
G\,357.967--0.163			&	17 41 20.26	&	--30 45 06.9	&	2008 Jun	&0.9	&	--2.9	&	--3.6,--2.5	&	0.2	\\
						&				&				&	2008 Dec	&	1	&	--3	&	--8.8,--2.8	&	0.4	 \\
G\,358.263--2.061			&	17 49 37.63	&	--31 29 18.0	&	2008 Jun	&0.5	&	1.9	&	1.5,1.9	&	0.1	\\
						&				&				&	2008 Dec	&	0.5	&	1.7	&	1.5,1.7	&	0.1 	\\
G\,358.371--0.468			&	17 43 31.95	&	--30 34 11.0	&	2008 Jun	&1.9	&	1.4	&	1,1.6	&	0.7	\\
						&				&				&	2008 Dec	&	8.1	&	1.3	&	0.8,2	&	3.2	\\
G\,358.721--0.126			&	17 43 02.31	&	--30 05 29.9	&	2008 Jun	&0.7	&	10.5	&	10.5,10.6	&	0.1	\\	
G\,358.809--0.085			&	17 43 05.40	&	--29 59 45.8	&	2008 Jun	&0.3	&	--56.2	&	--56.2,--56.1	&	0.1	\\
						&				&				&			&	0.3	&	--56.2	&	--56.3,--56.1	&	0.1 	\\
G\,358.841--0.737			&	17 45 44.29	&	--30 18 33.6	&	2008 Jun	&3.2	&	--20.7	&	--21.3,--20.4	&	1.4	\\	
G\,358.906+0.106			&	17 42 34.57	&	--29 48 46.8	&	2008 Jun	&$<$0.70	&		&	--150,190	&		\\
						&				&				&	2008 Dec	&	0.7	&	--17.3	&	--18.2,--17.2	&	0.1	\\
G\,359.436--0.104			&	17 44 40.60	&	--29 28 16.0	&	2008 Jun	&$<$0.75	&		&	--150,190	&		\\
						&				&				&	2008 Dec	&	0.5	&	--56.5	&	--56.5,--56.3	&	0.1\\
G\,359.615--0.243$^{2,8}$	&	17 45 39.09	&	--29 23 30.0	&	2008 Dec	&	5.7	&	19.3	&	19.1,24.5	&	6.4 \\
G\,0.092--0.663			&	17 48 25.90	&	--29 12 05.9	&	2008 Dec	&	6.3	&	23.5	&	21,23.9	&	3.9	 \\
G\,0.212--0.001$^{1}$		&	17 46 07.63	&	--28 45 20.9	&	2008 Jun	&0.8	&	49.4	&	48.9,49.6	&	0.5	\\	
G\,0.315--0.201$^{1,3}$		&	17 47 09.13	&	--28 46 15.7	&	2008 Jun	&2.4	&	18.2	&	17.7,20	&	1.9	\\	
G\,0.496+0.188				&	17 46 03.96	&	--28 24 52.8	&	2008 Jun	&18	&	1	&	--9.7,1.5	&	16	\\
G\,0.546--0.852$^{1,2,4,8}$	&	17 50 14.53	&	--28 54 31.2	&	2008 Jun	&3.9	&	18.4	&	13.7,19.2	&	2.5	\\	
G\,0.651--0.049$^{2,8}$		&	17 47 21.12	&	--28 24 18.2	&	2008 Jun	&2.1	&	48.1	&	47.8,48.5	&	1.0		\\	
G\,0.666--0.029			&	17 47 18.64	&	--28 22 54.6	&	2008 Jun	& 0.5	&	68.7	&	68.6,68.8	&	0.1	\\	
G\,0.667--0.034			&	17 47 19.87	&	--28 23 01.3	&	2008 Jun	&0.6	&	55.1	&	55.1,55.3	&	0.1\\	
G\,0.836+0.184				&	17 46 52.86	&	--28 07 34.8	&	2008 Jun	&$<$0.50	&		&	--150,190	&		\\
						&				&				&	2008 Dec	&	0.6	&	3.5	&	--2.2,3.6	&	0.1 	\\
G\,1.719--0.088			&	17 49 59.84	&	--27 30 36.9	&	2008 Jun	&0.8	&	--8	&	--8.1,--7.8	&	0.2	\\	
G\,2.536+0.198				&	17 50 46.48	&	--26 39 44.9	&	2008 Jun	&3.7	&	6.9	&	3.3,7.2	&	2.6	\\
G\,2.615+0.134				&	17 51 12.30	&	--26 37 37.2	&	2008 Jun	&0.6	&	94.2	&	94.1,94.3	&	0.1	\\
						&				&				&	2008 Dec	&	$<$0.55	&		&	--200,140	\\	
G\,2.703+0.040				&	17 51 45.98	&	--26 35 56.7	&	2008 Jun	&2.6	&	93.6	&	92.9,94.8	&	1.3	\\	
G\,3.253+0.018				&	17 53 05.96	&	--26 08 13.0	&	2008 Jun	&0.9	&	2.4	&	--0.3,2.6	&	0.6	\\	
G\,3.502--0.200			&	17 54 30.06	&	--26 01 59.4	&	2008 Jun	&$<$0.70	&		&	--150,190	&		\\
						&				&				&	2008 Dec	&	0.4	&	43.8	&	43.6,43.9	&	0.1	\\	
						&				&				&	2010 Mar	& $<$0.80 & & --140,200\\

\end{tabular}

\end{table*}
\clearpage

\begin{table*}\addtocounter{table}{-1}
  \caption{-- {\emph {continued}}}
  \begin{tabular}{lllllccclclllclllcll} \hline
\multicolumn{1}{c}{\bf Methanol maser} &\multicolumn{1}{c}{\bf RA} & \multicolumn{1}{c}{\bf Dec}  &{\bf Epoch}&  {\bf S$_{12.2}$} & {\bf Vp$_{12.2}$} & {\bf Vr$_{12.2}$} & \multicolumn{1}{c}{\bf I$_{12.2}$} \\
    \multicolumn{1}{c}{\bf ($l,b$)}&  \multicolumn{1}{c}{\bf (J2000)} & \multicolumn{1}{c}{\bf (J2000)}   & & {\bf (Jy)} & {\bf (\kmsns)} & {\bf  (\kmsns)} & \multicolumn{1}{c}{\bf (Jy \kmsns)}\\
    \multicolumn{1}{c}{\bf (degrees)}  & \multicolumn{1}{c}{\bf (h m s)}&\multicolumn{1}{c}{\bf ($^{o}$ $'$ $``$)}\\  \hline \hline
						
G\,4.393+0.079				&	17 55 25.77	&	--25 07 23.6	&	2008 Jun	&2.1	&	1.9	&	0.5,2.2	&	1.6	\\	
G\,5.618--0.082			&	17 58 44.78	&	--24 08 40.1	&	2008 Jun	&	$<$0.75$^{*}$ &			&	--145,190	&		\\	
						&				&				&	2008 Dec	&	$<$0.55$^{*}$ &			&	--200,140	&\\
						&				&				&	2010 Mar	& $<$0.78$^{*}$ & & --140,200\\
						&				&				&	Avg all	& 0.4$^{\alpha}$ & --27.0 & --27.1,--26.8	& 0.1\\
G\,5.630--0.294			&	17 59 34.60	&	--24 14 23.7	&	2008 Jun	&0.8	&	10.6	&	9.6,11	&	0.7	\\
G\,6.189--0.358			&	18 01 02.16	&	--23 47 10.8	&	2008 Jun	&5.5	&	--30.6	&	--34.4,--30.1	&	2.4	\\
G\,6.588--0.192			&	18 01 16.09	&	--23 21 27.3	&	2008 Jun	&2.9	&	4.9	&	4.3,5.9	&	2.7	\\
G\,6.610--0.082$^{1}$		&	18 00 54.03	&	--23 17 03.1	&	2008 Jun	&11	&	0.8	&	0.4,1.3	&	4.7	\\
G\,6.795--0.257$^{3}$		&	18 01 57.75	&	--23 12 34.9	&	2008 Jun	&18	&	15.6	&	14.9,27.4	&	16	\\
G\,7.166+0.131				&	18 01 17.48	&	--22 41 44.0	&	2008 Jun	&$<$0.70$^{*}$	&	&		--145,195	&	\\
						&				&				&	2008 Dec	&	$<$0.80$^{*}$	&		&	--200,140	& \\
						&				&				&	2010 Mar	&	$<$0.55$^{*}$ & &--140,200	\\
						&				&				&	Avg all	& 0.4$^{\alpha}$ & 80.1 & 80.0,80.3	&0.1\\
G\,7.632--0.109		      	&	18 03 11.63	&	--22 24 32.4	&	2008 Jun	&2.7	&	156.9	&	149.4,158.4	&	1.7	\\
G\,8.317--0.096		      	&	18 04 36.02	&	--21 48 19.6	&	2008 Jun	&1.1	&	47.6	&	46.7,48.5	&	1.1		\\
G\,8.669--0.356$^{1}$	      	&	18 06 18.99	&	--21 37 32.2	&	2008 Jun	&0.4	&	39.3	&	39.3,39.4	&	0.1	\\
						&				&				&	2008 Dec	&	$<$1.15	&		&	--200,140		\\
G\,8.683--0.368$^{1,2,4,8}$	&	18 06 23.49	&	--21 37 10.2	&	2008 Jun	&8.1	&	44.1	&	40.9,45.4	&	10		\\
G\,8.832--0.028			&	18 05 25.67	&	--21 19 25.1	&	2008 Jun	&37	&	--5.2	&	--5.6,4.3	&	55	\\
G\,9.215--0.202			&	18 06 52.84	&	--21 04 27.5	&	2008 Jun	&0.6	&	46.7	&	41.3,46.8	&	0.2	\\
						&				&				&	2008 Dec	&	$<$0.80	&		&	--200,140	&	\\
						&				&				&	2010 Mar	& $<$0.55 & & --140,200\\
G\,9.621+0.196$^{1,2,4,5,8,9}$&	18 06 14.67	&	--20 31 32.4	&	2008 Jun	&401	&	1.4	&	--3,5.8	&	203	\\
G\,9.619+0.193$^{1}$		&	18 06 14.92	&	--20 31 44.3	&	2008 Jun	&0.4	&	5.8	&	5.7,5.9	&	0.1		\\
G\,9.986--0.028$^{1}$		&	18 07 50.12	&	--20 18 56.5	&	2008 Jun	&1	&	47	&	44.2,47.2	&	0.5		\\ \hline

\end{tabular}

\end{table*}

\begin{figure*}
	\epsfig{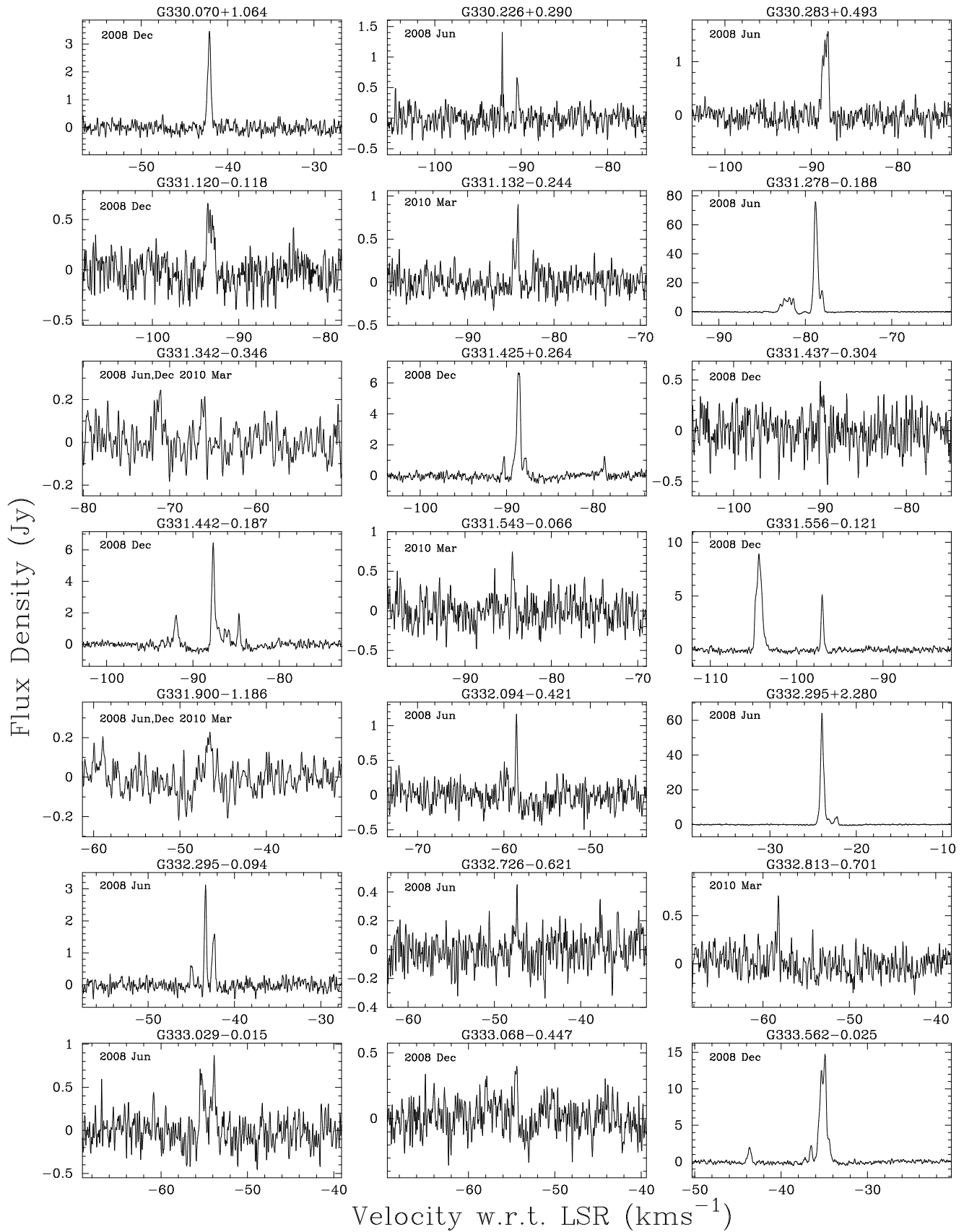}
\caption{Spectra of the 12.2-GHz methanol masers detected towards 6.7-GHz MMB sources.}
\label{fig:12MMB}
\end{figure*}

\begin{figure*}\addtocounter{figure}{-1}
	\epsfig{figure=12mmb_paper2.eps}
\caption{--{\emph {continuued}}}
\end{figure*}

\begin{figure*}\addtocounter{figure}{-1}
	\psfig{figure=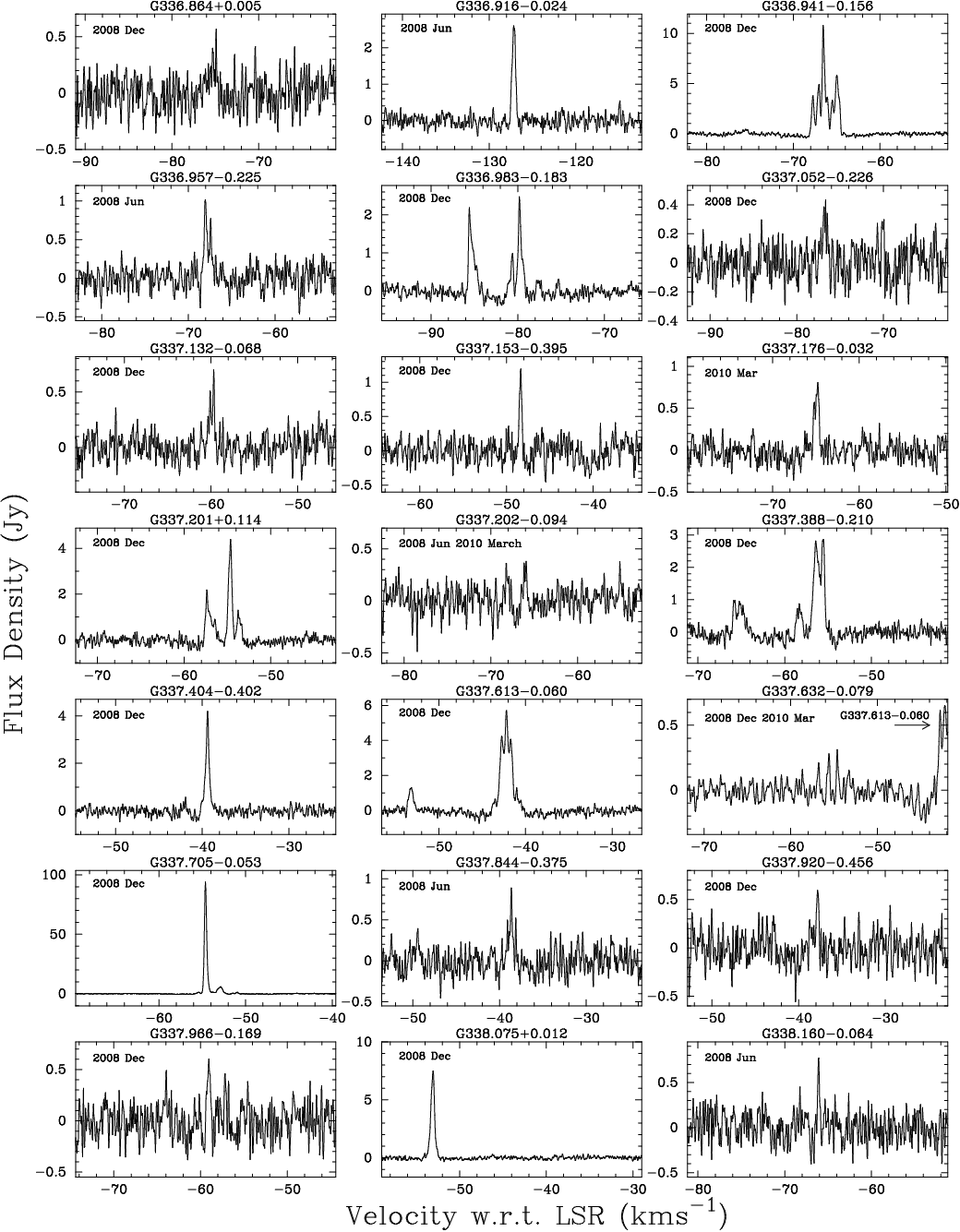}
\caption{--{\emph {continuued}}}
\end{figure*}

\begin{figure*}\addtocounter{figure}{-1}
	\psfig{figure=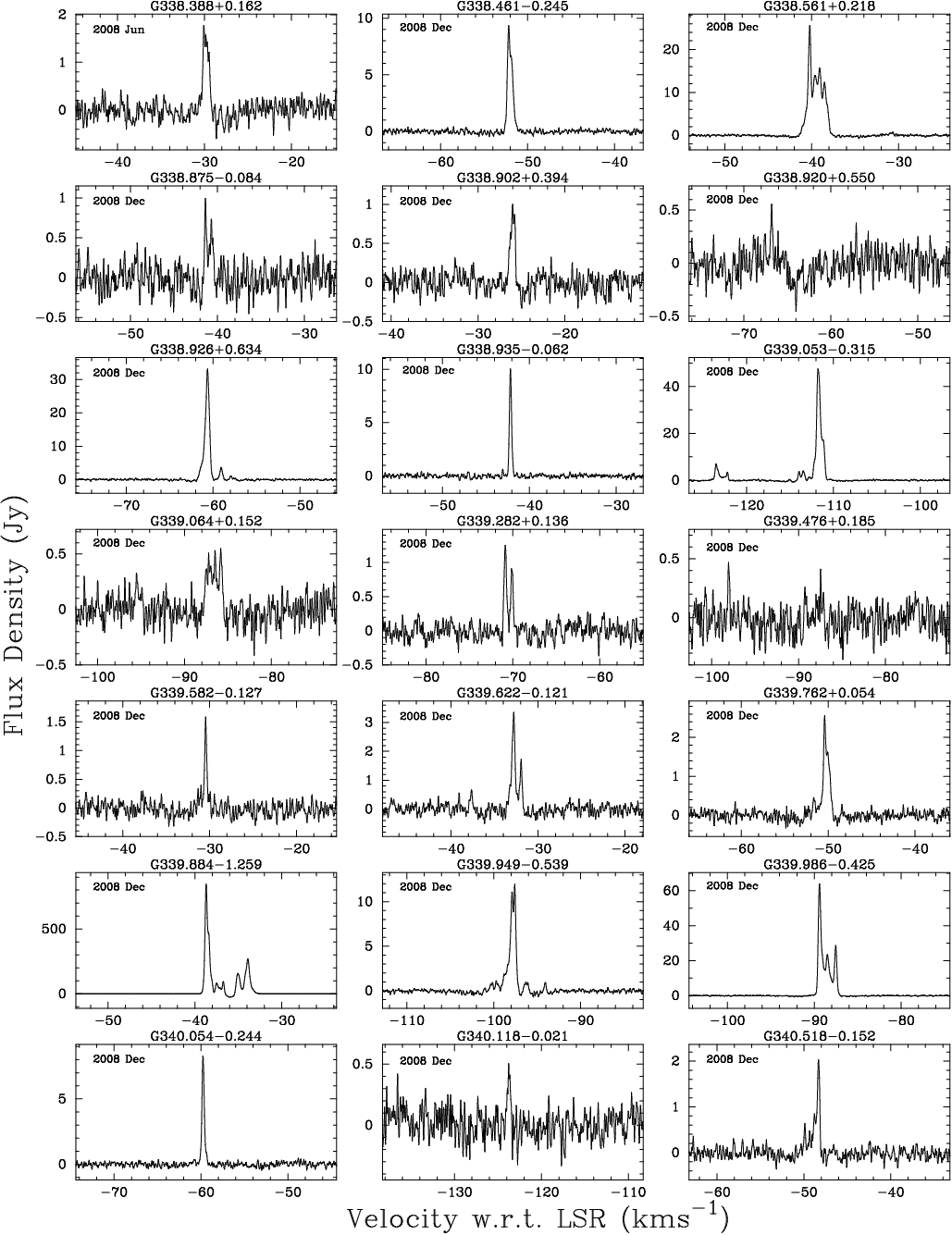}
\caption{--{\emph {continuued}}}
\end{figure*}

\begin{figure*}\addtocounter{figure}{-1}
	\psfig{figure=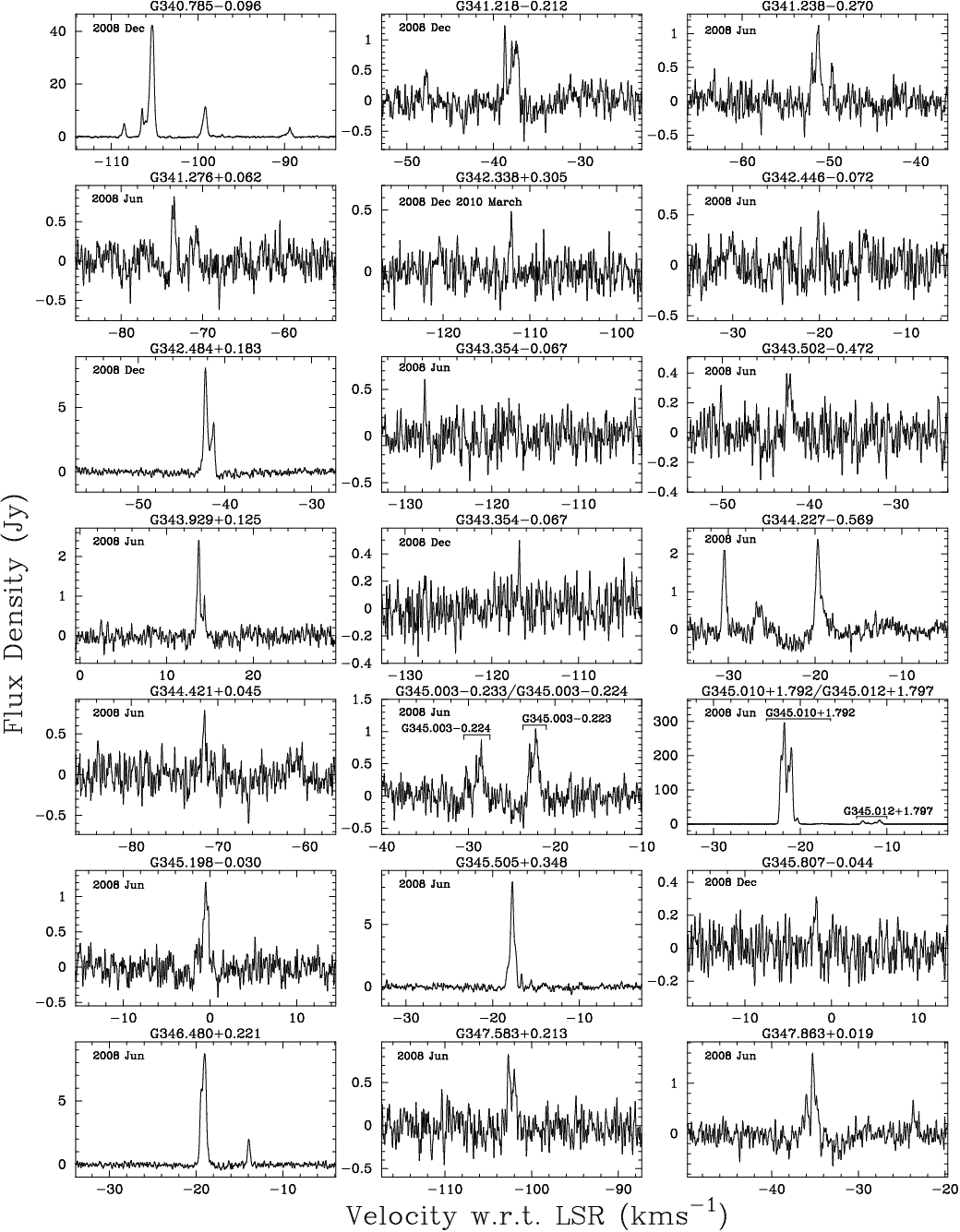}
\caption{--{\emph {continuued}}}
\end{figure*}

\begin{figure*}\addtocounter{figure}{-1}
	\psfig{figure=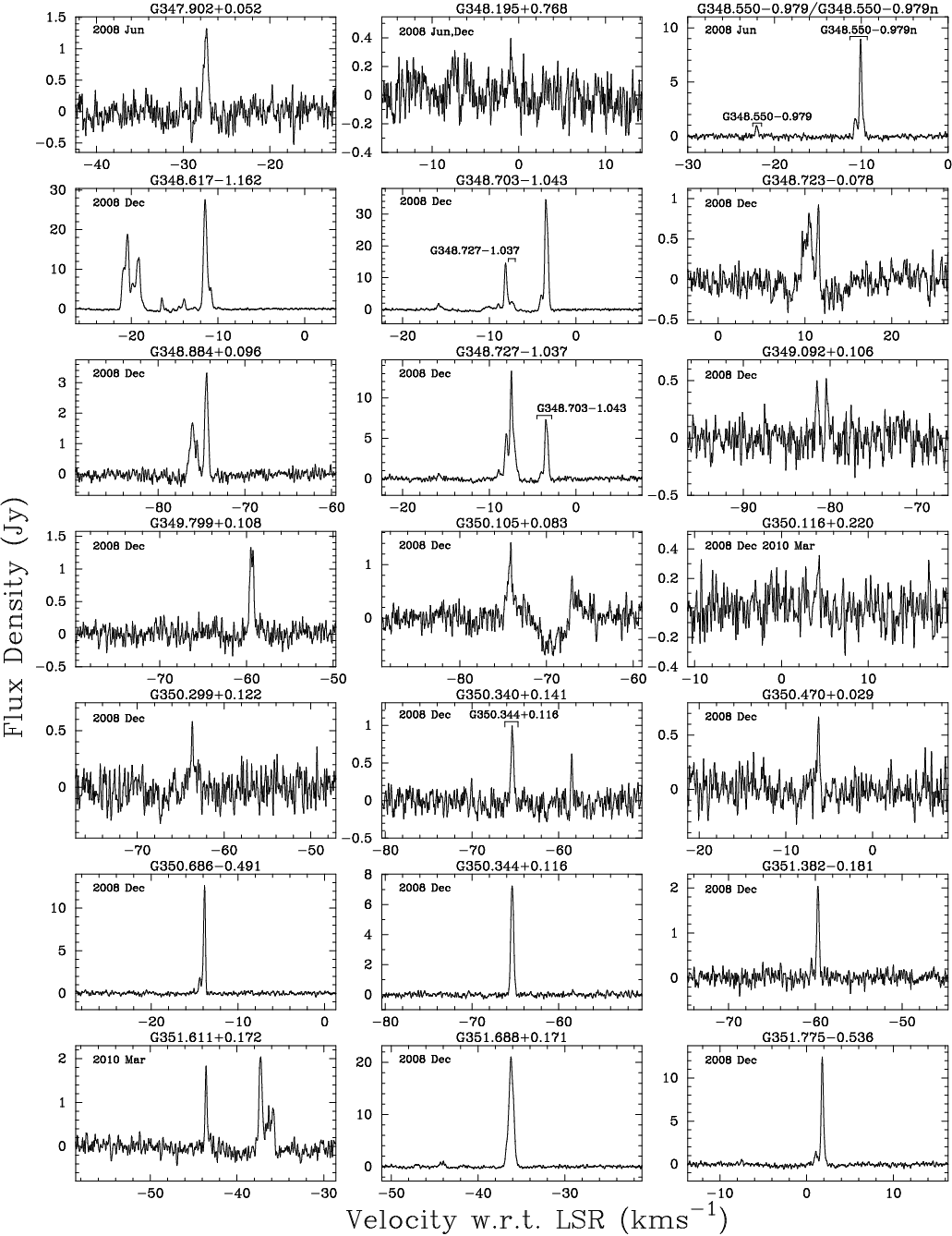}
\caption{--{\emph {continuued}}}
\end{figure*}

\begin{figure*}\addtocounter{figure}{-1}
	\psfig{figure=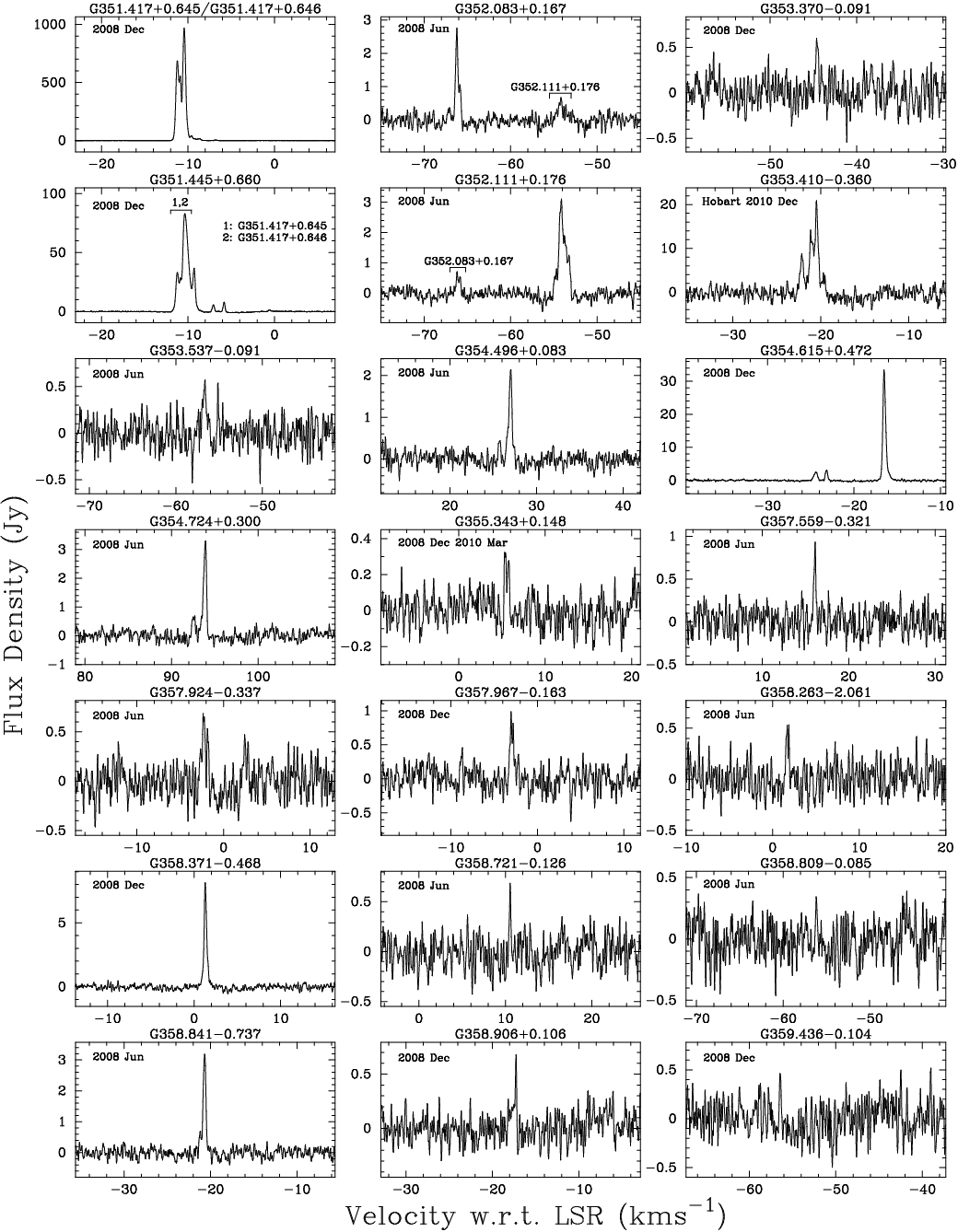}
\caption{--{\emph {continuued}}}
\end{figure*}

\begin{figure*}\addtocounter{figure}{-1}
	\psfig{figure=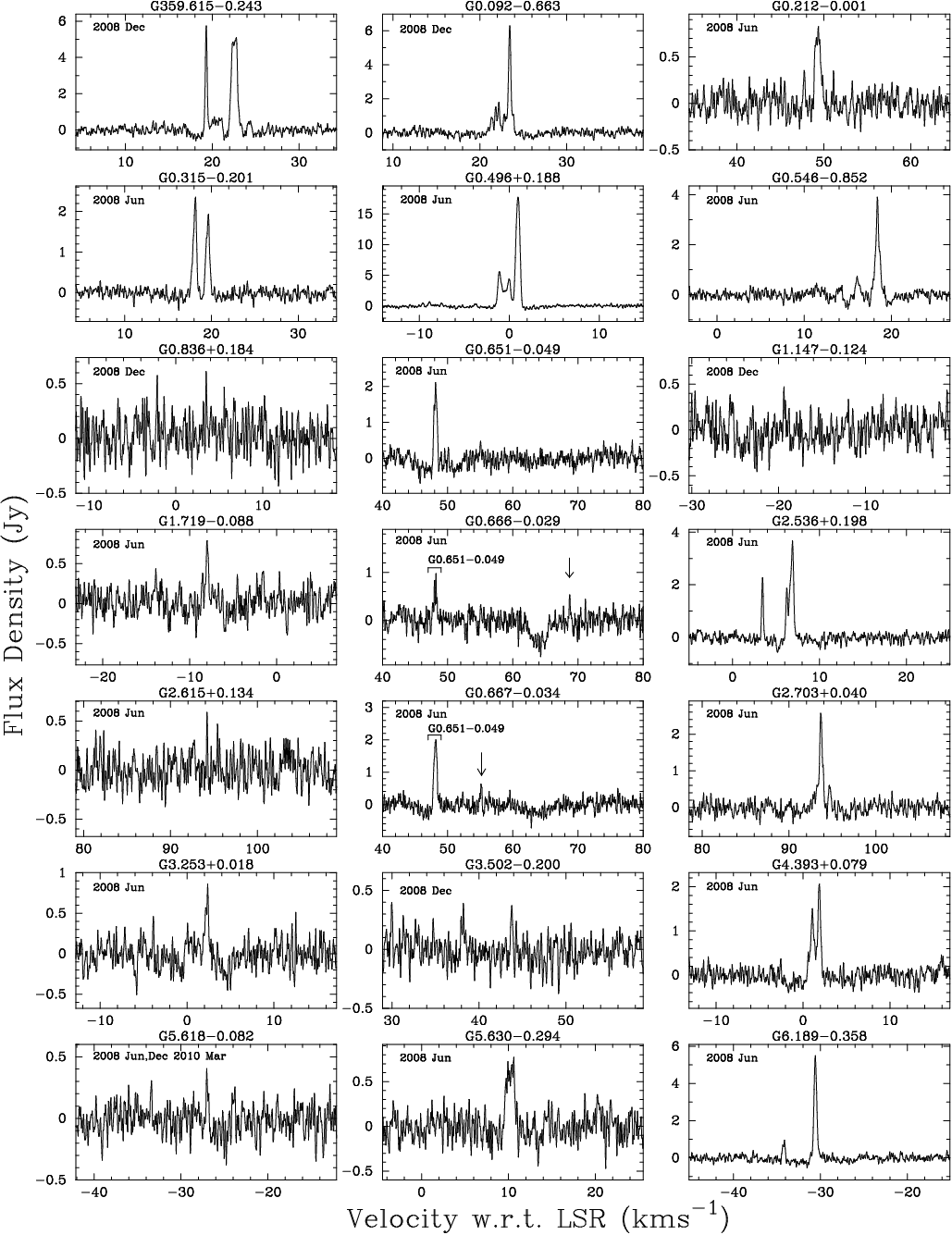}
\caption{--{\emph {continuued}}}
\end{figure*}

\begin{figure*}\addtocounter{figure}{-1}
	\psfig{figure=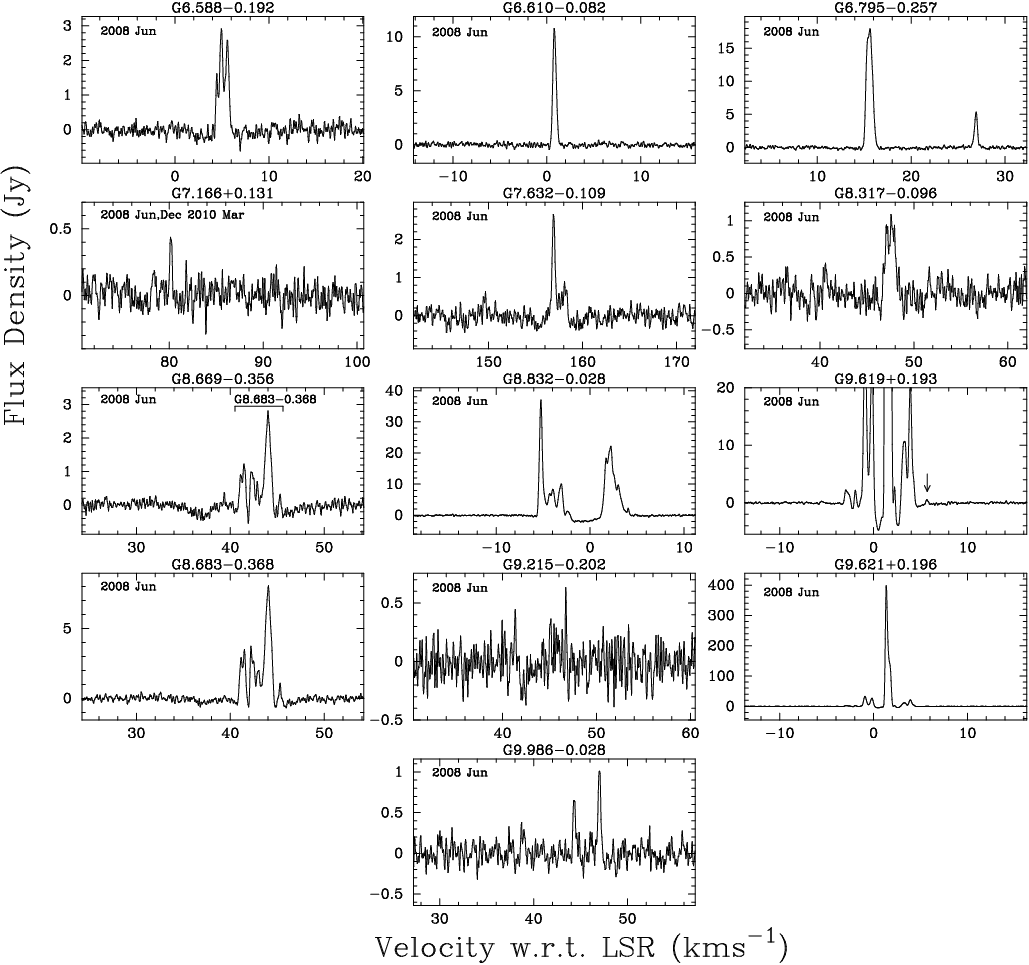}
\caption{--{\emph {continuued}}}
\end{figure*}

\section{Individual sources}\label{sect:12GHz_ind}

Here we draw attention to specific sources for which the information in Table~\ref{tab:12MMB} and Fig.~\ref{fig:12MMB} does not adequately describe them. Complex situations where more than one 12.2-GHz methanol maser is located within close proximity are discussed, particularly where there is no clear separation in velocity between sources. Variable sources (particularly those 12.2-GHz sources that have varied by more than 20~per~cent between 2008 June and December), sources with interesting histories and associations with other maser species are also highlighted. Instances where the flux density of the 12.2-GHz methanol emission surpasses that of the 6.7-GHz counterpart are of particular interest and are noted below. All comments relating to smoothing indicate levels that have been applied after the initial Hanning smoothing.

\subparagraph{G\,330.226+0.290.} 12.2-GHz emission was detected in 2008 June, with a peak flux density of 1.4~Jy at a velocity  of --92.2 \kmsns. At this velocity very weak ($\sim$0.2~Jy) 6.7-GHz emission was present in 2008 Aug 21. 12.2-GHz emission was also detected at the 6.7-GHz peak velocity ($\sim$0.7~Jy). No 12.2-GHz emission was detected towards this source in 2008 December (5-$\sigma$ detection limit of 0.75~Jy).

\subparagraph{G\,331.132--0.244.} No 12.2-GHz emission was detected at either epoch in 2008, but weak emission ($\sim$0.3~Jy) is present in the average spectrum. Observations carried out during 2010 March revealed stronger emission (0.9~Jy) as shown in Fig.~\ref{fig:12MMB}. Both water maser and OH maser emission have been detected at this site \citep{C98,Breen10b}. The methanol maser emission at 6.7-GHz shows strong evidence for periodicity \citep{Goed} and extrapolation of their light-curve is consistent with our 12.2-GHz observations.

\subparagraph{G\,331.278--0.188.} 12.2-GHz emission peaks at a velocity of --78.8 \kms and is similar to the spectrum reported by \citet{Cas93}, coincident with a 6.7-GHz secondary feature. At this velocity the 12.2-GHz emission is 76~Jy, stronger than the 46~Jy 6.7-GHz emission observed two months later. A spectrum comparing the two transitions is shown in Fig.~\ref{fig:12stronger} and is similar to the results of near-simultaneous observations made with Hobart. This methanol maser site is associated with OH and water maser emission (\citet{C98,Breen10b}). 

\subparagraph{G\,331.342--0.346.} 12.2-GHz observations were made during all three epochs. Weak emission was first identified in the average of the 2008 June and December spectra and confirmed in 2010 March. The spectrum presented in Fig.~\ref{fig:12MMB} is the average of all epochs and has had additional smoothing over four channels. This source shows weak (0.2~Jy) emission at the velocity of the strongest 6.7-GHz feature (86~Jy), and shows a slightly stronger feature (0.4~Jy) at the velocity of a very weak 6.7-GHz feature at $\sim$--71.3 \kms. Emission from both OH and water masers are also present at this site (\citet{C98,Breen10b}).


\subparagraph{G\,331.900--1.186.} Observations were carried out during all three epochs and no individual epoch revealed any discernible emission. However, the average of the three observations shows weak 12.2-GHz emission. The spectrum presented in Fig.~\ref{fig:12MMB} has been smoothed over an additional 4 channels. While the signal-to-noise in the averaged spectrum is still very low, the two spectral features are coincident in velocity with 6.7-GHz methanol maser features. Interestingly, the slightly weaker 12.2-GHz feature at --59.1 \kms is only 0.3~Jy at 6.7-GHz.
 
 
\subparagraph{G\,333.068--0.447.} 12.2-GHz emission was detected at a peak flux density of 0.4~Jy in 2008 December at the velocity of the 6.7-GHz peak, but no emission was detected at 12.2-GHz in 2008 June (5-$\sigma$ detection limit of 0.75~Jy). The spectrum shown in Fig.~\ref{fig:12MMB} is from 2010 March when the emission was at its strongest (0.7~Jy).
 
\subparagraph{G\,333.135--0.431.} No 12.2-GHz observations were directed towards this source. We derive emission upper limits (presented in Table~\ref{tab:6MMB}) from observations of G\,333.128--0.560 which is located 38 arcsec away. Since the nearby source has an overall integration time of 20 minutes, the epoch-averaged detection limit at the offset position is comparable to the majority of sources which had integration times of 5 to 10 minutes.
 
\subparagraph{G\,333.646+0.058.} Emission at 12.2-GHz (0.6~Jy) was detected in 2008 June and 2010 March at the velocity of the 6.7-GHz peak and the average of the two epochs is shown in Fig.~\ref{fig:12MMB}. However, observations in 2008 December showed no emission (5-$\sigma$ detection limit of 0.75~Jy). Emission from the 22-GHz transition of water is also found towards this site \citep{Breen10b}.
 
\subparagraph{G\,335.426--0.240.} This source exhibits multiple 12.2-GHz features with those between --53.4 and --52.4 \kms showing stronger emission at 12.2-GHz (1.9~Jy) than at 6.7-GHz ($\sim$0.2~Jy). The 6.7-GHz observations were carried out 4 months prior to the 12.2-GHz follow-up observations.
 
\subparagraph{G\,335.556--0.307.} The major feature at 12.2-GHz is at the velocity of the 6.7-GHz peak.  A secondary feature at --114.7 \kms has a peak flux density of 5.3~Jy compared to the corresponding 6.7-GHz feature of 4.4~Jy. The 6.7-GHz observations were carried out 4 months prior to the 12.2-GHz follow-up observations.
 
\subparagraph{G\,335.789+0.174.} The peak of the 12.2-GHz emission is at the velocity of a secondary feature in the 6.7-GHz spectrum and is marginally stronger at 12.2-GHz (162~Jy) than at 6.7-GHz (145~Jy). Near-simultaneous observations carried out with Hobart (discussed in Section~\ref{sect:weirdos}) confirm that this is not a consequence of variability between the Parkes 6.7- and 12.2-GHz epochs. \citet{Breen10b} detect a somewhat variable water maser at this location, with a 3~Jy detection during observations conducted in 2003, but no detectable emission above 0.2~Jy in 2004. \citet{C98} detected an OH maser at this location. Longer term variability at 12.2 GHz, since the observations of \citet{Cas93} has been small.
 
\subparagraph{G\,336.864+0.005 and G\,336.881+0.008.} 12.2-GHz emission (0.5~Jy) was detected towards relatively strong 6.7-GHz source G\,336.864+0.005 (59~Jy) in 2008 December. Observations in 2008 June failed to detect emission (5-$\sigma$ detection limit of 0.75~Jy). Observations conducted by ~\citet{Breen10b} revealed a water maser at this site and \citet{C98} detected an OH maser.

12.2-GHz observations of G\,336.881+0.008 were made at the half power point of the primary beam during observations targeting G\,336.864+0.005 (offset $\sim$1 arcmin) and we have adjusted our detection limit accordingly.

\subparagraph{G\,336.957--0.225.} 12.2-GHz emission consists of two main features, the first is at the velocity of the 6.7-GHz maser and is also the stronger 12.2-GHz feature. The second has a velocity of --67.4 \kms and is stronger at 12.2-GHz (0.8~Jy) than at 6.7-GHz (0.4~Jy). These 12.2-GHz observations were carried out 4 months after the 6.7-GHz observations.

\subparagraph{G\,337.132--0.068.} The peak of the 12.2-GHz emission detected in 2008 December is 0.7~Jy at --59.7 \kms compared with a mere 0.3~Jy at 6.7-GHz for this feature. Observations of the two transitions were separated by 4 months. No 12.2-GHz emission was detected in 2008 June (5-$\sigma$ detection limit of 0.7~Jy).


\subparagraph{G\,337.202--0.094.} 12.2-GHz observations were carried out towards this source in both 2008 June and 2010 March. While no emission was detectable at either epoch, weak emission is seen in the average of the two observations as shown in Fig.~\ref{fig:12MMB}. 

\subparagraph{G\,337.632--0.079.} The peak of the 12.2-GHz emission is at the velocity of a secondary feature in the 6.7-GHz spectrum, but weak emission is also present at the velocity of the 6.7-GHz peak. The peak 12.2-GHz flux density was 0.5~Jy in 2008 December, 0.6~Jy in 2010 March, but no emission was detected in 2008 June (5-$\sigma$ detection limit of 0.75~Jy). Figure~\ref{fig:12MMB} shows the average spectrum which has been smoothed over 4 channels.

\subparagraph{G\,337.703--0.053 and G\,337.705--0.053.} These methanol masers are separated by only 9 arcsec. Strong 12.2-GHz methanol maser emission is detected at the stronger of the two 6.7-GHz maser sites, G\,337.705--0.053, which is also the location of an OH maser \citep{C98,Caswell11}. It is difficult to determine if the weaker 6.7-GHz methanol maser site has any associated 12.2-GHz maser emission within the overlapping velocity range of the two sources, but, since there is no 12.2-GHz emission towards the strongest feature of G\,337.703--0.053 we list it as a non-detection. Spectral features at 12.2-GHz have changed little since the observations of \citet{Cas93} conducted 20 years ago.


\subparagraph{G\,338.075+0.012.} No 12.2-GHz emission is associated with the 6.7-GHz emission peak at --44 \kms (14~Jy) but relatively strong emission is detected towards a 6.7-GHz secondary feature at --53.1 \kms (8.3 Jy). We attribute the detected 12.2-GHz emission to this site 
rather than to companion site G\,338.075+0.009, offset by 9 arcsec and 
weaker at 6.7-GHz, since the velocities are in good agreement.  We note 
that the 12-GHz position measurement of \citet{Cas93}, although 
better corresponding with G\,338.075+0.009,  is not 
sufficiently precise to distinguish the correct counterpart.

The flux density of the 12.2-GHz emission rose by $\sim$27 per cent between 2008 June and December, from a flux density of 5.9 to 7.5~Jy. \citet{Breen10b} detect a weak water maser (0.5~Jy) towards this source (in 2003 but in 2004 the source was not detected) and \citet{C98} detected an OH maser at this position.


\subparagraph{G\,338.561+0.218.} This source exhibits numerous spectral features at both methanol transitions. The peak of the 12.2-GHz emission observed in 2008 December has a velocity of --40.2 \kms and flux density of 26~Jy while the 6.7-GHz emission at this velocity had a flux density of $\sim$15~Jy when measured in 2008 Aug 23. Observations of the two transitions made on consecutive days with Hobart are discussed in Section~\ref{sect:weirdos}. A very weak (0.22~Jy) water maser was detected at this location \citet{Breen10b}.

\subparagraph{G\,338.925+0.634 and G\,338.926+0.634.} This pair of sources are separated by less than 7 arcsec. ATCA observations at 6.7-GHz show that most of the emission in the velocity range --62 to --52 \kms can be attributed to the stronger site (G\,338.925+0.634) but that there is a contribution over this velocity range from the weaker site, G\,338.926+0.634 (see \citet{CasMMB102} for a more comprehensive explanation). Given the velocity range of the reasonably strong (33~Jy) 12.2-GHz source detected here covers the velocity range --61.9 to --52.8 \kmsns, we are unable to rule out possible 12.2-GHz contributions from G\,338.926+0.634. Due to this, we regard the two sources as a single site for the purposes of the analysis presented here, as well as that presented in \citet{Breen12stats}. 


\subparagraph{G\,339.476+0.185.} The peak of the 12.2-GHz emission is associated with the weakest 6.7-GHz emission feature located at a velocity of --98.1 \kmsns. Observations in 2008 December showed 12.2-GHz emission of 0.5~Jy at this velocity, whereas 6.7-GHz observations on 2008 Aug 23 showed a 0.3~Jy feature. Weak 12.2-GHz emission was also detected at the velocity of the 6.7-GHz methanol maser peak during the 2008 December observations. No 12.2-GHz emission was detected in 2010 March beyond the 5-$\sigma$ detection limit of 0.8~Jy.

\subparagraph{G\,339.622-0.121.}  The main 12.2-GHz feature is weaker by a factor of 5 than in 1988 \citep{Cas93}, 
and a periodicity of 202 days has been recognised in the 6.7-GHz emission 
\citep{Goed}.

\subparagraph{G\,342.338+0.305.} In 2008 December the 12.2-GHz peak was 0.8~Jy and coincident in velocity with the 6.7-GHz emission peak of 0.9~Jy. Further observations taken in 2010 March show a reduced flux density of $\sim$0.5~Jy. Due to unusually high noise in the initial spectrum we show the average of the two epochs in Fig.~\ref{fig:12MMB}.

\subparagraph{G\,342.446--0.072.} In 2008 December the 12.2-GHz at a velocity of --20.1~\kms is barely visible above the noise in the spectrum (3.3-$\sigma$ detection). Without the perfect correspondence with the velocity of the 6.7-GHz this source would not have been recognised. 

\subparagraph{G\,343.354--0.067.} 12.2-GHz spectra from both 2008 June and December are presented in Fig.~\ref{fig:12MMB} to show the variability between the two epochs. Each epoch shows a single feature source, with distinctly different velocities although both corresponding to the velocity of 6.7-GHz methanol maser spectral features.

\subparagraph{G\,345.010+1.792.} The peak of the 12.2-GHz surpasses the flux density of its 6.7-GHz counterpart by about a factor of two. The 12.2-GHz observations were carried out in 2008 June and the 6.7-GHz observations were conducted in 2008 Aug. Spectra of both transitions are shown in Fig.~\ref{fig:12stronger}. This unique source shows a multitude of other class II methanol maser transitions, yielding detections in at almost all of the transitions observed \citep[e.g.][and references therein]{Ellingsen04,Val'tts}. This site additionally harbours strong radio continuum and a weak, variable water maser \citep{Breen10b}, as well as an OH maser \citep{C98}.

Comparison with the 1988 spectrum of \citet{Cas93} shows that this source has varied quite markedly, unlike nearby source G\,345.012+1.797 which has remained more stable.


\subparagraph{G\,347.863+0.019.} 6.7-GHz observations in 2008 Aug revealed emission between --37.6 and --28.1 \kmsns, peaking at 6.4~Jy at a velocity of --34.7 \kmsns. 12.2-GHz observations during 2008 June (2 months prior to the 6.7-GHz observations) showed similar but weaker emission associated with the strong 6.7-GHz features, as well as an additional feature at --23.7 \kms (0.7~Jy), outside the velocity range of the 6.7-GHz emission. The nearby source G\,347.813+0.018 has 6.7-GHz emission over a velocity range covering this additional feature, but no feature at this velocity. Furthermore, no 12.2-GHz emission was detected toward G\,347.813+0.018. G\,347.863+0.019 is therefore another candidate that needs further observations to see if the 12.2-GHz emission truly dominates at the --23.7\kms feature.

\subparagraph{G\,348.195+0.768.} Weak 12.2-GHz emission was detected towards this 6.7-GHz source during both epochs at the velocity of the 6.7-GHz peak emission. Given the limited signal-to-noise, Fig.~\ref{fig:12MMB} shows the average spectrum.

\subparagraph{G\,348.550--0.979 and G\,348.550--0.979n.} These 6.7-GHz methanol masers are separated by only 2 arcsec and have some overlap in velocity range. There is 12.2-GHz emission associated with both sources which is easily separated since features are confined to the velocity ranges which are unique to each source. Both 12.2-GHz sources appear to have halved in flux density since their observation in 1988 \citep{Cas93}.

\subparagraph{G\,349.799+0.108.} A comparison between the 6.7- and 12.2-GHz emission is shown in Fig.~\ref{fig:12stronger} and reveals strong emission at both transitions. The feature between --20 and --21.5 \kms is twice as strong at 12.2-GHz as it is at the usually stronger 6.7-GHz transition (19~Jy compared with 10~Jy). 

\subparagraph{G\,350.105+0.083.} The prominent 12.2-GHz absorption towards this source, first recognised by \citet{Caswell95b}, is easily identifiable in the current spectrum. The strongest emission feature at --74.1 \kms has decreased in flux density by about a factor of 4 since the observations of \citet{Caswell95b} made in 1992.


\subparagraph{G\,351.417+0.645, G\,351.417+0.646 and G\,351.445+0.660.} G\,351.417+0.645 and G\,351.417+0.646 are 6.7-GHz sources separated by $\sim$2 arcsec. Because they are so close, strong and have overlapping velocity ranges, they are very difficult to separate using single dish data. Due to this, Table~\ref{tab:6MMB} presents 6.7-GHz velocity ranges as published in Caswell (2009) from ATCA observations. There is undoubtedly 12.2-GHz emission associated with both of these 6.7-GHz sources given the velocity of the detected spectral features. \citet{Norris93} present VLBI maps of the methanol masers associated with this region and show that the two sources are quite heavily blended together. 

Located near G\,351.417+0.645 and G\,351.417+0.646 lies a very strong water maser (1400~Jy) as well as an OH maser \citep{Breen10b,C98} and the H{\sc ii} region NGC6224F.

G\,351.445+0.660 is also confused by G\,351.417+0.645 and G\,351.417+0.646 at 6.7-GHz so the velocity range from Caswell (2009) is also used here. At 12.2-GHz this source is much less confused and can be clearly separated from the emission associated with G\,351.417+0.645 and G\,351.417+0.646.

\subparagraph{G\,351.611+0.172.} Observations conducted in 2010 March detected a 2~Jy 12.2-GHz source. It is possible that there is some contribution to the feature between --35 and --37 \kms from nearby source G\,351.688+0.171, but since there are no concurrent observations of the two sources it is difficult to confirm.


\subparagraph{G\,353.410--0.360.} No usable data were obtained towards this source during 2008 December with Parkes. Instead, Fig.~\ref{fig:12MMB} shows a new spectrum taken with the Hobart radio telescope. The detected emission has changed greatly since observations made in 1988 and 1992 \citep{Cas93,Caswell95b}. The current peak emission at --20.5 \kms has tripled in strength from the previous observations, and the previous peak at --22.3 \kms is now approximately one-third of its former flux density.


\subparagraph{G\,354.615+0.472.} This 12.2-GHz source exhibits three spectral features, two of which have remained stable since their observation by \citet{Cas93} in 1988. The third feature, at --23 \kmsns, exhibited quite significant variability during observations made by \citet{Cas93}, increasing in flux density from 10 to 20~Jy in just 5 weeks. We observed this variable feature with a flux density of 3~Jy during our observations in 2008 December.

\subparagraph{G\,355.343+0.148.} A 0.3~Jy 12.2-GHz methanol maser was identified in the average spectrum of the observations made in 2008 December and 2010 March. In order to increase the signal-to-noise in the spectrum presented in Fig.~\ref{fig:12MMB}, the average spectrum was further averaged with that of nearby source G\,355.344+0.147.  

\subparagraph{G\,357.965--0.164 and G\,357.967--0.163.} 12.2-GHz emission is detected at the velocity of  the peak 6.7-GHz feature towards G\,357.967--0.163 in both 2008 June and December. Observations of nearby source G\,357.965--0.164  in 2008 June revealed no 12.2-GHz emission. In 2008 December observations were conducted only towards G\,357.967--0.163, and show a marginal peak at --8.8~\kms which is outside the velocity range of the 6.7-GHz emission towards this source, but close to the peak velocity of the 6.7-GHz emission associated with G\,357.965--0.164. Further observations of G\,357.965--0.164 are needed to determine if there is weak, variable 12.2-GHz emission associated with this 6.7-GHz methanol maser.

Both G\,357.965--0.164 and G\,357.967--0.163 have associated water masers \citep{Breen10b}, the first of which showed extreme variation over a $\sim$10 month period reducing in flux density from 53 to 0.9~Jy. G\,357.967--0.163 shows a total velocity range of 181 \kms over the two epochs and moderate levels of variability, has an OH maser counterpart \citep{C98} and is associated with a weak 22-GHz continuum source \citep{Voronkov11}. The morphology of associated mid-infrared emission, as well as the location of a nearby class I methanol maser at 23.4-GHz is discussed in detail in \citet{Voronkov11}.

\subparagraph{G\,358.371--0.468.} 12.2-GHz observations in 2008 June were carried out towards a position offset from the accurate 6.7-GHz position by more than 1.2 arcmin, partially accounting for the change in flux density between epochs from 1.9~Jy in June to 8.1~Jy in December where observations were targeted towards the accurate position. Also coincident with this methanol maser site is a water maser \citep{Breen10b}.

\subparagraph{G\,358.809--0.085.} Very weak 12.2-GHz emission was detected in both 2008 June and December. While the emission was only detected at 0.3~Jy during both epochs, its velocity matches that of the 6.7-GHz emission peak. High noise in the 2008 December spectrum makes the average of the two epochs no better than the more sensitive observations conducted in 2008 June.


A very similar situation is observed for G\,359.436--0.104. Weak emission of 0.5~Jy was detected in 2008 December ,but not 2008 June, towards a weak secondary feature present in the 6.7-GHz spectrum.

\subparagraph{G\,359.615--0.243.} Observed in both 1988 and 1992 \citep{Cas93,Caswell95b}, this 12.2-GHz sources appears to have been steadily decreasing in flux density over the past 20 years. The measurement made in 2008 Dec, shows emission approximately one-third of the strength reported by the 1988 observations of \citet{Cas93}.

\subparagraph{G\,0.546--0.852.} The 12.2-GHz spectral feature detected at 14 \kms has steadily decreased in flux density since its observation in 1988 \citet{Cas93} (when it was the peak emission at 8~Jy), to observations in 1992 \citep{Caswell95b} and our current observations where we observed a 0.4~Jy feature at this velocity. The current peak emission near 18 \kms is almost identical to that observed in 1992 \citep{Caswell95b}.

\subparagraph{ G\,0.651--0.049, G\,0.666--0.029 and G\,0.667--0.034.} These three 6.7-GHz methanol masers are clustered near Sgr B2, which in total contains 10 6.7-GHz methanol masers within a 3 arcmin region. The 6.7-GHz targets in the Sgr B2 region were presented in \citet{CasMMB10}, however, positions and velocity ranges were taken from \citet{HW95} who conducted high sensitivity ATCA observations towards this region. We conducted 12.2-GHz observations towards all 10 6.7-GHz masers in this region, and detect emission towards three sources: G\,0.651--0.049, G\,0.666--0.029 and G\,0.667--0.034.

G\,0.651--0.049 is the strongest 12.2-GHz methanol maser detected in this region at 2.1~Jy with a peak velocity of 48.1 \kmsns. \citet{Cas93}
and \citet{Caswell95b} similarly detected a single peak of maser emission centred near G\,0.650--0.048
(J2000  position 17 47 20.8, --28 24 22 with uncertainty less than 10 
arcsec), with velocity of 48.0~\kms and peak intensity of 7.2~Jy.  Measurements of 
\citet{HW95} show a strong 6.7-GHz maser at G\,0.651--0.049 
(17 47 21.12, --28 24 18.2) and velocity 48.0 \kmsns, seen with peak 
intensity of 22 Jy in the MMB observations \citep{CasMMB10}. Within 
its position uncertainty, the 12.2-GHz maser thus coincides with the more 
precisely measured 6.7-GHz maser.  

Two weak (0.5 and 0.6~Jy), new 12.2-GHz methanol masers were detected towards G\,0.666--0.029 and G\,0.667--0.034. The 12.2-GHz emission associated with G\,0.666--0.029 falls at the velocity of a secondary 6.7-GHz methanol maser feature. Whiteoak et al. (1988) map a strong 12.2-GHz absorption feature which is clearly 
seen on our spectrum of G\,0.666-0.029. 

12.2-GHz emission detected towards G\,0.667--0.034 falls at the velocity of the peak 6.7-GHz methanol maser emission reported by \citet{HW95}. The flux density of the 6.7- and 12.2-GHz maser emission towards this source have comparable flux densities; \citet{HW95} report a 6.7-GHz flux density of 0.4~Jy while we detect 12.2-GHz maser emission with a peak flux density of 0.6~Jy. From inspection of the MMB spectrum \citep{CasMMB10} it appears that the 6.7-GHz emission has increased to more than 1~Jy since the observations of \citet{HW95}.


\subparagraph{G\,0.836+0.184.} 12.2-GHz observations carried out in 2008 December show alignment of weak (0.6~Jy), narrow emission with the 3.5 \kms 6.7-GHz peak, but also alignment of secondary feature at --2.2 \kms with 6.7-GHz emission. The correspondence in velocity lends support that this source is a genuine detection which is needed since the signal-to-noise in the spectrum is low. No 12.2-GHz emission was detected in 2008 June. 

\subparagraph{G\,1.147--0.124.} Possible 12.2-GHz emission was identified in 2008 December at --19.4 \kms due to its correspondence with a secondary feature in the 6.7-GHz spectrum.  At 0.5~Jy this feature is only a 3-$\sigma$ detection and the presence of spurious features of similar flux densities within the spectrum make the possible emission questionable. Further observations are required to determine whether this emission is genuine and due to this we do not list it as a detection but we include the spectrum in Fig.~\ref{fig:12MMB} .


\subparagraph{G\,3.502--0.200.} A weak 12.2-GHz source consisting of two features was detected in 2008 December with a peak flux density of 0.4~Jy. Observations in both 2008 June and 2010 March showed no discernible emission.


\subparagraph{G\,8.683-0.368 and G\,8.669--0.356.} The separation of these companions is 66 arcsec, and for our 12.2-GHz 
observations, each lies at about the half-power point of the other. All 
prominent 12.2-GHz features arise from G\,8.683-0.368, as suggested 
by our aligned spectra at the two positions, and confirmed 
from the \citet{Cas93} observations;  the spectrum has remained 
similar for more than 20 years, and closely matches that of 6.7-GHz at 
this position.  However, a weak feature of 0.4~Jy was observed in 2008 June towards G\,8.669--0.356 at 39.9 \kmsns, the velocity of the 10~Jy peak of this 6.7-GHz maser. Further observations completed in 2008 December were of uncharacteristically poor sensitivity, with a 5-$\sigma$ detection limit of 1.15~Jy, and it is therefore not surprising that no emission was detected at this epoch. Also located at G\,8.669--0.356 are a water maser, 22 GHz radio continuum \citep{Breen10b} and an OH maser \citet{C98}.

\subparagraph{G\,9.621+0.196 and G\,9.619+0.193.}
For G\,9.621+0.196 the coincidence of the main 12.2-GHz feature with its 6.7-GHz 
counterpart has been confirmed not only by Caswell et al. (1993), but also 
by Sanna et al. (2009).  The weaker 12.2-GHz emission from 9.619+0.193 has 
not previously been reported but its velocity supports our conclusion 
that it arises here rather than at the companion position offset by 12 arcsec.

\section{Discussion}

This catalogue of 12.2-GHz methanol masers represents the largest sensitive search of a complete sample of 6.7-GHz methanol masers achieved to date. The longitude range included in this portion of the catalogue is only a subset of the survey range, which, when complete, will cover the full range of the MMB survey.

Our current study uses 6.7-GHz maser targets which have subarcsec position 
uncertainties.  Very few 12.2-GHz masers have been positioned to similar 
accuracy, but where these accuracies have been achieved the spatial correspondence has been confirmed \citep[e.g.][]{Mos02,Goed05a}. In addition, a substantial number of 12.2-GHz sources, more than 50, were positioned with 10 
arcsec a ccuracy \citep{Cas93} in 1988, prior to the detection of 
6.7-GHz masers.  Subsequent comparisons show excellent coincidence 
with presumed 6.7-GHz counterparts, usually well within the expected 10 
arcsec uncertainties. Thus, despite the limited number of coincidences confirmed to high 
precision, these facts strongly suggest that most 12.2-GHz detections found 
in our targeted search do indeed coincide precisely with the 6.7-GHz 
targets, although additional observations to test this have not yet been 
made.

Global properties of the 12.2-GHz methanol masers detected towards the full sample of MMB sources south of declination --20$^{\circ}$ (580 sources) have already been given in \citet{Breen12stats}. This analysis confirmed that the 12.2-GHz methanol masers are principally detected towards the stronger 6.7-GHz methanol masers, and in 80 per cent of cases show their peak emission at the same velocity. Using the complete samples of 6.7- and 12.2-GHz methanol masers, \citet{Breen12stats} found that the detection rate of 12.2-GHz methanol masers is lower, at 43 per cent, than found in previous searches of comparable sensitivity. The authors concluded that the difference in detection rate is a consequence of the biases that were present in previous searches (which typically targeted slightly more evolved star formation regions). From this previous work, \citet{Breen12stats} found that the data support an evolutionary scenario whereby the 12.2-GHz methanol masers are present during the latter half of the 6.7-GHz methanol maser lifetime. Furthermore, the data show evidence that these methanol maser transitions increase in both flux density and velocity range as the sources evolve.

Here, we do not attempt to reproduce those results of \citet{Breen12stats} on the subset of data we present. Instead, the discussion that follows focuses on specific data, unusual sources and the completeness of our search.

\subsection{Basic statistics}

The 184 12.2-GHz methanol masers detected in the longitude range 330$^{\circ}$ to 10$^{\circ}$ exhibit velocity ranges between 0.1~\kms and 20~\kms and flux densities from 0.3~Jy to 976~Jy. The strongest newly discovered source is 64~Jy.

In the Galactic longitude range 330$^{\circ}$ to 10$^{\circ}$ we find 12.2-GHz methanol maser emission towards 46 per cent of the 400 6.7-GHz methanol masers searched. In comparison, the detection rate towards the full 580 6.7-GHz methanol masers south of declination --20$^{\circ}$ was 43 per cent \citep{Breen12stats}. The slightly higher detection rate in the portion of the Galactic plane presented in this catalogue is not unexpected since it encompasses the 335$^{\circ}$ to 340$^{\circ}$ longitude region, noted by \citet{Breen12stats} as having a significantly higher 12.2-GHz detection rate than other portions of the Galactic plane. 

\begin{figure}
	\epsfig{figure=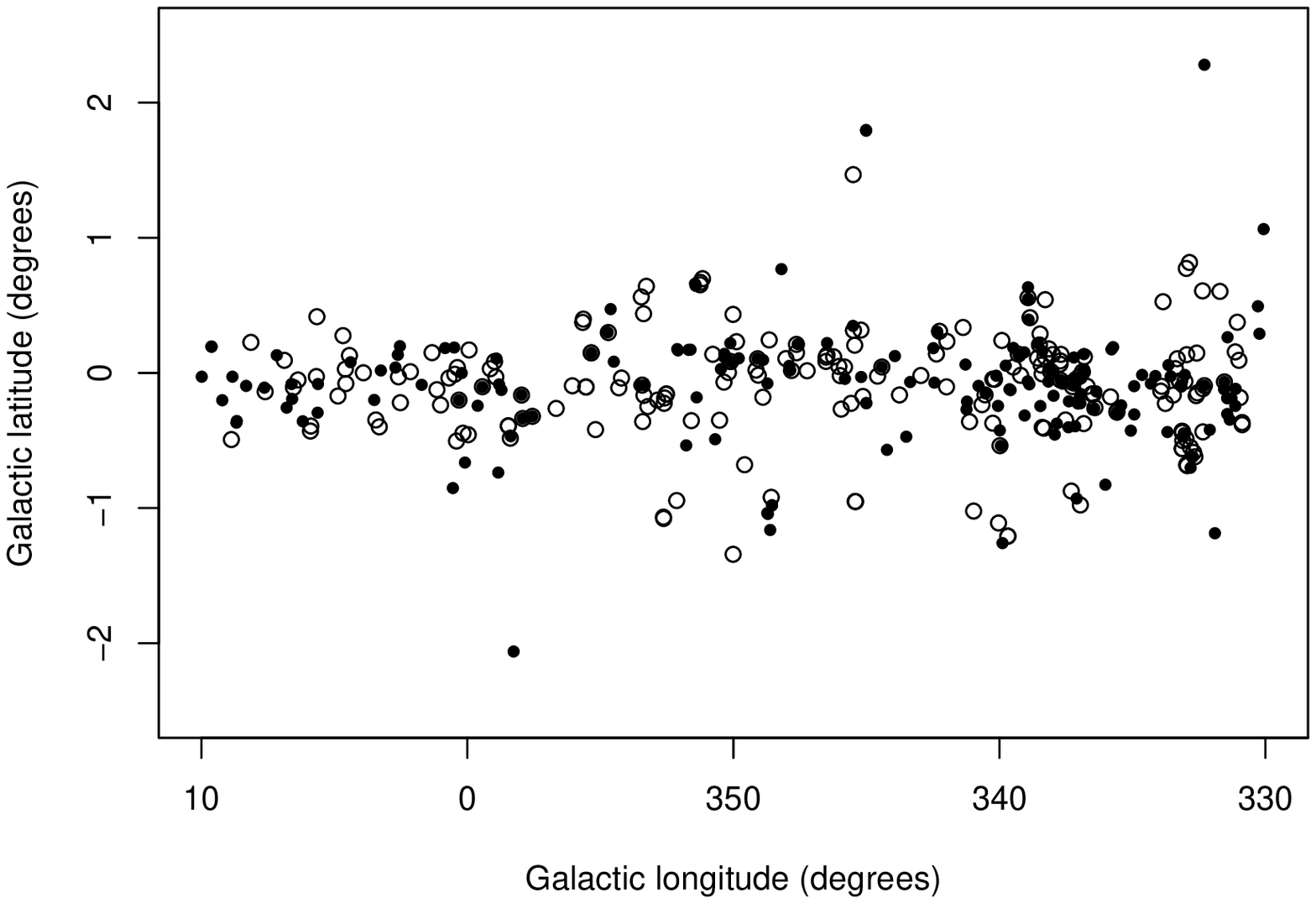,angle=270,width=9cm}\vspace{-1cm}
	\epsfig{figure=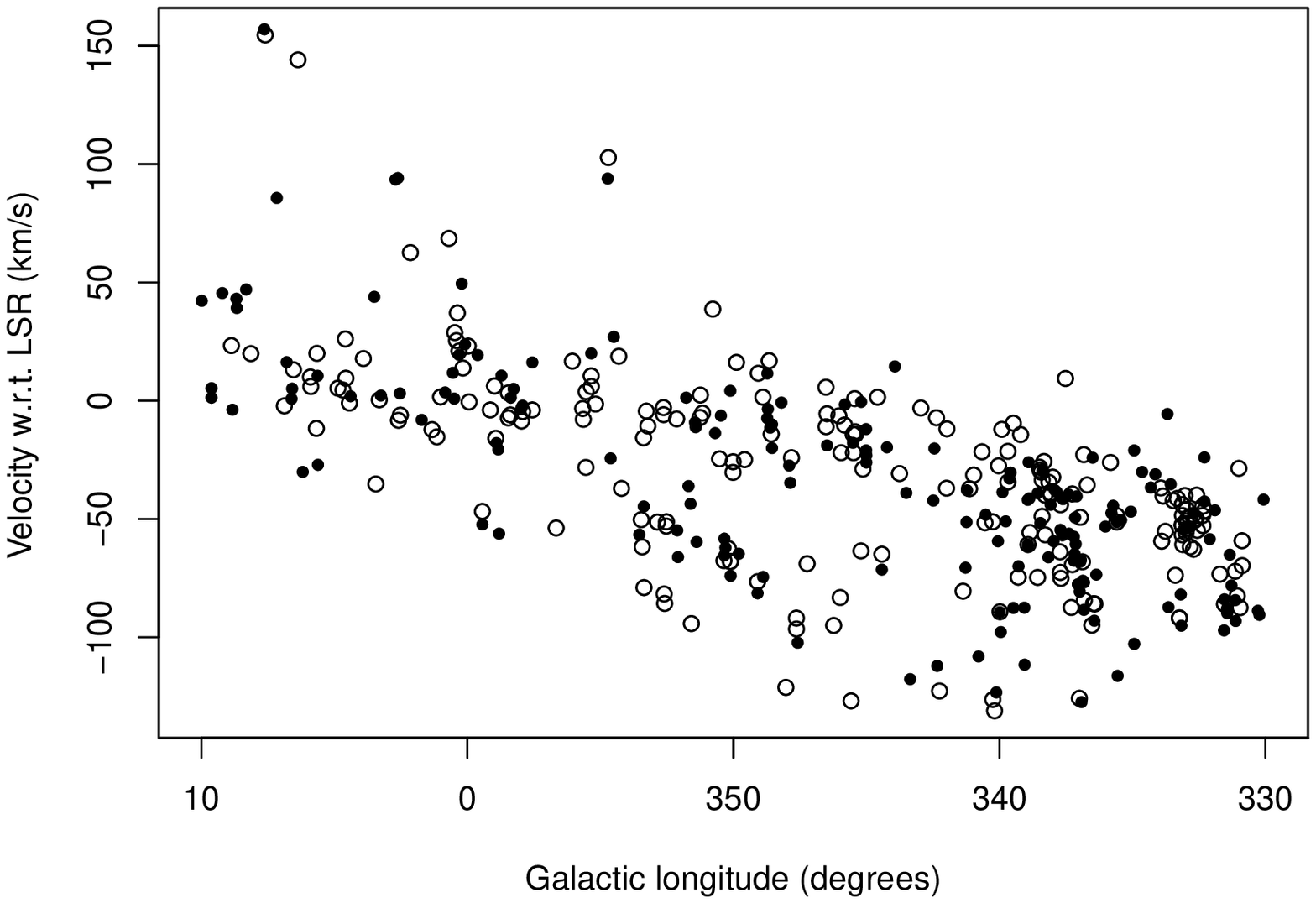,angle=270,width=9cm}
\caption{Distribution of the methanol masers in the Galactic longitude range 330$^{\circ}$ (through 360$^{\circ}$) to 10$^{\circ}$ (top) and distribution of the peak 6.7-GHz methanol maser velocity versus Galactic longitude (bottom). Dots mark the locations of 6.7-GHz methanol masers with 12.2-GHz counterparts, while open circles show those 6.7-GHz sources devoid of 12.2-GHz emission.}
\label{fig:dist}
\end{figure}

Fig.~\ref{fig:dist} shows the distribution of 6.7-GHz methanol masers in this portion of the Galactic plane, as well as the distribution of line-of-sight velocities versus Galactic longitude. In this figure, the 6.7-GHz sources with associated 12.2-GHz methanol masers have been distinguished from those without by the use of different symbols. The distribution of the 6.7-GHz methanol masers with and without 12.2-GHz counterparts appears to be approximately random, with the exception of the notable dominance of 12.2-GHz sources in the 330$^{\circ}$ to 340$^{\circ}$ longitude region (as described in \citet{Breen12stats}).

\subsection{Survey completeness}

\begin{figure}
	\epsfig{figure=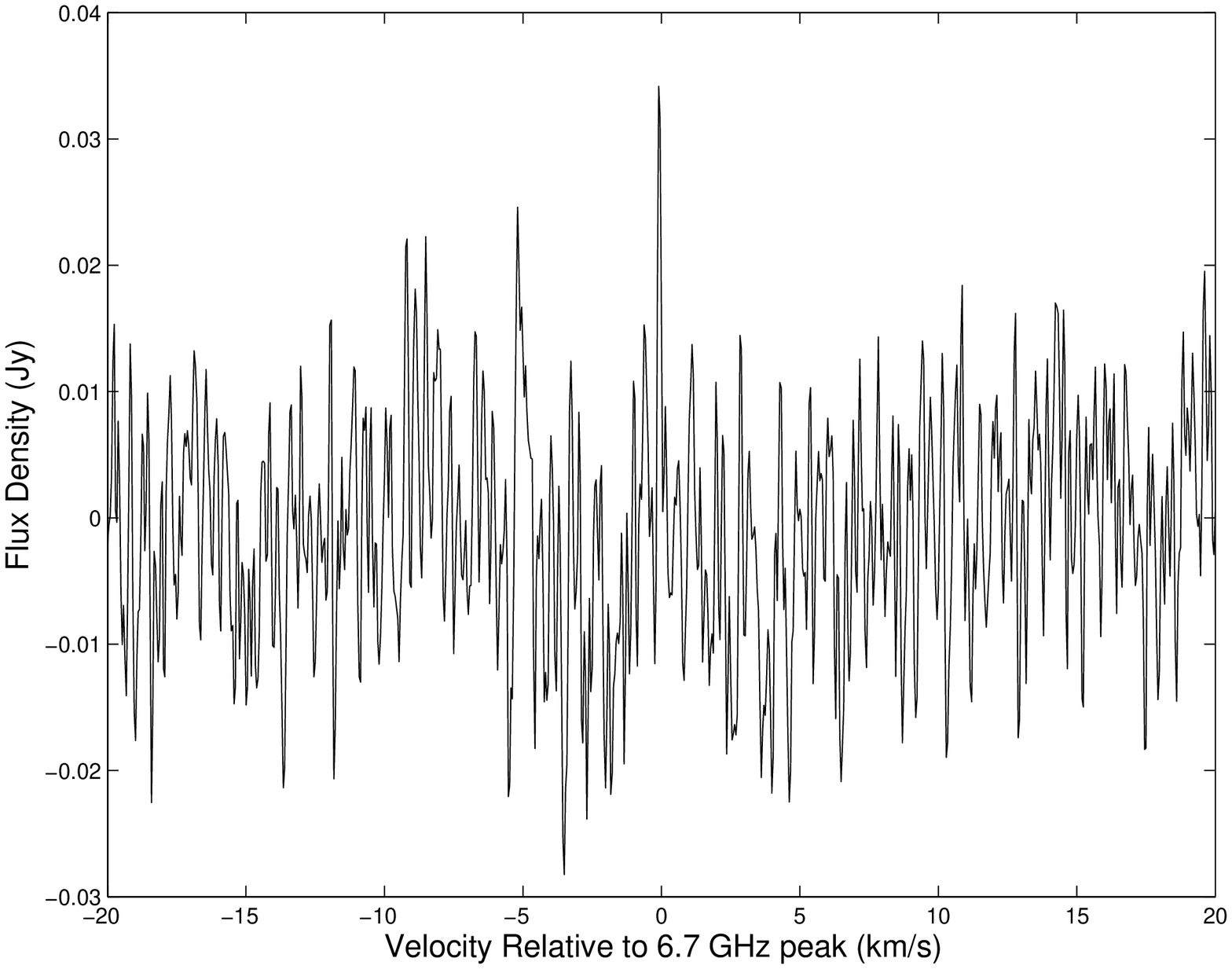,angle=0,width=9cm}
\caption{`Stacked' spectrum of all 12.2-GHz non-detections aligned at the velocity of the associated 6.7-GHz methanol maser peak. The rms of the spectrum is 0.009~Jy.}
\label{fig:comp}
\end{figure}

The completeness of our search can be meaningfully tested since we know the peak velocity of the 6.7-GHz methanol maser emission, and that 80 per cent of 12.2-GHz methanol masers share their peak velocity with their 6.7-GHz methanol maser counterpart \citep{Breen12stats}. Using the method known as `stacking' we have averaged all of our 12.2-GHz non-detections (from the 2008 June and December epochs) after first aligning them at their 6.7-GHz methanol maser peak velocity. Any non-detections that were contaminated by absorption or emission from neighbouring sources were excluded from this analysis. A total of 277 12.2-GHz spectra were `stacked' (for some sources spectra from both epochs have been included) and resulted in an averaged spectrum with an rms noise of 9~mJy, measured over a 100 \kms velocity range centred on the velocity of the 6.7-GHz methanol maser peak emission. The resultant `stacked' spectrum presented in Fig.~\ref{fig:comp} shows no emission beyond the 4-$\sigma$ level.

Since 12.2-GHz methanol masers are chiefly detected towards stronger 6.7-GHz sources, we have conducted further, luminosity-based tests of the completeness. For this test, only the 12.2-GHz non-detection spectra observed towards 6.7-GHz methanol masers with luminosities greater than the average luminosity (71 sources) were `stacked'. All luminosities were calculated using the near distance determined using the prescription of \citet{Reid09}. The resultant `stacked' spectrum had an rms of 0.015~Jy, and no emission above 0.04~Jy.


We conclude that the 12.2-GHz non-detections, on average, have no emission above a flux density of 0.09~Jy (10 times the rms of the stacked spectrum presented in Fig.~\ref{fig:comp}) at the velocity of the 6.7-GHz methanol maser peak. Therefore, we are confident that there are few 12.2-GHz methanol masers just below our detection limit, and consequently sensitivity biases do not have a large influence on the results presented here, or on the conclusions drawn in \citet{Breen12stats}. 

%
%

\subsection{Luminosity relationship of individual features}\label{sect:individual}

In \citet{Breen12stats} we presented a preliminary analysis of luminosity ratio of individual maser features as a function of 6.7-GHz methanol maser luminosity. In that analysis the near kinematic distances were assumed for the five sources presented, since the distance ambiguities were unable to be resolved. Here we repeat the analysis, using kinematic distances that have been resolved using H{\sc i} self absorption data \citep{GM11}. 

\citet{Breen10a} recognised a relationship between the ratio of 6.7:12.2-GHz peak flux densities with the luminosities of the respective data. They found that the ratio of 6.7:12.2-GHz peak flux density apparently increased with increasing 6.7-GHz methanol maser luminosity; implying that as the sources evolved the 6.7-GHz peak flux density increased more rapidly then the associated 12.2-GHz maser emission. \citet{Breen12stats} confirmed this trend with a more thorough analysis of a larger number of sources. Analysis of the 6.7:12.2-GHz peak flux density ratio as a function of 6.7-GHz methanol maser luminosity of the multiple individual features (present in both the 6.7- and 12.2-GHz spectra) of a group of sources has the potential to lend further insights into this phenomenon as the evolutionary stage of the exciting star will be common to individual maser features.

\begin{figure*}\vspace{-1cm}
	\psfig{figure=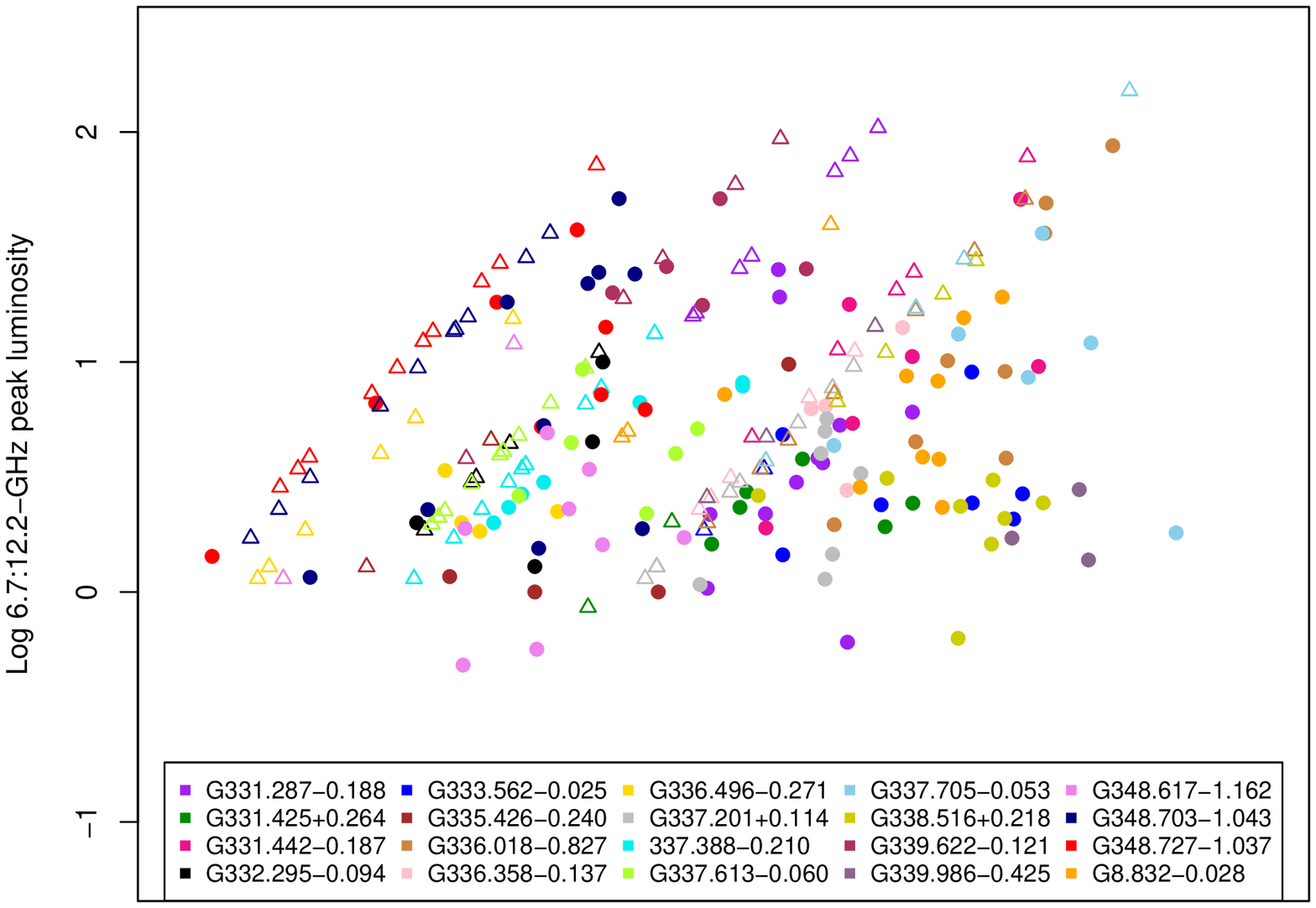,angle=270,width=14cm}\vspace{-3cm}
	\psfig{figure=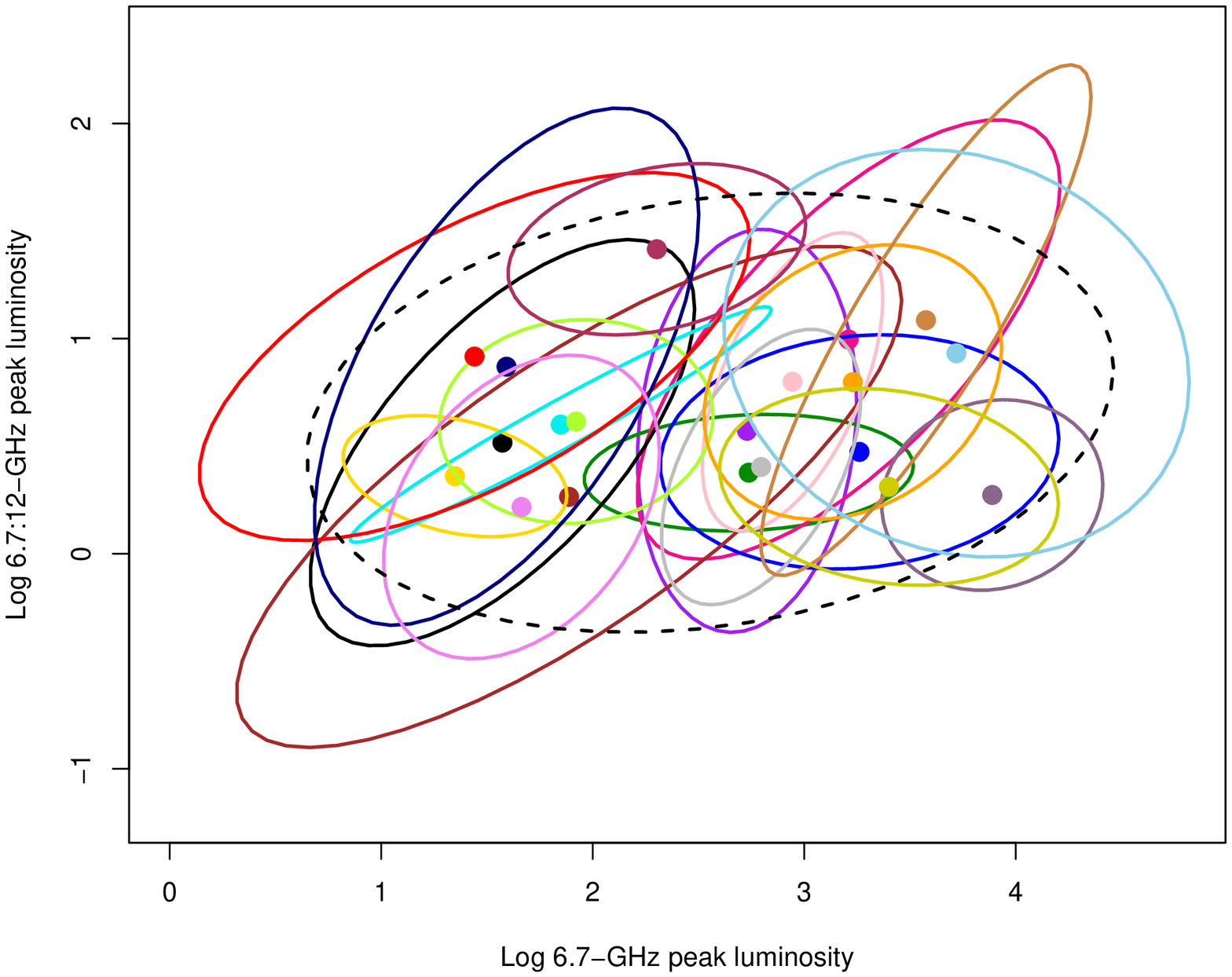,angle=270,width=14cm}
\caption{Log of the 6.7:12.2-GHz individual feature ratio versus the log 6.7-GHz peak luminosity for every spectral feature associated with 20 methanol masers.  Top: Dots represent 6.7 GHz features with a 12.2 GHz counterpart, while triangles show features with no detectable 12.2 GHz counterpart, thus showing the lower limit of the 6.7:12.2 GHz feature ratio. Colours distinguish between individual maser sites which are listed in the lower panel. Bottom: 80 per cent confidence ellipses for features detected at both methanol maser frequencies where large dots show the central point of each ellipse. Colour scheme as for the top figure. The dashed black ellipse shows the 90 per cent confidence ellipse of all spectral features detected at both frequencies associated with the 20 sources.}
\label{fig:ind}
\end{figure*}

We have selected 20 methanol masers from our sample which show four or more spectral features at both 6.7- and 12.2-GHz, and also have had their kinematic distance ambiguities resolved \citep{GM11}. Luminosities of individual features have been calculated by multiplying their respective peak flux density by the square of the distance (giving units of Jy~kpc$^2$). Fig.~\ref{fig:ind} displays the ratio of 6.7:12.2-GHz luminosity of individual features as a function of the 6.7-GHz luminosity. In the top portion of the figure, the features detected at both maser frequencies are represented by dots, and triangles show lower limits on the ratios of 6.7-GHz features with no detectable 12.2-GHz counterpart. Individual sources are colour-coded which shows, firstly, that all points associated with a single source are generally clustered together, and secondly, that the lower limits on ratios of features with no detected 12.2-GHz emission lie on the outskirts of the clustered dots associated with each individual source. 

%
%
%

The bottom half of Fig.~\ref{fig:ind} shows an ellipse for each of the 20 methanol maser sources presented in the top figure, representing the parameter space covered by all of the features associated with a single source. For each source we show the 80 per cent confidence ellipse and these were chosen as they do a good job of estimating the parameter space covered by all the features of each single source, even though a normally distributed data set is assumed. The 80 per cent confidence ellipse actually represents the region where we can be 80 per cent confident that the true centre of the data associated with each source lies, and is achieved by superimposing normal probability contours over the data associated with each source. Only 10 of the 131 data points associated with the 20 sources lie beyond the boundary of their associated ellipse, and in every case these data points fall only just beyond the boundary of the fitted ellipse. For comparison, we have also drawn the 90 per cent confidence ellipse associated with all of the data points associated with all of the 20 sources (which similarly encompasses all but 14 of the 131 data points). It is evident on comparison of the individual source ellipses with the entire dataset ellipse that the scatter is much less in individual sources than the whole dataset; in fact, the median area covered by the individual sources as a percentage of the total population ellipse is $\sim$20 per cent.


The data presented in Fig.~\ref{fig:ind} is consistent with the findings of \citet{Breen12stats}, and supports the idea that the luminosities of the sources are changing as a function of evolution. This is evident from the fact that the scatter in the values of individual sources is much less than is present in the full dataset comprising the 20 sources. The top part of the figure additionally shows that the non-detections of some features at 12.2-GHz are not primarily due to sensitivity limitations, since the ratio lower limits are located away from the detected features which are clustered together. 

\subsection{6.7-GHz sources with stronger 12.2-GHz features}\label{sect:weirdos}

\begin{figure*}
	\psfig{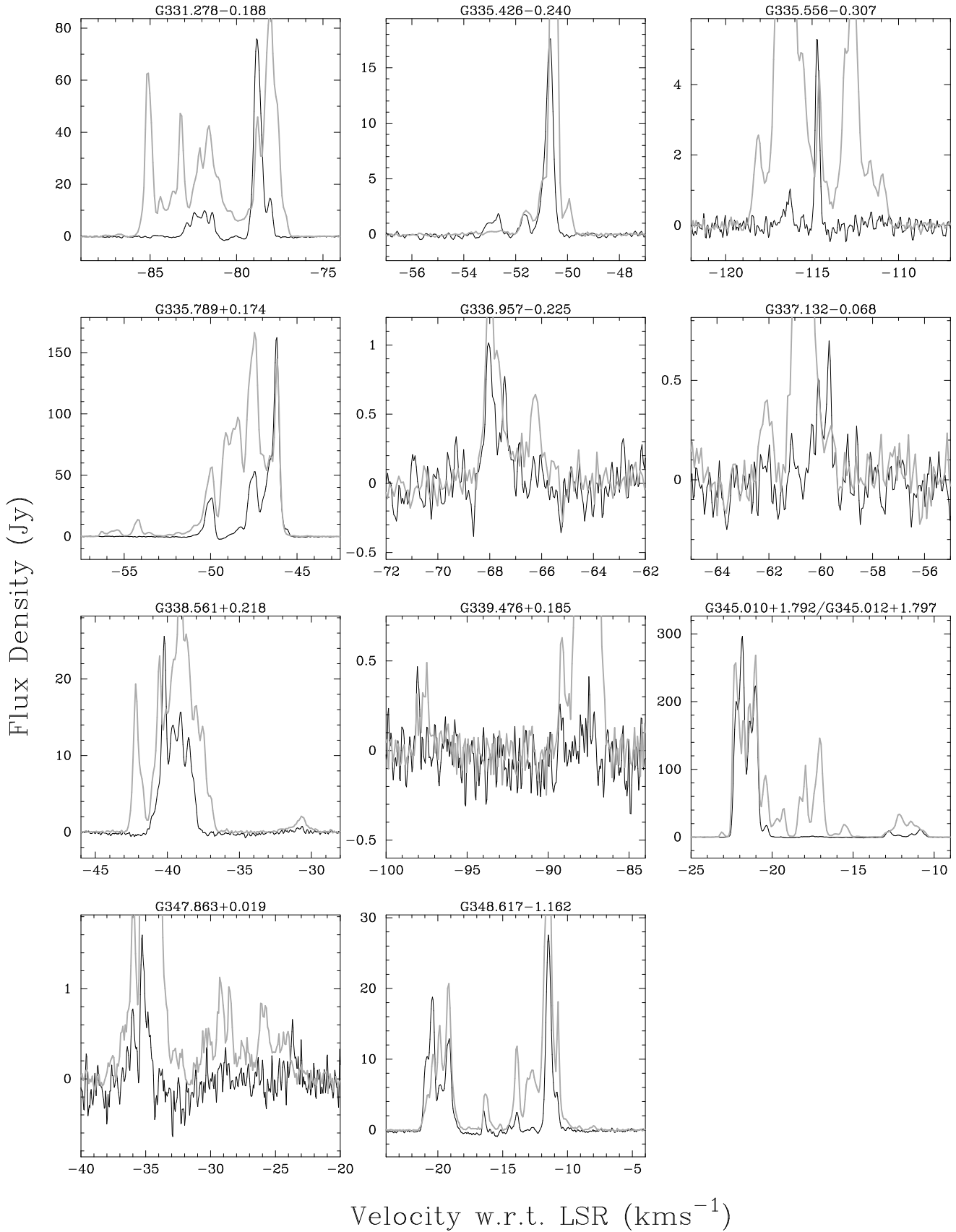}
\caption{Comparison between the 12.2-GHz emission (black) and 6.7-GHz emission (grey) for sources showing stronger emission at 12.2-GHz for at least one spectral feature.}
\label{fig:12stronger}
\end{figure*}

From previous observations and from maser pumping models, 12.2-GHz methanol masers are expected to be weaker than their 6.7-GHz counterparts. We find, of the 400 observed 6.7-GHz methanol maser sources, 11 exhibit stronger emission at 12.2-GHz than at 6.7-GHz, in at least one spectral feature and these are shown in Fig.~\ref{fig:12stronger} (i.e. 3 per cent of sources). Only one source, G\,345.010+1.792, has a higher peak flux density at 12.2-GHz. The analysis of the individual features of the 20 sources studied in Section~\ref{sect:individual} shows that 1.6 per cent of their individual features show larger flux densities at 12.2-GHz than at 6.7-GHz. In some cases, the apparently stronger 12.2-GHz may be a result of variability between the observations epochs, which were commonly separated by at least 3-4 months. 

In an attempt to investigate the relative age of these unusual sources, we have considered both the 6.7-GHz luminosities (which are thought to increase with evolution \citep[e.g.][]{Breen10a,Breen12stats,Wu10}), and associations with other tracers. We find that the average luminosity of these 6.7-GHz methanol masers is  comparable to the average luminosity of the full sample of 6.7-GHz methanol masers with associated 12.2-GHz sources. Comparison between these sources and the locations of extended green objects \citep{Cyg08} and OH masers \citep{C98} also provides no clear insights, since some are associated with neither tracer, some with both and others with either an extended green object or an OH maser. Therefore we find no evidence that these sources are associated with a particular evolutionary stage in the methanol maser lifetime.


%
%
%

%

In order to determine if it is variability that is responsible for the apparently stronger 12.2-GHz maser emission, we conducted near simultaneous observations (at both frequencies) of three sources, using the Hobart 26-m radio telescope. These three sources (G\,331.278--0.188, G\,335.789+0.174 and G\,338.561+0.218) were selected from the list of 11 sources which showed stronger emission in at least one spectral feature at 12.2-GHz, primarily on the basis of their large flux densities. Observations were completed during 2010 October 20 and 22 at 12.2- and 6.7-GHz respectively. Remarkably, all three sources show very similar spectra
in these observations as those presented in Fig.~\ref{fig:12stronger}. It is therefore unlikely that variability is a viable explanation for a large fraction of the 11 12.2-GHz methanol masers that show stronger emission in this transition than at 6.7-GHz.

%


Given that there are no systematic differences in the properties associated with these sources, it appears that the most likely explanation is that it is not remarkable to have $\sim$3 per cent of  sources showing stronger emission at 12.2-GHz than at 6.7-GHz in a single feature. In fact, a similar number of 6.7-GHz methanol masers with high luminosities show either no 12.2-GHz emission, or extremely weak 12.2-GHz emission relative to their emission at 6.7-GHz. Furthermore, since the integrated flux density of these 12.2-GHz masers is never larger than that of the 6.7-GHz methanol maser, perhaps the most plausible explanation lies in the stochasticity of the maser process; it is quite conceivable that these 12.2-GHz features are just experiencing more favourable path lengths than their 6.7-GHz counterparts.  We can further investigate the relationship between the relative intensities of the 6.7- and 12.2-GHz maser emission through the results shown in Figure~\ref{fig:ind}.  This shows that it is not uncommon for the intensity ratio between 6.7- and 12.2-GHz methanol masers at the same velocity to vary by more than an order of magnitude within an individual maser region.  This shows that although the two transitions are usually observed to be co-spatial on milliarcsecond scales \citep{Mos02} the variations in the physical conditions on scales of 10$^{17}$ cm (6000 au) \citep{Cas97} are sufficient to cause significant differences in the relative strength of the 6.7- and 12.2-GHz maser emission.  This suggests that the stochastic nature of the masing process causes fundamental limits in the ability of multi-transition modelling of masers to measure the physical conditions at very high angular resolutions, such as has been attempted by \citet{Cragg01} and \citet{Sutton01}.  The effect of stochastic processes (both variations in the physical conditions along the path and the turbulence of the velocity field) on the observed intensity ratios involving the rarer methanol transitions are likely to be even greater, as the degree of saturation is likely to be less than it is for the 6.7 and 12.2 GHz transitions. This source of uncertainty is likely to dominate compared to considerations such as variability due to non-simultaneous observations of the different transitions.

\section{Summary}

We present 184 12.2-GHz methanol masers detected towards the complete sample of 6.7-GHz methanol masers detected in the MMB survey in the longitude range 330$^{\circ}$ (through 360$^{\circ}$) to 10$^{\circ}$. This catalogue represents the largest, most sensitive, systematic search for 12.2-GHz methanol masers targeted towards a complete sample of 6.7-GHz methanol masers. Over this longitude range we find a detection rate of 46 percent and have discovered 117 new 12.2-GHz methanol maser sources. We have tested the completeness of our 12.2-GHz search and conclude that, on average, 12.2-GHz non-detections have no emission above a flux density of 0.09~Jy.

Individual interesting sources are highlighted and discussed, especially those which unexpectedly have stronger emission at 12.2-GHz than at 6.7-GHz. We suggest that the relatively small number of sources that show stronger emission at 12.2-GHz is consistent with expectations due to the stochasticity of the maser process, and find no evidence that these sources share a common evolutionary stage.

An investigation of the individual spectral features of 20 sources exhibiting both 6.7- and 12.2-GHz emission has revealed trends that are consistent with their luminosities as well as their flux density ratios being governed by the evolutionary stage of each source. This result is in agreement with previous studies.

Electronic versions of all of the 6.7-GHz MMB detections are available at www.astromasers.org or www.manchester.ac.uk/jodrellbank/mmb and we anticipate adding digital versions of the 12.2-GHz data in the near future, allowing for detailed comparisons between individual spectra.

\section*{Acknowledgments}

The Parkes telescope is part of the Australia Telescope which is funded by the Commonwealth of Australia for operation as a National Facility managed by CSIRO. Financial support for this work was provided by the Australian
Research Council. This research has made use of: NASA's Astrophysics
Data System Abstract Service; and  the SIMBAD data base, operated at CDS, Strasbourg,
France.


\begin{thebibliography}{}


%
%
%
%
%
%
%

 \bibitem[\protect\citeauthoryear{Batrla \etal}{1987}]{Batrla87}
Batrla W., Matthews H.\ E., Menten K.\, M., Walmsley C.\ M., 1987, Nature, 326, 49
  
  
%
%
%
%
%


\bibitem[\protect\citeauthoryear{B{\l}aszkiewicz \& Kus}{2004}]{Blas04}
 B{\l}aszkiewicz L., Kus A.\ J., 2004, A\&A, 413, 233

%
%
%
%

 \bibitem[\protect\citeauthoryear{Brand \& Blitz}{1993}]{BB93}
Brand J., Blitz L., 1993, A\&A, 275,67

%
%
%
%
%
%





\bibitem[\protect\citeauthoryear{Breen \etal}{2010a}]{Breen10a}
 Breen S.\ L., Ellingsen S.\ P., Caswell J.\ L., Lewis B.\ J., 2010a, MNRAS, 401, 2219

\bibitem[\protect\citeauthoryear{Breen \etal}{2010b}]{Breen10b}
 Breen S.\ L., Caswell J.\ L., Ellingsen S.\ P., Phillips C.\ J., 2010b, MNRAS, 406, 1487

\bibitem[\protect\citeauthoryear{Breen \etal}{2011}]{Breen12stats}
 Breen S.\ L., \etal\ 2011, ApJ, 733, 80







%
%
%
%
%
%
%
%
%
%
%
%
%
%

\bibitem[\protect\citeauthoryear{Caswell \etal}{1993}]{Cas93}
 Caswell J.\ L., Gardner F.\ F., Norris R.\ P., Wellington, K.\ J., McCutcheon W.\ H., Peng, R.\ S., 1993, MNRAS, 260, 425 
 

%

\bibitem[\protect\citeauthoryear{Caswell \etal}{1995}]{Caswell95b}
 Caswell J.\ L., Vaile R.\ A., Ellingsen S.\ P., Norris R.\ P., 1995, MNRAS, 274, 1126
 


\bibitem[\protect\citeauthoryear{Caswell}{1997}]{Cas97}
 Caswell J.\ L., 1997, MNRAS, 289, 79 
 
 
\bibitem[\protect\citeauthoryear{Caswell}{1998}]{C98}
 Caswell J.\ L., 1998, MNRAS, 297, 215

%
%
%
%
%
%
%
%
%
%
%


\bibitem[\protect\citeauthoryear{Caswell}{2009}]{C09}
 Caswell J.\ L., 2009, PASA, 26, 454
 

 


 \bibitem[\protect\citeauthoryear{Caswell \etal}{2010}]{CasMMB10}
Caswell J.\ L. et al., 2010, MNRAS, 404, 1029 


\bibitem[\protect\citeauthoryear{Caswell, Kramer \& Reynolds}{2011}]{Caswell11}
 Caswell J.\ L., Kramer B.\ H., Reynolds J.\ E., 2011, MNRAS, 415, 3872


 \bibitem[\protect\citeauthoryear{Caswell \etal}{2011}]{CasMMB102}
Caswell J.\ L. et al., 2011, MNRAS, {\em in press}


%
 

  \bibitem[\protect\citeauthoryear{Catarzi, Moscadelli \& Panella }{1993}]{Catarzi93}
Catarzi M., Moscadelli L., Panella D., 1993, A\&AS, 98, 127
 
 
%
%
%
%
%
%
%
%
%
%
%
%
%
%
%
%
%
%
%
%
%
\bibitem[\protect\citeauthoryear{Cragg \etal}{2001}]{Cragg01}
Cragg D.\ M., Sobolev A.\ M., Ellingsen S.\ P., Caswell J.\ L., Godfrey P.\ D., Salii S.\ V., Dodson R.\ G., 2001, MNRAS, 323, 939
  
 \bibitem[\protect\citeauthoryear{Cragg, Sobolev \& Godfrey}{2002}]{Cragg02}
  Cragg D.\ M., Sobolev A.\ M., Godfrey P.\ D., 2002, MNRAS, 331, 521
   
 
\bibitem[\protect\citeauthoryear{Cragg, Sobolev \& Godfrey}{2005}]{Cragg05}
  Cragg D.\ M., Sobolev A.\ M., Godfrey P.\ D., 2005, MNRAS, 360, 533


\bibitem[\protect\citeauthoryear{Cyganowski  \etal}{2008}]{Cyg08}
Cyganowski C.\ J. \etal, 2008, AJ, 136, 2391 

%
\bibitem[\protect\citeauthoryear{Cyganowski  \etal}{2009}]{Cyg09}
Cyganowski C.\ J., Brogan C. L., Hunter T.\ R., Churchwell E., 2009, ApJ, 702, 1615



%
%
%
%
%
%
%
%
%
%
%
%
%
%
%
\bibitem[\protect\citeauthoryear{Ellingsen \etal}{2004}]{Ellingsen04}
Ellingsen S.\ P., Cragg D.\ M., Lovell J.\ E.\ J., Sobolev A.\ M., Ramsdale P.\ D., Godfrey P. D., 2004, MNRAS, 354, 401
%
%
%
%

 
\bibitem[\protect\citeauthoryear{Ellingsen}{2006}]{Ellingsen06}
Ellingsen S.\ P., 2006, ApJ, 638, 241
    

\bibitem[\protect\citeauthoryear{Ellingsen \etal}{2007}]{Ellingsen07}
Ellingsen S.\ P., Voronkov M.\ A., Cragg D.\ M., Sobolev A.\ M., Breen S.\ L., Godfrey P.\ D., 2007, in Chapman J.\ M., Baan W.\ A., eds., Proc. IAU Symp., 242, Astrophysical Masers and their Environments. Cambridge Univ. Press, Cambridge, p. 213

 
 \bibitem[\protect\citeauthoryear{Ellingsen \etal}{2010}]{Ellingsen2010}
Ellingsen S.\ P., Breen S.\ L., Caswell J.\ L., Quinn L.\ J., Fuller G.\ A., 2010, MNRAS, 404, 779

%
%
%
%
%
%

\bibitem[\protect\citeauthoryear{Fontani, Cesaroni \& Furuya}{2010}]{Fontani10}
Fontani F., Cesaroni R., Furuya R.\ S., A\&A, 517, 56

\bibitem[\protect\citeauthoryear{Forster \& Caswell}{1989}]{FC89}
 Forster J.R., Caswell J.L., 1989, A\&A, 213, 339 

%
%
%
%


\bibitem[\protect\citeauthoryear{Gallaway \etal}{2010}]{Gallaway10}
Gallaway M., Thompson M.\ A., Lucas P.\ W., Fuller G.\ A., Caswell J.\ L., Green J.\ A., Voronkov M.\ A., Breen, S.\ L., Quinn L., Ellingsen S.\ P., Avison A., Ward-Thompson D., Cox J, 2010, MNRAS, in prep.

%
%
%
%
 \bibitem[\protect\citeauthoryear{Gaylard, MacLeod \& van der Walt}{1994}]{Gay}
 Gaylard M.\ J., MacLeod G.\ C., van der Walt D.\ J., 1994, MNRAS, 269, 257 
 
%
%
%
%
%
%
%
   
 \bibitem[\protect\citeauthoryear{Goedhart, Gaylard \& van der Walt}{2004}]{Goed}
 Goedhart S., Gaylard M.\  J., van der Walt D.\ J., 2004, MNRAS, 355, 553


 \bibitem[\protect\citeauthoryear{Goedhart \etal}{2005a}]{Goed05a}
 Goedhart S., Minier V., Gaylard M.\  J., van der Walt D.\ J., 2005a, MNRAS, 356, 839 

 \bibitem[\protect\citeauthoryear{Goedhart \etal}{2005b}]{Goed05b}
 Goedhart S., Gaylard M.\ J., Walt D.\ J., 2005b, A\&AS, 295, 197

 \bibitem[\protect\citeauthoryear{Goedhart \etal}{2009}]{Goed09}
 Goedhart S., Langa M. C., Gaylard M.\  J., van der Walt D.\ J., 2009, MNRAS, 398, 995 





%

 \bibitem[\protect\citeauthoryear{Green \etal}{2008}]{GreenLMC}
Green J.\ A. et al., 2008, MNRAS, 385, 948


 \bibitem[\protect\citeauthoryear{Green \etal}{2009}]{Green09}
 Green J.\ A. et al., 2009, MNRAS, 392, 783
 

 \bibitem[\protect\citeauthoryear{Green \etal}{2010}]{GreenMMB10}
Green J.\ A. et al., 2010, MNRAS, 409, 913


 \bibitem[\protect\citeauthoryear{Green \& McClure-Griffiths}{2011}]{GM11}
Green J.\ A., McClure-Griffiths, N., 2011, MNRAS, ({\em submitted})


%
%
\bibitem[\protect\citeauthoryear{Hill \etal}{2005}]{Hill05}
  Hill T., Burton M.\ G., Minier V., Thompson M.\ A., Walsh A.\ J., Hunt-Cunningham M., Garay G., 2005, MNRAS, 363, 405
%
%
%
%
%
\bibitem[\protect\citeauthoryear{Houghton \& Whiteoak}{1995}]{HW95}
Houghton S., Whiteoak J.\ B., 1995, MNRAS, 273, 1033 
%
%
%
%
%
%
%
%
%
%
%
%
%
\bibitem[\protect\citeauthoryear{Kemball, Gaylard \& Nicolson}{1988}]{Kemball88}
Kemball A.\ J., Gaylard M. J., Nicolson G.\ D., 1988, ApJ, 331, L37
%
%

\bibitem[\protect\citeauthoryear{Koo \etal}{1988}]{Koo88}
Koo B.\ C., Williams D.\ R.\ D., Heiles C., Backer D.\ C, 1988, ApJ, 326, 931

%
%
%
%
%

\bibitem[\protect\citeauthoryear{Lewis}{2007}]{Lewis07} 
  Lewis B.\ E., 2007, Honours Thesis University of Tasmania
 
%
%
%
%
%
%
%

 \bibitem[\protect\citeauthoryear{MacLeod, Gaylard \& Kemball}{1993}]{MacLeod93}
MacLeod G.\ C., Gaylard M.\ J., Kemball A.\ J., 1993, MNRAS, 262, 343  
%
%
%
%
%
%
%
%
%
%
%
%
%
%
%
%
%
%
%
%
\bibitem[\protect\citeauthoryear{Minier \etal}{2003}]{Minier03}
 Minier V., Ellingsen S.\ P., Norris R.\ P., Booth R.\ S., 2003, A\&A, 403, 1095
%
%
%
\bibitem[\protect\citeauthoryear{Moscadelli \etal}{2002}]{Mos02}
Moscadelli L. et al., 2002, ApJ., 564, 813
%
%
%
 \bibitem[\protect\citeauthoryear{M{\"u}ller, Menten \& M{\"a}der}{2004}]{Muller04} 
 M{\"u}ller H.\ S.\ P., Menten K.\ M., M{\"a}der H., 2004, A\&A, 428, 1019
%
%
%
%
%

 \bibitem[\protect\citeauthoryear{Norris \etal}{1987}]{Norris1987}
Norris R.\ P., Caswell J.\ L., Gardner F.\ F., Wellington K.\ J., 1987, ApJ, 321, L159
%
 \bibitem[\protect\citeauthoryear{Norris \etal}{1993}]{Norris93}  
Norris R.\ P., Whiteoak J.\ B., Caswell J.\ L., Wieringa M.\ H., Gough R.\ G., 1993. ApJ, 412, 222
%
%
%
%
%
%
 \bibitem[\protect\citeauthoryear{Pandian \etal}{2009}]{Pandian09}
Pandian J.\ D., Menten K.\ M., Goldsmith P.\ F., 2009, ApJ, 706, 1609


 \bibitem[\protect\citeauthoryear{Pestalozzi, Minier \& Booth}{Pestalozzi \etal}{2005}]{Pestalozzi05}
   Pestalozzi M.\ R., Minier V., Booth R.\ S., 2005, A\&A, 432, 737

%
%
%
%
%
%
%
%
%
%
%
%
%
%

\bibitem[\protect\citeauthoryear{Reid \etal}{2009}]{Reid09}
Reid M.\ J., Menten K.\ M., Zheng X.\ W., Brunthaler A., Moscadelli L., Xu Y., Zhang B., Sato M., Honma M., Hirota T., Hachisuka K., Choi Y.\ K., Moellenbrock G.\ A., Bartkiewicz A., 2009, ApJ, 700, 137
%
%
%
%
%
%
%
%
%
%
\bibitem[\protect\citeauthoryear{Sault}{2003}]{Sault03} 
 Sault R.\ J., 2003,``ATCA flux density scale at 12mm", (http://www.atnf.csiro.au/observers/docs/12mm/calibrators.html)
%
%
%
%
%
%
%
%
%
%
%
%
%
%
%
%
%
%
\bibitem[\protect\citeauthoryear{Sutton \etal}{2001}]{Sutton01}
Sutton E.\ C., Sobolev A.\ M., Ellingsen S.\ P., Cragg D.\ M., Mehringer D.\ M., Ostrovskii A.\ B., Godfrey P.\ D., 2001, ApJ, 554, 173
%
%
%
%
%
%
%
%
%
%
%
%
%
%
%
%

 \bibitem[\protect\citeauthoryear{Val'tts}{1998}]{Val'tts}
Val'tts I.\ E., 1998, Astron. Lett., 24, 788

 
%
%
%
%
%
%
%
%
%
%
%

\bibitem[\protect\citeauthoryear{Voronkov \etal}{2011}]{Voronkov11} 
Voronkov M.\ A., Walsh A.\ J., Caswell J.\ L., Ellingsen S.\ P., Breen S.\ L., Longmore S.\ N., Purcell C.\ R., Urquhart J.\ S., 2011, MNRAS, 413, 2339

\bibitem[\protect\citeauthoryear{Walsh \etal}{1997}]{Walsh97}
 Walsh A.\ J., Hyland A.\ R., Robinson G., Burton M.\ G., 1997, MNRAS, 291, 261

\bibitem[\protect\citeauthoryear{Walsh \etal}{1998}]{Walsh98}
 Walsh A.\	J., Burton M.\ G., Hyland A.\ R., Robinson G., 1998, MNRAS, 301, 640
 
 \bibitem[\protect\citeauthoryear{Whiteoak \etal}{1988}]{Whiteoak88}
 Whiteoak J.\ B., Gardner F.\ F., Caswell J.\ L., Norris R.\ P., Wellington K.\ J., Peng R.-S., 1988, MNRAS, 235, 655

 \bibitem[\protect\citeauthoryear{Wu \etal}{2010}]{Wu10}
Wu Y.\ W., Xu Y., Pandian J.\ D., Yang J., Henkel C., Menten K.\ M., Zhang S.\ B., 2010, ApJ, 720, 392

%
%
%
%
%
%
%
%
%
%
%
%
%
%
\bibitem[\protect\citeauthoryear{Xu \etal}{2008}]{Xu08}
 Xu Y., Li J.\ J., Hachisuka K., Pandian J.\ D., Menten K.\ M., Henkel C., 2008, A\&A, 485, 729
%
%
%
%
%
%
%
%
%
\end{thebibliography}
\end{document}